\documentclass{ecai}
\usepackage{graphicx}
\usepackage{latexsym}

\usepackage{amsmath}
\usepackage{amssymb}
\usepackage{mathtools}
\usepackage{amsthm}
\usepackage{bbm}
\usepackage{bm}
\usepackage{autobreak}

\usepackage{algorithm}
\usepackage{algorithmic}

\usepackage{microtype}
\usepackage{graphicx}
\usepackage{subfigure}
\usepackage{booktabs} 
\usepackage{multirow}
\usepackage{utfsym}

\theoremstyle{plain}
\newtheorem{assumption}{Assumption}
\newtheorem{thm}{\bf Theorem}
\newtheorem{lemm}{\bf Lemma}[section]
\newtheorem{fact}{\bf Fact}

\newtheorem{remark}{\bf Remark}
\newtheorem{coro}{\bf Corollary}
\allowdisplaybreaks[4]

\begin{document}

\begin{frontmatter}

\title{Decentralized Local Updates with Dual-Slow Estimation and Momentum-based Variance-Reduction for Non-Convex Optimization}

\author[A]{\fnms{Kangyang}~\snm{Luo}}
\author[B]{\fnms{Kunkun}~\snm{Zhang}}
\author[B]{\fnms{Shengbo}~\snm{Zhang}} 
\author[B]{\fnms{Xiang}~\snm{Li}\thanks{Corresponding Author.}}
\author[B]{\fnms{Ming}~\snm{Gao}}

\address[A, B]{East China Normal University,
Shanghai, China}
\address[]{{\tt\small \{52205901003, 51205903092, 51205901087\}@stu.ecnu.edu.cn, \{xiangli, mgao\}@dase.ecnu.edu.cn}}

\begin{abstract}
Decentralized learning~(DL) has recently employed local updates to reduce the communication cost for general non-convex optimization problems.
Specifically, local updates require each node to perform multiple update steps on the parameters of the local model before communicating with others.
However, most existing methods could be highly sensitive to data heterogeneity (i.e., non-iid data distribution) and adversely affected by the stochastic gradient noise. 
In this paper, we propose DSE-MVR to address these problems.  
Specifically, DSE-MVR introduces a dual-slow estimation strategy that utilizes the gradient tracking technique to estimate the global accumulated update direction for handling the data heterogeneity problem; also for stochastic noise, the method uses the mini-batch momentum-based variance-reduction technique. 
We theoretically prove that DSE-MVR can achieve optimal convergence results for general non-convex optimization in both iid and non-iid data distribution settings. 
In particular, the leading terms in the convergence rates derived by DSE-MVR are independent of the stochastic noise  for large-batches or large partial average intervals (i.e., the number of local update steps). 
Further, we put forward DSE-SGD and theoretically justify the importance of the dual-slow estimation strategy in the data heterogeneity setting.
Finally, we conduct extensive experiments to show the superiority of DSE-MVR against other state-of-the-art approaches. 
We provide our code here:
https://anonymous.4open.science/r/DSE-MVR-32B8/.
\end{abstract}

\end{frontmatter}

\section{Introduction}
Rapid dataset scaling~\cite{krizhevsky2012imagenet, sun2017revisiting} is the main issue that has attracted significant attention in the field of data science.
Distributed learning,
which is built on data-parallel training, has become an effective approach 
to handle the problem.
In distributed learning, 
each node can access all or part of the training data 
and then collaboratively communicate the model updates with others.
The classical distributed learning paradigm is \emph{centralized learning} with a parameter server~\cite{li2014communication}, 
where each node sends local 
parameter
updates to the server 
for aggregation, and then the server returns the aggregated results to each node. 
In contrast, 
\emph{decentralized learning} (DL)~\cite{lian2017can,  xin2021hybrid, wang2021cooperative} does not need the
server and each node in the network simply communicates its updates with directly-connected neighbors. 

Compared with centralized learning, 
DL
has been shown to effectively reduce the communication overhead, 
and  
improve the computation efficiency and system robustness~\cite{assran2019stochastic, guo2021hybrid, sun2020improving, wang2019slowmo}. 
As a result, 
DL has gained significant attention from both academia and industry in recent years,
which has become a promising paradigm for distributed learning. 
However, 
the high synchronization cost for communication remains a key challenge that 
restricts the wide applicability of decentralized learning~\cite{koloskova2020unified, nadiradze2021asynchronous}. 
To solve the issue,
some recent works~\cite{wang2019slowmo, koloskova2020unified, nadiradze2021asynchronous} have been proposed to show that the utilization of \emph{local updates} in DL 
can reduce the communication cost. 
Specifically,
local updates require each node to perform multiple update steps on the parameters of the local model before communicating with others.
In particular,
there are methods~\cite{li2019communication, koloskova2020unified, wang2021cooperative, qin2021communication} that directly use SGD to perform local updates under the DL framework.
Despite the success,
it has been pointed out that 
SGD could inject stochastic noise in training~\cite{johnson2013accelerating, nesterov2003introductory, wang2019spiderboost} and is very sensitive to data heterogeneity (i.e., non-iid data distribution)~\cite{guo2021hybrid, karimireddy2020scaffold, rothchild2020fetchsgd},
which could adversely affect the model performance. 


\begin{table*}[h]
    \centering
    \caption{The comparison between our proposed methods and other state-of-the-arts. For non-convex problems, our analysis improves all prior convergence rates in the iid and non-iid data distribution settings. Due to the space limitation, we use comm. to denote the communication rounds. Moreover, $\bigtriangleup$~($\bigtriangledown$) denotes the convergence rates under the non-iid (iid) setting. Note that, a.) $\|\nabla f_i(\bm{x}, \bm{\xi}^{(i)})\|^2 \leq G$ for any $\bm{x}\in \mathbbm{R}^d$. b.) Here, SlowMo uses Local-SGD as the inner optimizer. c.) $\Lambda_1 = \lambda^2(1-\lambda^2)^{-3/2}$, $\Lambda_2 = \lambda^2(1-\lambda^2)^{-2}$, $F_0 = F(\bm{\overline{x}}_{0}) - F^\star$. d.) $\tau=\mathcal{O}(1)>1$, i.e., the algorithms only perform a few local update steps.}  %
    \resizebox{1.8\columnwidth}{!}{
    \begin{tabular}{llcl}
        \toprule
        Methods & Convergence to $\epsilon$-accuary  & Non-iid. &  Comm. \\
        \midrule
        DSGD~\cite{lian2017can}  & $\Tilde{\mathcal{O}}\left(\sigma^4N^{-1}\epsilon^{-2}+N(\sigma^2+\varsigma^2)(1-\lambda)^{-2}\epsilon^{-1} \right)$  & $\bigtriangleup$  &  $\mathcal{O}(T)$    \\
        GT-DSGD~\cite{xin2021improved}  & $\Tilde{\mathcal{O}}\left(\sigma^4N^{-1}\epsilon^{-2}+N\sigma^2\Lambda_1^2\lambda^{-2}\epsilon^{-1}\right)$  & $\bigtriangledown$  &  $\mathcal{O}(T)$  \\
        DLSGD~\cite{li2019communication}  & $\Tilde{\mathcal{O}}\left(\sigma^4N^{-1}\epsilon^{-2}+N(\sigma^2+\varsigma^2)\epsilon^{-1} \right)$  & $\bigtriangleup$  &  $\mathcal{O}(T/\tau)^{d.}$  \\
        \multirow{2}{*}{DSE-SGD(Ours)} 
        & $\Tilde{\mathcal{O}}\left(\sigma^4N^{-1}\epsilon^{-2}+N(\sigma^2+\varsigma^2)\epsilon^{-1}+N\sigma^2\left(\Lambda_1^2+\Lambda_2^2\right)\epsilon^{-1}\right)$ & $\bigtriangleup$  & $\mathcal{O}(T/\tau)^{d.}$      \\
        & $\Tilde{\mathcal{O}}\left(\sigma^4+\varsigma^4(1+\Lambda_2^4)N\epsilon^{-2}+N^{1/3}\sigma^{8/3}(1+\Lambda_1^{8/3})\epsilon^{-4/3}\right)$ & $\bigtriangleup$  &  $\mathcal{O}(T^{3/4}N^{3/4})$ \\
        \midrule
        GT-HSGD~\cite{xin2021hybrid}  & $\Tilde{\mathcal{O}}\left(\sigma^3N\epsilon^{-3/2}+N^{1/2}\sigma^{3/2}\Lambda_1^{3/2}\lambda^{-3/2}\epsilon^{-3/4} \right)$  & $\bigtriangledown$  & $\mathcal{O}(T)$   \\
        DSE-MVR(Ours)     & $\Tilde{\mathcal{O}}\left(F_0^{3/2}N\epsilon^{-3/2}+ \sigma^{3/2}N^{-1/4}\epsilon^{-3/4} + N\sigma\left(1+ \Lambda_1\right) \epsilon^{-1/2} \right)$ & $\bigtriangledown$  & $\mathcal{O}(T/\tau)^{d.}$ \\ 
       \midrule
        QG-DSGDm\cite{lin2021quasi}     & $\Tilde{\mathcal{O}}\left(\sigma^2N^{-1}\epsilon^{-2}+\sigma^2+\varsigma^2(1-\lambda^2)^{-1}\epsilon^{-3/2}\right)$  & $\bigtriangleup$  &  $\mathcal{O}(T)$ \\
       DecentLaM~\cite{yuan2021decentlam} & $\Tilde{\mathcal{O}}\left(\sigma^4N\epsilon^{-2}+N\varsigma^2\epsilon^{-1} \right)$ & $\bigtriangleup$  &  $\mathcal{O}(T)$ \\
        PD-SGDM$^{a.}$~\cite{gao2020periodic}     & $\Tilde{\mathcal{O}}\left(\sigma^4N^{-1}\epsilon^{-2}+G^4(1-\lambda)^{-4}N^{-2(2v-1)}\epsilon^{-2} \right)(v>0)$       & $\bigtriangledown$  &  $\mathcal{O}(T^{3/4}N^v)$ \\
        ADmSGD~\cite{yu2019linear} &  $\Tilde{\mathcal{O}}\left((\sigma^4+\varsigma^4)N^{-1}\epsilon^{-2}+N^{1/3}\sigma^{8/3}\epsilon^{-4/3} \right)$  & $\bigtriangleup$  &  $\mathcal{O}(T^{3/4}N^{3/4})$ \\
        SlowMo$^{b.}$ \cite{wang2019slowmo}     & $\Tilde{\mathcal{O}}\left(\sigma^4N^{-1}\epsilon^{-2}+N(\sigma^2+\varsigma^2)\epsilon^{-1}\right)$  & $\bigtriangleup$  &  $\mathcal{O}(T/\tau)^{d.}$ \\
        \multirow{2}{*}{DSE-MVR $^{c.}$ (Ours)}   & $\Tilde{\mathcal{O}}\left(F_0^{2}N^{-1}\epsilon^{-2}+\sigma N^{-1/2}\epsilon^{-1/2} + N^{1/3}\sigma^{2/3} (1+\Lambda_1^{2/3})\epsilon^{-1/3} + N\varsigma^2\Lambda_2^2\epsilon^{-1} \right)$       & $\bigtriangleup$  & $\mathcal{O}(T/\tau)^{d.}$ \\
        & $\Tilde{\mathcal{O}}\left((F_0^2 + \varsigma^4(1 +\Lambda_2^4))N^{-1}\epsilon^{-2}+\sigma^{8/7}N^{-1}\epsilon^{-4/7}  + \sigma^{8/9}(1+ \Lambda_1^{8/9})N^{-5/9}\epsilon^{-4/9}\right)$       & $\bigtriangleup$  &  $\mathcal{O}(T^{3/4}N^{3/4})$ \\
        \bottomrule
    \end{tabular}}
    \label{tab_rate:}
\end{table*}

In this paper, 
we aim to 
mitigate the influence of data heterogeneity and stochastic gradient noise with local updates in DL.
Specifically, we focus on a decentralized network $\mathcal{G}=(\mathcal{V}, \mathcal{E})$, where $\mathcal{V}$ is a set of nodes and $\mathcal{E}$ is a set of edges that represent the connectivity between nodes.
Each node can send/receive information only from its adjacent neighbors.
Generally, the communication between nodes is parameterized by a mixing matrix $\bm{W}$~(e.g., a weighted adjacency matrix of $\mathcal{G}$) for simplicity, 
where $w_{ij}>0$ implies that there exist communications between nodes $i$ and $j$; $w_{ij}=0$, otherwise.
In our setting, all nodes jointly solve a smooth non-convex decentralized optimization problem: 
\begin{equation}
    \label{prob1:}
    \min_{\bm{x} \in \mathbbm{R}^{d}} F(\bm{x})=\frac{1}{N} \sum_{i=1}^{N} \underbrace{\mathbbm{E}_{\bm{\xi}_{r}^{(i)} \sim \mathcal{D}_{i}} [f_{i}(\bm{x}; \bm{\xi}^{(i)}) ]}_{:={f_{i}(\bm{x})}},
\end{equation}
where $N$ is the number of nodes, 
and $f_{i}: \mathbbm{R}^{d} \rightarrow \mathbbm{R}$ denotes the local loss function of the $i$-th node. 
The stochastic function $f_{i}(\bm{x})$ can generate the stochastic gradients $\nabla f_{i}(\bm{x}; \bm{\xi}_{r}^{(i)})$, where $\mathcal{D}_i$ is a local training data distribution on the $i$-th node.  
Note that $\mathcal{D}_i$ is a uniform distribution over the local training data on the $i$-th node, which could be different on various nodes. 
In the offline setting,  $f_i$ is a deterministic function when $f_i(\bm{x}):=\frac{1}{n_i}\sum_{r=1}^{n_i}f_i(\bm{x},\bm{\xi}_{r}^{(i)})$ holds,  
where $n_i$ is the number of local samples at the $i$-th node.  

To solve problem (\ref{prob1:}),
in this paper, we propose a new decentralized local-updates-based method with \textbf{D}ual-\textbf{S}low \textbf{E}stimation and \textbf{M}omentum-based \textbf{V}ariance \textbf{R}eduction, namely DSE-MVR. 
To mitigate the effect of data heterogeneity in communication rounds, DSE-MVR introduces a {dual-slow estimation strategy}: \emph{slow gradient tracking (SGT)} and \emph{slow partial averaging (SPA)}.
Specifically, SGT uses the gradient tracking technique ~\cite{xin2021improved} to estimate the global average  accumulated gradient descent direction for each node after it completes local updates, while SPA employs the estimates to re-update the local model parameters of the previous communication round for each node and communicates the updates with connected neighbors.
On the other hand, to alleviate the influence of stochastic noise, DSE-MVR estimates the gradient descent directions of local models by sampling with multiple replacements and using the momentum-based variance-reduction method~\cite{cutkosky2019momentum}. 
To further justify the importance of the dual-slow estimation strategy in the non-iid data distribution setting, we replace MVR with SGD and put forward DSE-SGD, which uses mini-batch SGD to estimate the local update direction.

In a nutshell, we summarize our main contributions as follows:
\begin{itemize}
    \item  We propose 
    DSE-MVR, which aims to solve the smooth non-convex decentralized optimization problem (\ref{prob1:}). The method can handle the challenges of data heterogeneity and stochastic noise that arise in DL with local updates for low communication cost. For completeness, we provide the convergence analysis of DSE-MVR under general assumptions about data distribution and decentralized network topology. To the best of our knowledge, DSE-MVR convergence results are optimal compared with other current state-of-the-art methods. 
    \item  We justify the importance of the dual-slow estimation strategy. 
    For fairness, we replace MVR with SGD to remove the benefits introduced by MVR and put forward DSE-SGD.  
    Our theoretical analysis on DSE-SGD shows that the dual-slow estimation strategy can adapt well to data heterogeneity.
    \item We conduct extensive  experiments to validate our theoretical analysis. Experimental results on two commonly used benchmark datasets MNIST and CIFAR-10 show that DSE-MVR is highly competitive compared with other state-of-the-art baselines.
\end{itemize}

\section{Related Work}

Decentralized learning (DL) has been widely studied in the fields of optimization~\cite{yuan2016convergence}, 
signal processing~\cite{nedic2020distributed} and control community ~\cite{nedic2014distributed} for decades. 
Recently,
DL in deep learning has also gained much attention~\cite{lian2017can}. 
Despite the success,
high communication cost~\cite{ wang2021cooperative, qin2021communication} and data heterogeneity~\cite{vogels2021relaysum, koloskova2020unified} are the two main issues in DL. 
Further,
some methods (e.g., decentralized SGD~\cite{lian2017can}) 
use SGD to perform local updates.
However, it has been pointed out that 
SGD could inject stochastic gradient noise in training~\cite{johnson2013accelerating,  wang2019spiderboost}, 
which leads to a slow convergence speed and is also a concern in DL.

To mitigate the influence of stochastic noise induced by SGD, 
some methods~\cite{gao2020periodic, yu2019linear, lin2021quasi, yuan2021decentlam} apply
various SGD variants to DL, 
such as momentum techniques (MT)~\cite{nesterov2003introductory, kingma2014adam}, 
variance reduction (VR)~\cite{johnson2013accelerating,   nguyen2017sarah} 
and momentum-based variance-reduction techniques (MVR)~\cite{cutkosky2019momentum}, 
to improve the model training efficiency.
For example,
DmSGD~\cite{assran2019stochastic} introduces momentum SGD for DL to estimate local updates
and ADmSGD~\cite{yu2019linear} further improves the performance 
by additionally synchronizing local momentum buffers.
These approaches can achieve a linear speedup (i.e., $\mathcal{O}(N^{-1}\epsilon^{-2})$) in the stochastic non-convex heterogeneous setting that matches the centralized mini-batch SGD~\cite{dekel2012optimal}.
Further, 
there are also 
methods~\cite{xin2021hybrid, xin2020variance,  sun2020improving}
that leverage gradient tracking techniques~\cite{xin2021improved, nedic2017achieving} to improve the model efficiency.
For example,
D-GET~\cite{sun2020improving}, 
and GT-HSGD~\cite{xin2021hybrid} have been shown to achieve lower oracle complexity in the non-convex homogeneous setting by
combining gradient tracking techniques with SGD variants. 


Further,
it has been shown in~\cite{tang2018d, koloskova2021improved,  zhang2019decentralized, pu2021distributed} 
that
gradient tracking can also be used to handle the data heterogeneity problem in DL.
For example,
GT~\cite{koloskova2021improved}
uses the gradient tracking technique to estimate
the global gradient direction and 
adds it to decentralized SGD.
Also,
there exist methods base on other techniques.
For example,
Lin et al. \cite{lin2021quasi} proposed Quasi-Global momentum, which locally approximates the global update direction with the goal of mitigating the effects of the non-iid data distribution.
DecentLaM~\cite{yuan2021decentlam}
mitigates the effect of data heterogeneity by eliminating the momentum-incurred bias in decentralized momentum SGD.

To reduce the communication cost, 
most existing methods can be roughly  divided into three categories.
Specifically, the first type of methods~\cite{koloskova2019decentralized, vogels2020practical} use compression techniques to reduce the communication amount, while that in the second type~\cite{li2019communication, wang2021cooperative, qin2021communication, gao2020periodic} perform local updates to reduce the communication frequency.
There are also methods~\cite{singh2021squarm, nadiradze2021asynchronous} that combine the two techniques.
In this paper, we focus on those based on local updates.
For example, 
~\cite{li2019communication} 
and 
~\cite{qin2021communication}
studied decentralized local SGD 
for local updates
in the strongly convex homogeneous and non-convex heterogeneous settings, respectively.  
PD-SGDM~\cite{gao2020periodic} further integrates momentum SGD into decentralized local SGD and obtains the convergence result $\mathcal{O}(N^{-1}\epsilon^{-2})$ in the non-convex heterogeneous setting.
A recent work~\cite{koloskova2020unified} provides a unified analysis framework for decentralized SGD with local updates and changing network topologies. 
SlowMo~\cite{wang2019slowmo}
performs a slow momentum update on the local model parameters 
in DL 
after multiple local update steps.
While these methods use local updates to 
reduce the communication cost in DL, 
most of them 
only provide convergence analysis on the heterogeneous setting but 
fail to alleviate the adverse effect of data heterogeneity.
To our best knowledge, 
our work is the first to
mitigate the influence of 
data heterogeneity and stochastic gradient noise with local updates.
Finally, we summarize our proposed methods and other state-of-the-arts in terms of convergence rates~(see Table~\ref{tab_rate:}).


\section{Preliminaries}
This section
describes notations used in this paper and general assumptions for theoretical proofs.
\subsection{Notations}
We use lowercase bold letters to denote vectors and uppercase ones to denote matrices.  
The $\bm{1}_N$ is the $N$-dimensional vector of all ones. 
Given any
positive integer $N \in \mathbbm{N}$, we denote $[N] : = \{1,2,\cdots, N\}$. We use $\|\cdot\|$ to denote the Euclidean norm of a vector or the spectrum norm of a matrix.  
The Frobenius norm of a matrix is denoted as $\|\cdot\|_F$. 
Let $\tau(t)=\max \left\{l:l\leq t \ {\rm and} \ {\rm mod} {(l,\tau)} = 0 \right\}$ denote the previous communication round closest to iteration $t$, 
where $\tau$ is partial average interval.


\subsection{Assumptions}
\label{Assum:}
Following~\cite{lian2017can, 
koloskova2020unified}, 
our theoretical results
are based on the 
following assumptions for problem (\ref{prob1:}). 
\begin{assumption}
\label{Global_Function_Below_Bounds:}
{\rm (Global Function Below Bounds).} 
Set $F^*=\inf_{\bm{x}\in \mathbbm{R}^d} F(\bm{x})$ and $F^*>-\infty$.
\end{assumption}

\begin{assumption}
\label{L_smooth:}

{\rm ($L$-smooth).}
$\forall i \in [N]$,
the local function $f_i(\cdot)$ and local stochastic function $f_i(\cdot, \bm{\xi}^{(i)})$ with $\bm{\xi}^{(i)} \sim  \mathcal{D}_i$ are differentiable, and 
there exist constant $L$ such that for any $\bm{x},\bm{y} \in \mathbbm{R}^d$,
\begin{equation}
    \label{LG}
    \left \| \nabla f_i(\bm{x})-\nabla f_i(\bm{y}) \right \| \leq L \|\bm{x}-\bm{y}\|,
\end{equation}
and 
\begin{equation}
    \label{LSG}
    \mathbbm{E}\left \| \nabla f_i(\bm{x},\bm{\xi}^{(i)})-\nabla f_i(\bm{y},\bm{\xi}^{(i)}) \right \| \leq L \mathbbm{E}\|\bm{x}-\bm{y}\|.
\end{equation}
\end{assumption}

\begin{assumption}
\label{UG_BV:} 
{\rm (Unbiased gradients and bounded stochastic noise).}
The local stochastic gradients computed at node $i$ are unbiased: for some constant $\sigma^2 > 0$ and any $i \in [N]$,
\begin{equation}
    \label{Unbiased_Gradient}
    \mathbbm{E}[\nabla f_i(\bm{x},\bm{\xi}^{(i)})-\nabla f_i(\bm{x})]=0, \bm{\xi}^{(i)} \sim  \mathcal{D}_i,
\end{equation}
and the stochastic noise is bounded: 
\begin{equation}
    \label{intra-node_bounded_variance}
    \mathbbm{E}\|\nabla f_i(\bm{x},\bm{\xi}^{(i)})-\nabla f_i(\bm{x})\|^2 \leq \sigma^{2}, \bm{\xi}^{(i)} \sim  \mathcal{D}_i.
\end{equation}
\end{assumption}

\begin{assumption}
\label{UG_BV:1} 
{\rm (Data heterogeneity).}
The degree of heterogeneity of the data distribution across nodes can be quantified as follows: for some constant $ \varsigma^2 \geq 0$,
\begin{equation}
    \label{inter-node_bounded_variance}
    \frac{1}{N}\sum_{i=1}^{N}\|\nabla f_i(\bm{x})-\nabla F(\bm{x})\|^2 \leq \varsigma^{2}, \forall i \in [N].
\end{equation}
\end{assumption}

\begin{assumption}
\label{CN:} 
{\rm (Communication Network).} The network is strongly connected and the mixing matrix $\bm{W} \in \mathbbm{R}^{N \times N}$ admits a doubly stochastic weight matrix, i.e., $\bm{W}\bm{1}_N = \bm{1}_N$, $\bm{1}_N^T\bm{W} = \bm{1}_N^T$. We set $\lambda = \|\bm{W}-\bm{Q}\|$ where $\bm{Q} = \frac{1}{N} \bm{1}_N \bm{1}^T_N$. 
Note that $\lambda \in (0,1)$. 
Then for any matrix $\bm{X} \in \mathbbm{R}^{d \times N}$ and $\overline{\bm{X}} = \bm{X} \bm{Q}$, 
the following inequality about the consensus distance 
holds:
\begin{equation}
   \label{consenus_dist:}
   \|\bm{XW}-\overline{\bm{X}}\|_F^2 \leq \lambda^2 \|\bm{X}-\overline{\bm{X}}\|_F^2.
\end{equation}
\end{assumption}

Note that 
Assumptions \ref{L_smooth:} and \ref{CN:} 
are commonly used in the analysis of 
decentralized learning \cite{lian2017can, vogels2021relaysum,  koloskova2020unified}.
In
Assumption \ref{L_smooth:},
the 
inequality (\ref{LSG}) requires the stochastic gradient functions $\nabla f_i(\cdot, \bm{\xi}^{(i)})$ 
w.r.t. the input parameters $\bm{x}$ and $\bm{y}$ are $L$-smooth on average, 
which is stronger than
the inequality (\ref{LG}). 
Moreover,
the inequality
(\ref{consenus_dist:})
in Assumption~\ref{CN:} 
ensures that the consensus distance 
decreases linearly after each communication step. 
The bounds (\ref{intra-node_bounded_variance}) and (\ref{inter-node_bounded_variance}) in Assumptions \ref{UG_BV:} and  \ref{UG_BV:1} quantify  intra- and inter-node variances,
which 
characterize 
stochastic noise and 
data heterogeneity, respectively~\cite{chen2021accelerating, wang2020tackling, bottou2018optimization}. 

\section{Algorithm and Convergence Results}

\subsection{Algorithm  description}
We first introduce the DSE-MVR algorithm.
Recall that
DSE-MVR adopts the dual-slow estimation strategy to alleviate the adverse effect of data heterogeneity.
Meanwhile,
it applies the 
momentum-based variance-reduction (MVR) technique to mitigate the influence of stochastic gradient noise injected by SGD.
The pseudocode of DSE-MVR is given in 
Algorithm~\ref{alg_momen:} and 
the overall procedure is summarized as follows.
For each node in the network,
we first perform $\tau$ local update steps (lines 12-15).
Specifically,
in the $t$-th step (iteration),
the $i$-th node updates the local model parameters and computes $\bm{x}_{t+1}^{(i)}$,
which 
is an estimate of the stationary point of the loss function $F$.
Then it computes the local update direction $\bm{v}_{t+1}^{(i)}$ by MVR based on the mini-batch gradients at 
$\bm{x}_{t+1}^{(i)}$ and $\bm{x}_{t}^{(i)}$ (lines 13-15). 
After $\tau$ local steps (i.e., when 
${\rm mod} (t+1,\tau) = 0$),
each node communicates its local updates with directly-connected neighbors.
For each node,
it first
calculates the accumulated gradient descent direction $\bm{h}_{t+1}^{(i)}$ 
in the previous $\tau$ steps
(line 7), 
which is further used to compute slow buffer $\bm{y}_{\tau(t)}^{(j)} + \bm{h}_{t+1}^{(j)}-\bm{h}_{\tau(t)}^{(j)}$ (line 8).
Note that $\tau(t)$ is the previous communication round closest to $t$.
Then
each node receives and aggregates the slow buffer from its neighbor nodes, which generates $\bm{y}_{t+1}^{(i)}$, the estimate of the global average accumulated descent direction. 
This process is slow gradient tracking (SGT).
After that, 
based on $\bm{y}_{t+1}^{(i)}$,
each node updates its local model parameters $\bm{x}_{t+1}^{(i)}$
by 
aggregating $\bm{x}_{\tau(t)}^{(i)}-\bm{y}_{t+1}^{(i)}$ from its neighbors (line 9). 
We call this process slow partial average (SPA). 
Before stepping into the next local update step, 
we reset the update direction for each node by calculating the full gradient on $\bm{x}_{t+1}^{(i)}$ (line 10). 
We repeat the above local updates and update communication process until the model converges or reaches pre-defined $T$ iterations.

To further show the effectiveness of the dual-slow estimation strategy,
we replace MVR with SGD to remove the gains brought by MVR
and put forward the DSE-SGD algorithm.
In particular,
DSE-SGD can be considered as a special case of DSE-MVR by setting $\alpha_{t+1} = 1$ and removing the full gradient calculation in the communication round.
The pseudocode is summarized in Alg.~\ref{alg:}
of Appendix~\ref{code_DSE_SGD:}.

\begin{algorithm}[!t]
  \caption{The DSE-MVR algorithm }
  \label{alg_momen:}
\begin{algorithmic}[1]
  \STATE {\bfseries Input:}
   Learning rate $\gamma_t > 0$, control parameter $\alpha_t \in (0, 1)$, weight matrix $\bm{W}$, batch size $b$, partial average interval $\tau>0$.
  \STATE Initial state $\bm{x}_{0}^{(i)}=\bm{x}_0$, $\bm{y}_{0}^{(i)}=\bm{h}_{0}^{(i)}=\bm{0}$, $i\in [N]$. 
  \STATE $\bm{v}_{0}^{(i)} = \frac{1}{n_i} \sum \limits_{r=1}^{n_i} \nabla f_i(\bm{x}_{0}^{(i)};\bm{\xi}_{r}^{(i)})$
  \FOR{$t=0,\cdots, T-1$}
    \FOR{$i=1,\cdots, N$}
        \STATE $\bm{x}_{t+\frac{1}{2}}^{(i)}= \bm{x}_{t}^{(i)}-\gamma_{t} \bm{v}_{t}^{(i)}$ 
        \IF{$\mod(t+1,\tau) = 0$} 
            \STATE $\bm{h}_{t+1}^{(i)} =  \bm{x}_{\tau(t)}^{(i)}-\bm{x}_{t+\frac{1}{2}}^{(i)}$ 
            \STATE $\bm{y}_{t+1}^{(i)}= \sum \limits _{j \in \mathcal{N}_{i}} w_{ij} (\bm{y}_{\tau(t)}^{(j)} + \bm{h}_{t+1}^{(j)}-\bm{h}_{\tau(t)}^{(j)})$  
            \STATE $\bm{x}_{t+1}^{(i)}= \sum \limits _{j \in \mathcal{N}_{i}} w_{ij} (\bm{x}_{\tau(t)}^{(j)} - \bm{y}_{t+1}^{(j)})$ 
            \STATE $\bm{v}_{t+1}^{(i)} = \frac{1}{n_i} \sum \limits_{r=1}^{n_i} \nabla f_i(\bm{x}_{t+1}^{(i)};\bm{\xi}_{r}^{(i)})$ 
            
        \ELSE
            \STATE $\bm{x}_{t+1}^{(i)}=\bm{x}_{t+\frac{1}{2}}^{(i)}$
            \STATE $\bm{g}_{t+1}^{(i)} = \frac{1}{b} \sum \limits_{r=1}^b \nabla f_i(\bm{x}_{t+1}^{(i)};\bm{\xi}_{{r}}^{(i)})$,  $\bm{\xi}_{{r}}^{(i)} \sim \mathcal{D}_i$ 
            \STATE $\bm{g}_{t}^{(i)} = \frac{1}{b} \sum \limits_{r=1}^b \nabla f_i(\bm{x}_{t}^{(i)};\bm{\xi}_{{r}}^{(i)})$,  $\bm{\xi}_{{r}}^{(i)} \sim \mathcal{D}_i$ 
            \STATE $\bm{v}_{t+1}^{(i)}=  \bm{g}_{t+1}^{(i)} + (1-\alpha_{t+1})(\bm{v}_{t}^{(i)}-\bm{g}_{t}^{(i)})$ 
        \ENDIF
    \ENDFOR
  \ENDFOR
\STATE {\bfseries Output:}
  $ \overline{\bm{x}}_q$, where $q$ chosen uniformly randomly from $[0,\cdots,T-1]$. \\
\STATE {\bfseries Note that}  $\tau(t)=\max \left\{l:l\leq t \ {\rm and} \ {\rm mod} {(l,\tau)} = 0 \right\}$
\end{algorithmic}
\end{algorithm}

\subsection{Convergence Results for DSE-MVR}

\begin{thm}
    \label{thm:DSE_MVR}
    Under the Assumptions \ref{Global_Function_Below_Bounds:}-\ref{CN:}, 
    if $\forall t \in [0,\cdots,T-1]$, 
    we set 
    $\gamma_t = \gamma \leq \min \left\{ \frac{1}{8L\tau}, \frac{(1-\lambda^2)^2}{64\sqrt{6}\lambda^2L\tau}\right\}, \ \alpha_t=\alpha=\frac{32L^2\gamma^2}{Nb},$
    the iterates $\overline{\bm{x}}_q$ generated by algorithm \ref{alg_momen:} 
    satisfy that 
    \begin{align*}
        \mathbbm{E}&[\|\nabla F(\bm{\overline{x}}_{q})\|^{2}] \\
            & \overset{}{\leq} 
                \frac{2(F(\bm{\overline{x}}_{0})-F^\star)}{\gamma T} +\frac{8\alpha^2 (\tau-1)  \sigma^2}{Nb} +\frac{64 L^2 \gamma^2(\tau-1)  \varsigma^2}{Nb} \notag \\
                & + \frac{32\alpha^2L^2\gamma^2\tau(\tau-1)^2\sigma^2}{b} + 96L^2\gamma^2\tau(\tau-1)\varsigma^2 \notag \\
                &   + \frac{4096\lambda^4\alpha^2L^2\gamma^2\tau(\tau-1)^2  \sigma^2}{(1-\lambda^2)^3b} +\frac{8192\lambda^4L^2\gamma^2\tau^2 \varsigma^2}{(1-\lambda^2)^4},
    \end{align*}
    where the expectation $\mathbbm{E}[\cdot]$ is w.r.t. the stochasticity of the algorithm and that $\mathbbm{E}\|\nabla F(\bm{\overline{x}}_{q})\|^{2} = \frac{1}{T}\sum_{t=0}^{T-1}\mathbbm{E}\|\nabla F(\bm{\overline{x}}_{t})\|^{2}$ since $q$ is chosen uniformly randomly from $[0,\cdots,T-1]$.
\end{thm}
The detail proof of Theorem \ref{thm:DSE_MVR} can be found in Appendix~\ref{CP_alg_momen:}. 
Next we discuss Theorem \ref{thm:DSE_MVR} by setting specific values of the input parameters $\gamma$, $\alpha$, $b$ and $\tau$ in Algorithm \ref{alg_momen:} under the settings of iid data distribution and non-iid data distribution, respectively. 
We first consider the case of iid data distribution, i.e., $\varsigma^2=0$ (see Assumption~\ref{UG_BV:1}) and derive Corollary~\ref{Th2_coro_2:}. 
Note that $\Lambda_1 = \lambda^2/(1-\lambda^2)^\frac{3}{2}$ and $\Lambda_2 = \lambda^2/(1-\lambda^2)^{2}$.

\begin{coro}
    \label{Th2_coro_2:}
    Under the Assumptions \ref{Global_Function_Below_Bounds:}-\ref{CN:},
    
    {\rm 1)} if we set $\gamma = N^\frac{2}{3}L^{-1}T^{-\frac{1}{3}}$, $\alpha = N^\frac{1}{3}T^{-\frac{2}{3}}$, $b = 1$, $\tau = \mathcal{O}\left(1\right)>1$ in Algorithm \ref{alg_momen:}, then for any $T \geq \max \{512N^2\tau^3, \frac{192^3N^2\lambda^6\tau^3}{(1-\lambda^2)^6}\}$, we have:
    $\Tilde{\mathcal{O}}\left(\frac{F_0^\frac{3}{2}}{N \epsilon^\frac{3}{2}} + \frac{\sigma^\frac{3}{2}}{N^\frac{1}{4}\epsilon^\frac{3}{4}}   + \frac{N\sigma\left(1+ \Lambda_1\right)}{\epsilon^\frac{1}{2}}  \right)$ iterations are needed to make $\mathbbm{E}\|\nabla F(\bm{\overline{x}}_{q})\|^{2}\leq \epsilon$ hold;
     
     {\rm 2)} if we set $\gamma = b^\frac{1}{2}N^\frac{1}{2} L^{-1}T^{-\frac{1}{2}}$, $\alpha = T^{-1}$ in Algorithm \ref{alg_momen:}, 
     then for any $T \geq \max \{64Nb\tau^2, \frac{192^2N\lambda^4b\tau^2}{(1-\lambda^2)^4}\}$, we have:
     
     for $b=1$ and $\tau =\mathcal{O}(T^\frac{1}{2}N^{-\frac{2}{3}})$,
    $\Tilde{\mathcal{O}}\left(\frac{F_0^2}{N \epsilon^{2}}+\frac{ \sigma^\frac{4}{3}}{N^\frac{10}{9}\epsilon^\frac{2}{3}}+\frac{\sigma^\frac{4}{3} (1+\Lambda_1^\frac{4}{3} )}{N^\frac{2}{3}\epsilon^\frac{2}{3}}\right)$
    iterations are needed to make $\mathbbm{E}\|\nabla F(\bm{\overline{x}}_{q})\|^{2}\leq \epsilon$ hold, and

    for $b=\mathcal{O}(T^\frac{1}{3}N^{-1})$ and  $\tau = \mathcal{O}(1) > 1$, 
     $\Tilde{\mathcal{O}}\left(\frac{F_0^\frac{3}{2}}{\epsilon^\frac{3}{2}}+\frac{\sigma^\frac{6}{7} }{\epsilon^\frac{3}{7}}+ \frac{N^\frac{1}{3}\sigma^\frac{2}{3} (1+ \Lambda_1^\frac{2}{3} )}{\epsilon^\frac{1}{3}}\right)$
    iterations are needed to make $\mathbbm{E}\|\nabla F(\bm{\overline{x}}_{q})\|^{2}\leq \epsilon$ hold
    , where $F_0 = F(\bm{\overline{x}}_{0}) - F^*$.
\end{coro}

\begin{remark}
    From Corollary~\ref{Th2_coro_2:}, we see that when the iteration $T$ is large enough, the leading terms are all independent of the stochastic noise $\sigma^2$ and can achieve the optimal convergence results to our best knowledge (see Table~\ref{alg_momen:}). Specifically,  when $b=1$ and $\tau = \mathcal{O}\left(1\right)>1$,
     the convergence result $\Tilde{\mathcal{O}}(\epsilon^{-\frac{3}{2}} + \epsilon^{-\frac{3}{4}} + \epsilon^{-\frac{1}{2}})$ of DSE-MVR outperforms that of GT-HSGD~\cite{xin2021hybrid}. 
     Note that DSE-MVR cannot be simply regarded as a local-update-version of GT-HSGD, because the convergence analysis in GT-HSGD cannot be directly extended to multiple local update settings. 
     In this paper, we utilize the gradient tracking technique to handle accumulated gradient updates and present novel convergence analysis. This is also the challenge of our paper.

\end{remark}

In the following, 
we show the convergence results of 
DSE-MVR in the non-iid 
setting, i.e., $\varsigma^2>0$.
\begin{coro}
    \label{Th2_coro_3:}
    Under the Assumptions \ref{Global_Function_Below_Bounds:}-\ref{CN:},
      if we choose $\gamma = b^\frac{1}{2}N^\frac{1}{2}L^{-1}T^{-\frac{1}{2}}$, $\alpha = T^{-1}$ in Algorithm \ref{alg_momen:}, then for any $T \geq \max \{64Nb\tau^2, \frac{192^2N\lambda^4b\tau^2}{(1-\lambda^2)^4}\}$, we have :
     
     for $b =1$ and $\tau = \mathcal{O}\left(1\right)\geq1$, $ \Tilde{\mathcal{O}}\bigg(\frac{F_0^2}{N\epsilon^{2}} + \frac{N\varsigma^2\left((\tau-1) +\Lambda_2^2\right)}{\epsilon} + \frac{   \sigma(\tau-1)}{N^\frac{1}{2}\epsilon^\frac{1}{2}} + \frac{N^\frac{1}{3}\sigma^\frac{2}{3}(\tau-1) (1+\Lambda_1^\frac{2}{3})}{\epsilon^\frac{1}{3}} \bigg)$
    iterations are needed to make $\mathbbm{E}\|\nabla F(\bm{\overline{x}}_{q})\|^{2}\leq \epsilon$ hold, and

    for $b =1$ and $\tau = \mathcal{O}(T^\frac{1}{4}N^{-\frac{3}{4}})$,
    $\Tilde{\mathcal{O}}\left(\frac{F_0^2 + \varsigma^4(1 +\Lambda_2^4)}{N\epsilon^{2}} +\frac{\sigma^\frac{8}{7}}{N\epsilon^\frac{4}{7}}  + \frac{\sigma^\frac{8}{9}(1   + \Lambda_1^\frac{8}{9})}{N^\frac{5}{9}\epsilon^\frac{4}{9}} \right)$
    iterations are needed to make $\mathbbm{E}\|\nabla F(\bm{\overline{x}}_{q})\|^{2}\leq \epsilon$ hold, and
    
    for $b=\mathcal{O}(T^\frac{1}{3}N^{-1})$ and $\tau = \mathcal{O}\left(1\right)\geq1$,
    $\Tilde{\mathcal{O}}\bigg(\frac{F_0^\frac{3}{2} + \varsigma^3((\tau-1)+\Lambda_2^3)}{\epsilon^\frac{3}{2}} + \frac{ \sigma^\frac{6}{7}(\tau-1)}{\epsilon^\frac{3}{7}} + \frac{N^\frac{1}{3}\sigma^\frac{2}{3}(1 + \Lambda_1^\frac{2}{3} )(\tau-1)}{\epsilon^\frac{1}{3}} \bigg)$
    iterations are needed to make $\mathbbm{E} \|\nabla F(\bm{\overline{x}}_{q})\|^{2} \leq \epsilon$ hold
    , where $F_0 = F(\bm{\overline{x}}_{0}) - F^*$.
\end{coro}

\begin{remark}
    From 
    Corollary~\ref{Th2_coro_3:}, 
    DSE-MVR outperforms or performs comparably with the state-of-the-art methods (see Table \ref{tab_rate:}) in terms of convergence rates for various choices of  $b$ and $\tau$. This is mainly
    because the leading terms derived by DSE-MVR are not affected by 
    the 
    stochastic noise $\sigma^2$.
\end{remark}


\subsection{Convergence Results for DSE-SGD}

\begin{thm}
    \label{thm:alg}
    Under the Assumptions \ref{Global_Function_Below_Bounds:}-\ref{CN:}, if $\forall t \in [0,\cdots,T-1]$, 
    we set 
    $\gamma_t = \gamma \leq \min \left\{ \frac{1}{4\sqrt{2}L\tau}, \frac{(1-\lambda^2)^2}{32\sqrt{6}\lambda^2L\tau}\right\},$
     the iterates $\overline{\bm{x}}_q$ generated by Algorithm \ref{alg:}
     satisfy that
    \begin{align*}
        \mathbbm{E}&\|\nabla F(\bm{\overline{x}}_{q})\|^{2} \overset{}{\leq} \frac{2(F(\bm{\overline{x}}_{0}) - F^\star)}{\gamma T}  + \frac{L\gamma\sigma^2}{Nb} \\
        & \quad \quad \quad +  \frac{12L^2\gamma^2(\tau-1) \sigma^2}{b} +24L^2\gamma^2\tau(\tau-1)\varsigma^2 
           \\
         & \quad \quad \quad \quad +  \frac{768\lambda^4 L^2 \gamma^2 \tau \sigma^2}{(1-\lambda^2)^3b} +\frac{1536\lambda^4L^2\gamma^2\tau^2 \varsigma^2}{(1-\lambda^2)^4},   
    \end{align*}
    where the expectation $\mathbbm{E}[\cdot]$ is w.r.t. the stochasticity of the algorithm and that $\mathbbm{E}\|\nabla F(\bm{\overline{x}}_{q})\|^{2} = \frac{1}{T}\sum_{t=0}^{T-1}\mathbbm{E}\|\nabla F(\bm{\overline{x}}_{t})\|^{2}$ since $q$ chosen uniformly randomly from $[0,\cdots,T-1]$.
\end{thm}

The proof of Theorem \ref{thm:alg} can be found in Appendix~\ref{con-sgd:}.
Next, we discuss the statement of Theorem \ref{thm:alg} by setting specific values for the input parameters $\gamma$, $b$ and $\tau$ in Algorithm \ref{thm:alg}.
\begin{coro}
    \label{Th1_coro_1:}
    Under the Assumptions \ref{Global_Function_Below_Bounds:}-\ref{CN:},  if we choose  $\gamma = b^\frac{1}{2}N^\frac{1}{2}T^{-\frac{1}{2}}$ in Algorithm \ref{alg:}, 
    then for any $T \geq \max \{32NL^2b\tau^2, \\ \frac{6144N\lambda^4L^2b\tau^2}{(1-\lambda^2)^4}\}$, we have :
    
    for $b=1$ and  $\tau = \mathcal{O}(1) \geq 1$, 
    $\Tilde{\mathcal{O}}\bigg(\frac{F_0^2+\sigma^4}{N\epsilon^2}+\frac{N(\sigma^2+\varsigma^2)(\tau-1)}{\epsilon}+ \\ \frac{N\sigma^2\left(\Lambda_1^2+\Lambda_2^2\right)}{\epsilon} \bigg)$
     iterations are needed to make $\mathbbm{E}\|\nabla F(\bm{\overline{x}}_{q})\|^{2}\leq \epsilon$ hold, and
     
    for $b=1$ and $\tau =\mathcal{O}(T^\frac{1}{4}N^{-\frac{3}{4}})$,
    $\Tilde{\mathcal{O}}\left(\frac{F_0^2+\sigma^4+\varsigma^4(1+\Lambda_2^4)}{N\epsilon^{2}}+\frac{N^\frac{1}{3}\sigma^\frac{8}{3}\left(1+\Lambda_1^\frac{8}{3}\right)}{\epsilon^\frac{4}{3}}\right)$
    iterations are needed to make $\mathbbm{E}\|\nabla F(\bm{\overline{x}}_{q})\|^{2}\leq \epsilon$ hold, and
    
    for $b=\mathcal{O}(T^\frac{1}{3}N^{-1})$ and  $\tau = \mathcal{O}(1) \geq 1$, 
    $\Tilde{\mathcal{O}}\bigg(\frac{F_0^\frac{3}{2}+\sigma^3+\varsigma^3((\tau-1)+\Lambda_2^3)}{\epsilon^\frac{3}{2}}+ \frac{N\sigma^2\left((\tau-1)+\Lambda_1^2\right)}{\epsilon}\bigg)$
    iterations are needed to make $\mathbbm{E}\|\nabla F(\bm{\overline{x}}_{q})\|^{2}\leq \epsilon$ hold, where $F_0 = F(\bm{\overline{x}}_{0}) - F^*$.
\end{coro}

\begin{remark}
     From these results, we see that the term about $\varsigma^2$ in the convergence rate of DSE-SGD matches with the optimal convergence estimation (see Table \ref{tab_rate:}) when $b=1$ and  $\tau = \mathcal{O}(1) \geq 1$.
     This shows that the dual-slow estimation strategy can adapt well to data heterogeneity, which is also demonstrated by the convergence results of DSE-MVR 
    from Corollary~\ref{Th2_coro_3:}.
\end{remark}

\section{Outline of the Convergence Analysis}
In this section, we outline the proof of Theorem~\ref{thm:DSE_MVR}, while the proof details can be found in the Appendix~\ref{CP_alg_momen:}. 
Note that the proof of Theorem ~\ref{thm:alg} can be regarded as a simplified version of Theorem~\ref{thm:DSE_MVR}, and we don't repeat it, cf. Appendix~\ref{con-sgd:}.
Noting that throughout the section, we assume that Assumptions \ref{Global_Function_Below_Bounds:} to \ref{CN:} hold.

\begin{lemm}
\label{al2_di_con_di:}
 For $\bm{V}_t = [ \bm{v}_t^{(1)}, \bm{v}_t^{(2)},\cdots, \bm{v}_t^{(N)}]$ where $\bm{v}_t^{(i)}$ for any $i\in[N]$ and $t\in[0,\cdots,T]$ is generated according to Algorithm \ref{alg_momen:}, we have:
\begin{align*}
    \mathbbm{E}\| \bm{V}_{t}-\overline{\bm{V}}_{t}\|_F^2   &\overset{}{\leq} 2\sum_{i=1}^{N} \mathbbm{E}\|\bm{e}_{t}^{(i)}\|^2  +2\mathbbm{E}\left\|\partial f(\bm{X}_{t})-\overline{\partial f}(\bm{X}_{t})\right\|_F^2,
\end{align*}
where the expectation $\mathbbm{E}[\cdot]$ is w.r.t the stochasticity of the algorithm.
$\overline{\bm{V}}_{t}=[ \overline{\bm{v}}_t, \overline{\bm{v}}_t,\cdots, \overline{\bm{v}}_t]$ where $\bm{\overline{v}}_{t}= \frac{1}{N}\sum _{i=1}^{N}\bm{v}_t^{(i)}$, 
$\partial f(\bm{X}_t)=[\nabla f_1(\bm{x}_t^{(1)}), \nabla f_2(\bm{x}_t^{(2)}),\cdots, \nabla f_N(\bm{x}_t^{(N)})]$ and $\overline{\partial f}(\bm{X}_t)=\bigg[ \overline{\nabla f}(\bm{X}_t), \overline{\nabla f}(\bm{X}_t),\cdots, \overline{\nabla f}(\bm{X}_t)\bigg]$ where $\overline{\nabla f}(\bm{X}_t) = \frac{1}{N}\sum _{i=1}^N\nabla f_i(\bm{x}_{t}^{(i)})$.
\end{lemm}

\begin{lemm} 
\label{al2_lemm_GEC:}
 For $t \in [0, \cdots,T -1]$, then the iterates generated by Algorithm \ref{alg_momen:} with $\gamma_t = \gamma \leq \frac{1}{8L\tau}$ and $\alpha_t=\alpha=\frac{32L^2\gamma^2}{Nb}$ satisfy that
\begin{align*}
    & \sum_{i=1}^{N}\mathbbm{E}\|\bm{e}_{t}^{(i)}\|^2 
     \overset{}{\leq}\frac{8 L^2  \gamma^2 (\tau-1) }{b}\|\partial f(\bm{X}_{t})-\overline{\partial f}(\bm{X}_{t})\|_F^2    \\
    & \quad \quad \quad \quad \quad \quad \quad \quad + \frac{4N L^2  \gamma^2(\tau-1)}{b}  \mathbbm{E}\|\overline{\bm{v}}_{t}\|^2 + \frac{2N\alpha^2(\tau-1) \sigma^2}{b},\\
    & \mathbbm{E}\|\bm{\overline{e}}_{t}\|^2 
    \overset{}{\leq}\frac{16  L^2  \gamma^2 (\tau-1)}{N^2b} \mathbbm{E}\|\partial f(\bm{X}_{t})-\overline{\partial f}(\bm{X}_{t})\|_F^2 \\
    & \quad \quad \quad \quad \quad \quad \quad \quad + \frac{8  L^2  \gamma^2 (\tau-1)}{Nb}  \mathbbm{E}\|\overline{\bm{v}}_{t}\|^2 +\frac{4\alpha^2 (\tau-1) \sigma^2}{Nb},
\end{align*}
where the expectation $\mathbbm{E}[\cdot]$ is w.r.t the stochasticity of the algorithm. 
\end{lemm}

\begin{lemm}
\label{al1_G_A_A_D_CD_main:}
For $\bm{Y}_{t^\prime}=[\bm{y}_{t^\prime}^{(1)}, \bm{y}_{t^\prime}^{(2)},\cdots, \bm{y}_{t^\prime}^{(N)}]$ where $\bm{y}_{t^\prime}^{(i)}$ for any $i \in [N]$ and $t^\prime \in [\tau, \cdots, T-\tau]$ with $t^\prime \in \mathcal{T}$ is generated by Algorithm~\ref{alg:}, we have:
\begin{align*}
    &\sum_{t^\prime=\tau}^{T-\tau}\mathbbm{E}\|\bm{Y}_{t^{\prime}}- \bm{\overline{Y}}_{t^{\prime}}\|_F^2
    \overset{}{\leq}    \frac{16N\lambda^2 \gamma^2 T\sigma^2}{(1-\lambda^2)b} +  \\ 
    &\quad \quad \quad \quad \quad \quad  \quad \quad \frac{16\lambda^2\gamma^2\tau}{(1-\lambda^2)^2} \sum_{t=0}^{T-\tau-1} \mathbbm{E}\left\|\partial f(\bm{X}_t)-\overline{\partial f}(\bm{X}_t))\right\|_F^2,
\end{align*}
where the expectation $\mathbbm{E}[\cdot]$ is w.r.t the stochasticity of the algorithm, and $\overline{\bm{Y}}_{t^\prime}=[\overline{\bm{y}}_{t^\prime}, \overline{\bm{y}}_{t^\prime},\cdots, \overline{\bm{y}}_{t^\prime}]$ where $\overline{\bm{y}}_{t^\prime}=\frac{1}{N} \sum_{i=1}^{N} \bm{y}_{t^\prime}^{(i)}$.
\end{lemm}

\begin{lemm}
\label{al_2_CD:}
For $\bm{X}_{t}=[\bm{x}_{t}^{(1)}, \bm{x}_{t}^{(2)},\cdots, \bm{x}_{t}^{(N)}]$ where $\bm{x}_{t}^{(i)}$ for any $i \in [N]$ and $t \in [0, \cdots, T-1]$ is generated by Algorithm \ref{alg_momen:} with $\gamma_t = \gamma \leq \min \left\{ \frac{1}{8L\tau}, \frac{(1-\lambda^2)^2}{64\sqrt{6}\lambda^2L\tau}\right\}$, we have:

\begin{align*}
    & \mathbbm{E}\|\bm{X}_{t}-\overline{\bm{X}}_{t}\|_F^2 
     \overset{}{\leq} \\
     & \quad \quad \quad \quad  \frac{32N L^2 \gamma^4(\tau-1)^2}{b} \left(1 + \frac{64\lambda^4 \tau}{(1-\lambda^2)^3}\right) \mathbbm{E}\|\overline{\bm{v}}_{t}\|^2  \\ 
    & \quad \quad \quad + \frac{1024N\lambda^4\alpha^2\gamma^2\tau(\tau-1)^2 \sigma^2}{(1-\lambda^2)^3b} +\frac{2048N\lambda^4\gamma^2\tau^2 \varsigma^2}{(1-\lambda^2)^4} \\
    & \quad \quad \quad \quad \quad + \frac{8N\alpha^2\gamma^2\tau(\tau-1)^2\sigma^2}{b} + 24N\gamma^2\tau(\tau-1) \varsigma^2, 
\end{align*}
where the expectation $\mathbbm{E}[\cdot]$ is w.r.t the stochasticity of the algorithm, and $\overline{\bm{X}}_{t}=[\overline{\bm{x}}_{t}, \overline{\bm{x}}_{t},\cdots, \overline{\bm{x}}_{t}]$ where $\bm{\overline{x}}_{t}=\frac{1}{N} \sum_{i=1}^{N} \bm{x}_t^{(i)}$.
\end{lemm}

\begin{table*}[h]
  \centering
  \caption{Top-1 test accuracy~($\%$) and training loss overview given partial difference settings.}
    \resizebox{2.0\columnwidth}{!}{
    \begin{tabular}{llcccccccccc}
    \toprule[2pt]
    \multirow{2}[2]{*}{Datasets} & \multirow{2}[2]{*}{Settings} & \multicolumn{2}{c}{DLSGD} & \multicolumn{2}{c}{SLOWMo-D} & \multicolumn{2}{c}{PD-SGDM} & \multicolumn{2}{c}{DSE-SGD} & \multicolumn{2}{c}{DSE-MVR} \\
          &       & test accuracy & training loss & test accuracy & training loss & test accuracy & training loss & test accuracy & training loss & test accuracy & training loss \\
    \midrule
    \multirow{3}[2]{*}{MNIST, $\omega=0.5$} & $b=64$ & 97.34$\pm$0.23 & 0.050$\pm$0.008 & \textbf{97.89$\pm$0.19} & 0.029$\pm$0.006 & 97.84$\pm$0.31 & 0.029$\pm$0.011 & 97.76$\pm$0.24 & 0.032$\pm$0.012 & 97.89$\pm$0.55 & \textbf{0.019$\pm$0.012} \\
          & $b=128$ & 97.36$\pm$0.31 & 0.045$\pm$0.007 & 98.09$\pm$0.41 & 0.019$\pm$0.006 & 97.94$\pm$0.43 & 0.035$\pm$0.014 & 97.89$\pm$0.28 & 0.031$\pm$0.011 & \textbf{98.49$\pm$0.46} & \textbf{0.016$\pm$0.009} \\
          & $b=256$ & 97.47$\pm$0.17 & 0.041$\pm$0.007 & 98.23$\pm$0.22 & 0.016$\pm$0.005 & 98.05$\pm$0.46 & 0.025$\pm$0.014 & 97.94$\pm$0.41 & 0.026$\pm$0.012 & \textbf{98.53$\pm$0.36} & \textbf{0.009$\pm$0.006} \\
    \midrule
    \multirow{3}[2]{*}{MNIST, $\omega=10$} & $\tau=3$ & 97.96$\pm$0.03 & 0.026$\pm$0.005 & 98.46$\pm$0.08 & 0.015$\pm$0.005 & 98.60$\pm$0.18 & 0.009$\pm$0.004 & 98.38$\pm$0.19 & 0.016$\pm$0.008 & \textbf{99.02$\pm$0.09} & \textbf{0.005$\pm$0.002} \\
          & $\tau=7$ & 97.90$\pm$0.05 & 0.027$\pm$0.005 & \textbf{98.37$\pm$0.01} & 0.018$\pm$0.022 & 98.16$\pm$0.54 & 0.023$\pm$0.028 & 97.95$\pm$0.58 & 0.026$\pm$0.018 & 98.29$\pm$0.27 & \textbf{0.016$\pm$0.009} \\
          & $\tau=20$ & 97.75$\pm$0.06 & 0.035$\pm$0.005 & 97.93$\pm$0.09 & 0.027$\pm$0.002 & 97.89$\pm$0.17 & 0.032$\pm$0.006 & 97.88$\pm$0.19 & 0.033$\pm$0.006 & \textbf{97.98$\pm$0.42} & \textbf{0.026$\pm$0.009} \\
    \midrule
    \multirow{3}[2]{*}{CIFAR-10, $\omega=0.5$} & $b=256$ & 79.59$\pm$1.22 & 0.419$\pm$0.056 & 84.21$\pm$0.49 & 0.192$\pm$0.042 & 84.31$\pm$0.83 & 0.209$\pm$0.113 & 82.19$\pm$0.98 & 0.309$\pm$0.052 & \textbf{84.65$\pm$0.57} & \textbf{0.185$\pm$0.052} \\
          & $b=512$ & 80.46$\pm$1.03 & 0.367$\pm$0.021 & 85.28$\pm$0.39 & 0.134$\pm$0.026 & 84.96$\pm$0.44 & 0.129$\pm$0.046 & 83.15$\pm$0.59 & 0.294$\pm$0.066 & \textbf{85.30$\pm$0.41} & \textbf{0.107$\pm$0.047} \\
          & $b=1024$ & 81.69$\pm$0.21 & 0.339$\pm$0.022 & \textbf{86.03$\pm$0.54} & 0.105$\pm$0.031 & 85.33$\pm$0.37 & 0.160$\pm$0.033 & 83.42$\pm$0.43 & 0.279$\pm$0.069 & 85.83$\pm$0.56 & \textbf{0.099$\pm$0.023} \\
    \midrule
    \multirow{3}[2]{*}{CIFAR-10, $\omega=10$} & $\tau=4$ & 85.05$\pm$0.36 & 0.220$\pm$0.043 & 88.29$\pm$0.41 & 0.017$\pm$0.008 & 88.01$\pm$0.62 & 0.017$\pm$0.006 & 87.22$\pm$0.59 & 0.026$\pm$0.004 & \textbf{88.54$\pm$0.39} & \textbf{0.011$\pm$0.017} \\
          & $\tau=8$ & 84.78$\pm$0.23 & 0.210$\pm$0.030 & 88.25$\pm$0.41 & 0.031$\pm$0.016 & 87.98$\pm$0.33 & 0.038$\pm$0.001 & 86.85$\pm$0.23 & 0.105$\pm$0.059 & \textbf{88.41$\pm$0.35} & \textbf{0.013$\pm$0.012} \\
          & $\tau=20$ & 84.49$\pm$0.45 & 0.214$\pm$0.109 & 86.41$\pm$0.54 & 0.088$\pm$0.050 & 86.65$\pm$0.43 & 0.097$\pm$0.078 & 85.87$\pm$0.78 & 0.169$\pm$0.029 & \textbf{86.83$\pm$0.64} & \textbf{0.059$\pm$0.046} \\
    \bottomrule
    \bottomrule
    \end{tabular}}%
  \label{table_per_ov:}%
\end{table*}%

\begin{figure*}[t]
\centering
\includegraphics[width=0.72\textwidth]{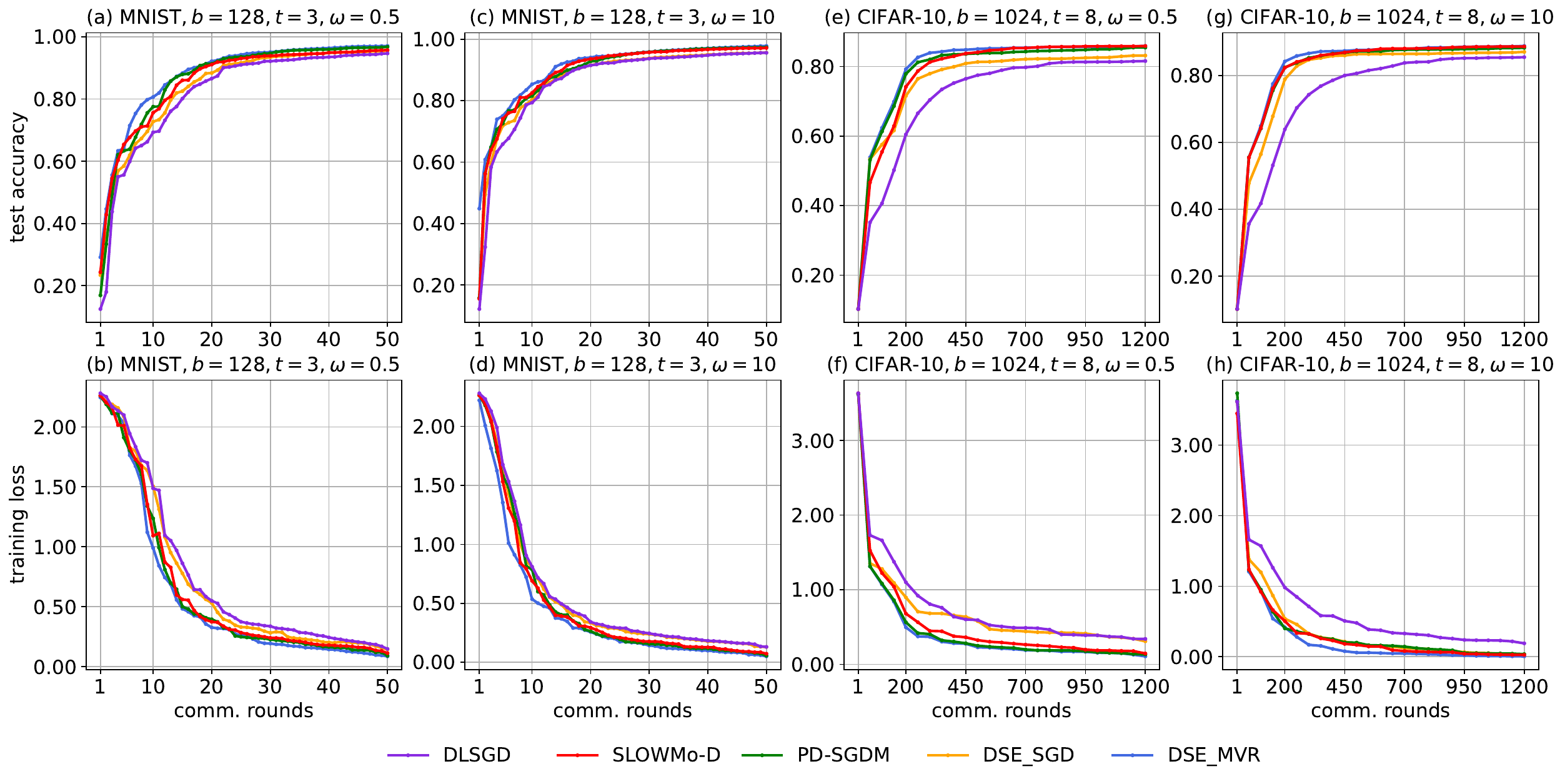} 
\caption{Varying $\omega$: learning curves selected over MNIST and CIFAR-10 datasets, which are averaged over 3 random seeds.}
\label{vary_alpha:}
\end{figure*}

\begin{figure*}[h]
\centering
\includegraphics[width=0.9\textwidth]{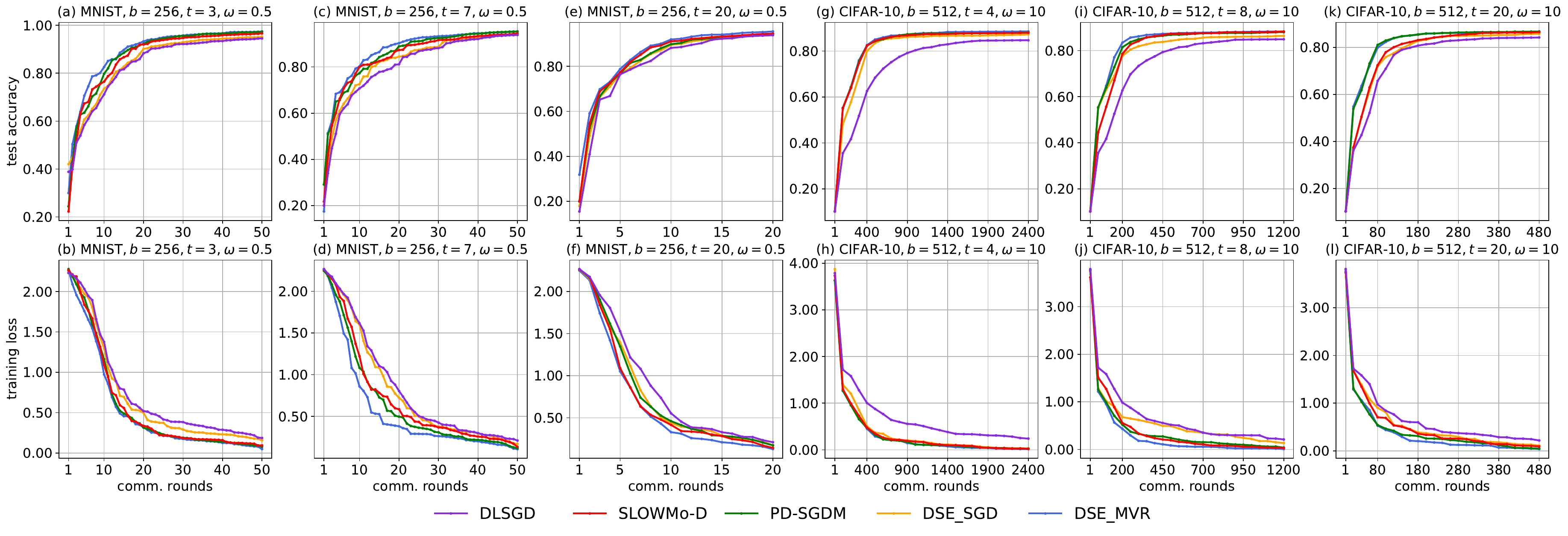} 
\caption{Varying $\tau$: learning curves selected over MNIST and CIFAR-10 datasets, which are averaged over 3 random seeds.}
\label{vary_tau:}
\end{figure*}

\begin{figure*}[t]
\centering
\includegraphics[width=0.9\textwidth]{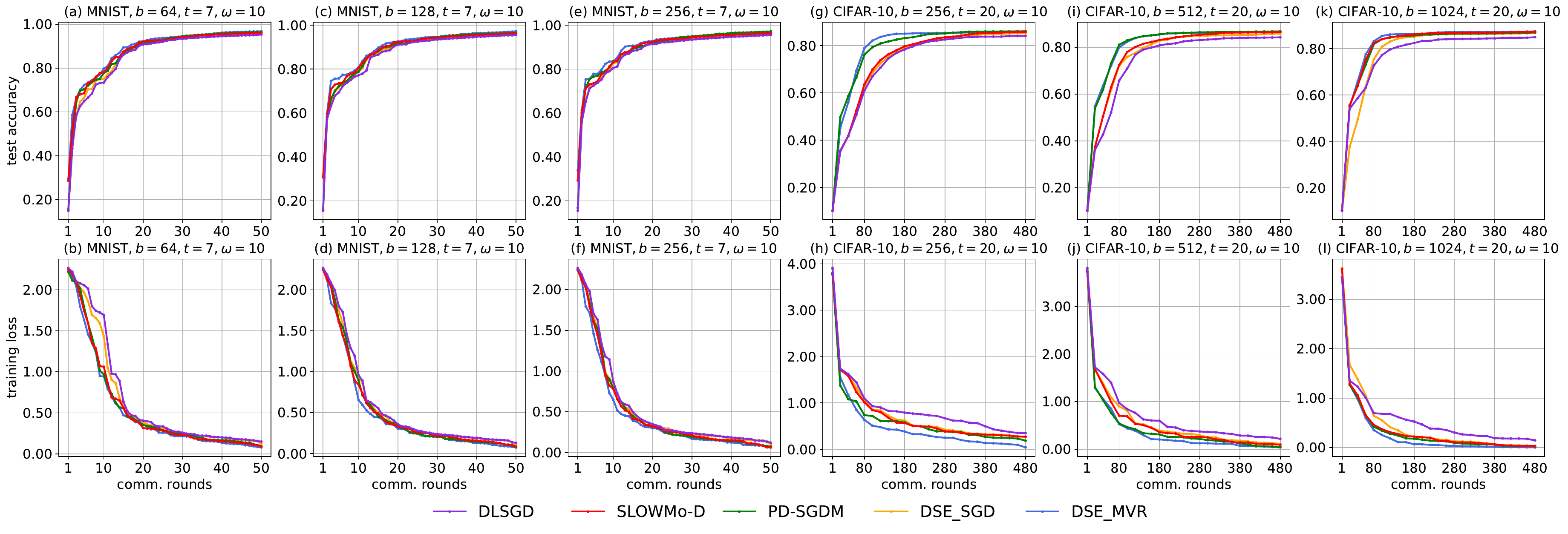} 
\caption{Varying $b$: learning curves selected over MNIST and CIFAR-10 datasets, which are  averaged over 3 random seeds.}
\label{vary_batch:}
\end{figure*}

\section{Experiments}
\label{experiment:}

In this section, 
we validate our theoretical results by comparing our methods
with other baselines
on image classification tasks. 
Specifically, 
we compare DSE-MVR with
local-update-based methods PD-SGDM~\cite{gao2020periodic}
and SLOWMo-D,
which extends SLOWMo~\cite{wang2019slowmo} to DL.
Further, 
to justify the importance of the dual-slow estimation strategy, 
we compare DSE-SGD with DLSGD~\cite{li2019communication}, 
which can be considered as
local-update-based
extensions of the non-local-update-based methods GT-DSGD~\cite{xin2021improved} and DSGD~\cite{lian2017can}, respectively. 

\textbf{Experimental Setup:} We conduct 10-class image classification on MNIST~\cite{lecun1998gradient} and CIFAR-10~\cite{krizhevsky2009learning} datasets.
For both datasets, we consider common network topology, i.e., {\it ring graph} and use the Metropolis-Hasting mixing matrix $W$~\cite{koloskova2020unified}, i.e., $w_{ij}=w_{ji}=\frac{1}{deg(i)+1}=\frac{1}{deg(j)+1}$ for any edge $(i,j) \in \mathcal{E}$, to parameterize the communication.  
For MNIST, 
a convolutional neural network (CNN) with two convolutional hidden layers plus two linear layers is implemented for each node. 
We set $T = 400$
and 
fine-tune the learning rate from $\{0.1, 0.2, 0.3, 0.4, 0.5\}$,
the batch size  
$b$ from $\{64, 128, 256\}$ and 
the partial average interval 
$\tau$ from $\{3, 7, 20\}$. 
Note that
we divide the learning rate 
by $2$ at iterations $0.5\cdot T$ and $0.75\cdot T$.  
Moreover,  
the control parameter $\alpha$ is tuned from $\{0.01,  0.05\}$,
which 
is decayed with a decay weight $0.99$.
For CIFAR-10, 
each node implements a Resnet-20-BN~\cite{he2016deep} architecture. 
We fix $T$ to $10000$ and 
schedule 
the learning rate (the control parameter $\alpha$) setting as 
$0.01 (0.002)$, $0.1 (0.02)$, $0.01 (0.002)$, and $0.001 (0.0002)$ at iterations $0\cdot T$, $0.1\cdot T$, $0.75\cdot T$, and $0.9\cdot T$, respectively.
Further, we fine-tune the batch size  
$b$ from $\{256, 512, 1024\}$ and 
the partial average interval 
$\tau$ from $\{4, 8, 20\}$. 
We use  Dirichlet process $Dp(\omega)$~\cite{vogels2021relaysum, lin2021quasi} to strictly partition training data across 20 (40) nodes for MNIST (CIFAR-10), where the scaling parameter $\omega$ controls the data heterogeneity
across nodes. 
For both datasets,
we set $\omega = 0.5$ and $\omega = 10$ to generate the non-iid and iid settings, respectively.
For fairness, 
we compare all the methods 
under uniform data heterogeneity settings w.r.t. the best training loss and test accuracy.
All the algorithms are implemented by PyTorch 1.11.0.
Due to the space limitation, we report only partial results in the main paper. 
The comprehensive results and detailed information on the computing devices and platforms used to perform the full experiments can be found in Appendix E.

\textbf{Performance Overview:} 
Table~\ref{table_per_ov:} shows the partial results of top-1 test accuracy($\%$) and training loss 
on MNIST and CIFAR-10 datasets. See Table~\ref{tab:full_test_train_ov} in Appendix E for full results.
Specifically,
we study the performance of all the methods 
with varying 
$\tau$ and $b$ in both non-iid ($\omega=0.5$) and iid ($\omega=10$) settings.
In our experiments, 
we vary one hyper-parameter with others fixed,
and calculate the mean and standard deviation of all the methods over 3 trials. From the table, 
our method DSE-MVR outperforms other two local-update-based methods SLOWMo-D and PD-SGDM in most cases. 
Meanwhile, 
DSE-SGD also consistently achieves better performance than 
DLSGD in all the cases.
This shows that the double-slow estimation can adapt well to a variety of settings, including small batch size and high partial-average interval in both non-iid and iid settings.
Generally, 
\textbf{DSE-MVR}$>$\textbf{SLOWMo-D}$>$\textbf{PD-SGDM}$>$\textbf{DSE-SGD}$>$\textbf{DLSGD} in terms of both the testing and training performances. Although SLOWMo-D and PD-SGDM enjoy better test accuracy and training loss than DSE-SGD and DLSGD, 
owing to the advantages induced from MVR~\cite{cutkosky2019momentum}, 
DSE-MVR can benefit each local node more by mitigating the stochastic noise.

\textbf{Impact of data heterogeneity:}
From Table~\ref{table_per_ov:}, we can clearly observe that the testing and training performances of each method in iid setting uniformly outperform that in non-iid setting.
Further, 
we show learning curves of all the methods on the MNIST (CIFAR-10) dataset regarding different data heterogeneity under $b=128$ ($1024$) and $\tau=3$ ($8$) settings, 
and the learning curves are averaged over 3 random seeds as shown in Fig.~\ref{vary_alpha:}.
See Appendix E for the comprehensive results.
From the figure, 
compared with other methods,
DSE-MVR has the most rapid learning curves to reach a given test accuracy (training loss). 

\textbf{Impact of partial average interval:} 
We next explore the impact of different partial average intervals $\tau$ on MNIST and CIFAR-10.  
A higher $\tau$ means longer synchronization delays before communication across nodes.
From Table~\ref{table_per_ov:}, 
we can see that 
the performances of all the methods uniformly deteriorate as $\tau$ increases 
on both datasets. 
Meanwhile, Fig.~\ref{vary_tau:} shows that the learning efficiency of DSE-MVR consistently outperforms other methods for each $\tau$.

\textbf{Impact of batch size:} 
We also conduct experiments on both MNIST and CIFAR-10 datasets 
using different batch sizes $b$. 
As shown in 
Table~\ref{table_per_ov:}, 
the test accuracy and training loss of all the methods degenerate 
as $b$ decreases on both datasets. 
Also, 
from Fig.~\ref{vary_batch:}, we can see that the 
superior learning efficiency 
of DSE-MVR is consistent across three different batch sizes. 
Concretely, 
our method requires much less communication rounds to reach a given performance, 
regardless of values of $b$ (See Appendix E for comprehensive results).

\section{Conclusion}

In this paper, 
we studied decentralized local updates and 
proposed DSE-MVR, 
which uses the dual-slow estimation strategy to handle the data heterogeneity problem  
and the mini-batch momentum-based variance-reduction method to alleviate the adverse effect of stochastic gradient noise.
We theoretically showed that DSE-MVR can achieve optimal convergence results for general non-convex optimization in both iid and non-iid settings.
We further put forward DSE-SDG,
based on which we theoretically justified the importance of the dual-slow estimation strategy. 
Finally,
we conducted extensive experiments to validate our theoretical results and show the superiority of our proposed methods against other state-of-the-arts.


\bibliography{ecai}

\onecolumn
\clearpage
\appendix
\section*{Appendix}
\subsection{Pseudocodes}
\label{code_DSE_SGD:}
We summarize the pseudocodes of DSE-SGD in Algorithm~\ref{alg:}.
\begin{algorithm}[h]
  \caption{The DSE-SGD algorithm}
  \label{alg:}
  \hspace*{0.02in} {\bf Input:}
   Learning rate $\gamma_t > 0$, weight matrix $\bm{W}$, batch size $b$, partial average interval $\tau>0$.
\begin{algorithmic}[1]
  \STATE Initial state $\bm{x}_{0}^{(i)}=\bm{x}_0$, $\bm{y}_{0}^{(i)}=\bm{h}_{0}^{(i)}=\bm{0}$, $i\in [N]$. 
  \STATE $\bm{g}_{0}^{(i)} = \frac{1}{b} \sum \limits_{r=1}^{b} \nabla f_i(\bm{x}_{0}^{(i)};\bm{\xi}_{r}^{(i)})$,  $\bm{\xi}_{{r}}^{(i)} \sim \mathcal{D}_i$ 
  \FOR{$t=0,\cdots, T-1$}
    \FOR{$i=1,\cdots, N$}
        \STATE $\bm{x}_{t+\frac{1}{2}}^{(i)}= \bm{x}_{t}^{(i)}-\gamma_{t} \bm{g}_{t}^{(i)}$ \hfill \% model parameter update
        \IF{$\mod(t+1,\tau) = 0$} 
            \STATE $\bm{h}_{t+1}^{(i)} =  \bm{x}_{\tau(t)}^{(i)}-\bm{x}_{t+\frac{1}{2}}^{(i)}$ \hfill \%   accumulated descent directions
            \STATE $\bm{y}_{t+1}^{(i)}= \sum \limits _{j \in \mathcal{N}_{i}} w_{ij} (\bm{y}_{\tau(t)}^{(j)} + \bm{h}_{t+1}^{(j)}-\bm{h}_{\tau(t)}^{(j)})$  \hfill \%   slow gradient tracking
            \STATE $\bm{x}_{t+1}^{(i)}= \sum \limits _{j \in \mathcal{N}_{i}} w_{ij} (\bm{x}_{\tau(t)}^{(j)} - \bm{y}_{t+1}^{(j)})$ \hfill \% slow partial average
        \ELSE
            \STATE $\bm{x}_{t+1}^{(i)}=\bm{x}_{t+\frac{1}{2}}^{(i)}$
        \ENDIF
        \STATE $\bm{g}_{t+1}^{(i)} = \frac{1}{b} \sum \limits_{r=1}^b \nabla f_i(\bm{x}_{t+1}^{(i)};\bm{\xi}_{{r}}^{(i)})$,  $\bm{\xi}_{{r}}^{(i)} \sim \mathcal{D}_i$ \hfill \%
            $b$ stochastic gradients on $\bm{x}_{t+1}^{(i)}$
    \ENDFOR
  \ENDFOR
\end{algorithmic}
\hspace*{0.02in} {\bf Output:} 
  $ \overline{\bm{x}}_q$, where $q$ chosen uniformly randomly from $[0,\cdots,T-1]$. \\
\hspace*{0.02in} {\bf Note that}  
  $\tau(t)=\max \left\{l:l\leq t \ {\rm and} \ {\rm mod} {(l,\tau)} = 0 \right\}$
\end{algorithm}

\subsection{Preliminary}
\label{Pre:}
\subsubsection{Notations and definitions}

We use the following notations and definitions to complete theoretical analysis:
\begin{itemize}
    \item $T$ and $N$ denote the total number of iterations and nodes, respectively;
    \item $n_i$, $i \in [N]$ denotes the sample size of node $i$;
    \item $\|\cdot\|$ denotes the Euclidean norm of a vector or the spectral norm of a matrix depending on the argument;
    \item $\|\cdot\|_F$ denotes the matrix Frobenius norm;
    \item $\nabla F(\cdot)$ denotes the full gradient of the loss function $F$, and  $\nabla F(\bm{x}) = \frac{1}{N} \sum_{i=1}^{N} f_i(\bm{x})$ for any $\bm{x} \in \mathbbm{R}^d$;
    \item $\bm{1}_N$ denotes the column vector in $\mathbbm{R}^N$ with $1$ for all elements. And we set $\bm{Q}=\frac{1}{N}\bm{1}_N\bm{1}_N^T$;
    \item $\bm{W} \in \mathbbm{R}^{N \times N}$ is a symmetric doubly stochastic matrix, which determines the topology of communication. See Assumption \ref{CN:} for the specific definition;
    \item $\bm{X}_{t}=[\bm{x}_{t}^{(1)}, \bm{x}_{t}^{(2)},\cdots, \bm{x}_{t}^{(N)}] \in \mathbbm{R}^{d \times N}$, $\bm{\overline{x}}_{t}=\frac{1}{N} \sum_{i=1}^{N} \bm{x}_t^{(i)}$ and $\bm{\overline{X}}_{t}= \left[\bm{\overline{x}}_t, \bm{\overline{x}}_t,\cdots, \bm{\overline{x}}_t\right]$;
    \item $\bm{Y}_{t}=[ \bm{y}_t^{(1)}, \bm{y}_t^{(2)},\cdots, \bm{y}_t^{(N)}] \in \mathbbm{R}^{d \times N}$, $\bm{\overline{y}}_{t}=\frac{1}{N} \sum_{i=1}^{N} \bm{y}_t^{(i)}$ and $\bm{\overline{Y}}_{t}= \left[\bm{\overline{y}}_t, \bm{\overline{y}}_t,\cdots, \bm{\overline{y}}_t\right]$;
    \item $\bm{V}_{t}=[ \bm{v}_t^{(1)}, \bm{v}_t^{(2)},\cdots, \bm{v}_t^{(N)}] \in \mathbbm{R}^{d \times N}$, $\bm{\overline{v}}_{t}= \frac{1}{N}\sum _{i=1}^{N}\bm{v}_t^{(i)}$ and $\bm{\overline{V}}_{t}= \left[\bm{\overline{v}}_t, \bm{\overline{v}}_t,\cdots, \bm{\overline{v}}_t\right]$;
    \item $\bm{H}_{t}=[ \bm{h}_t^{(1)}, \bm{h}_t^{(2)},\cdots, \bm{h}_t^{(N)}] \in \mathbbm{R}^{d \times N}$, $\bm{\overline{h}}_{t}= \frac{1}{N}\sum _{i=1}^{N}\bm{h}_t^{(i)}$ and $\bm{\overline{H}}_{t}= \left[\bm{\overline{h}}_t, \bm{\overline{h}}_t,\cdots, \bm{\overline{h}}_t\right]$;
    \item $\bm{G}_{t}=[ \bm{g}_{t}^{(1)}, \bm{g}_{t}^{(2)},\cdots, \bm{g}_{t}^{(N)}] \in \mathbbm{R}^{d \times N}$, $\bm{g}_{t}^{(i)} = \frac{1}{b}\sum_{r=1}^b\nabla f_i(\bm{x}_t^{(i)},\bm{\xi}_{r}^{(i)})$,$\bm{\xi}_{{r}}^{(i)} \sim \mathcal{D}_i$ and $\bm{\overline{G}}_{t}= \left[\bm{\overline{g}}_{t}, \bm{\overline{g}}_{t},\cdots, \bm{\overline{g}}_{t}\right]$ where $\bm{\overline{g}}_{t}= \frac{1}{N}\sum _{i=1}^{N}\bm{g}_{t}^{(i)}$;
    \item $\partial f(\bm{X}_t)=[ \nabla f_1(\bm{x}_t^{(1)}), \nabla f_2(\bm{x}_t^{(2)}),\cdots, \nabla f_N(\bm{x}_t^{(N)})] \in \mathbbm{R}^{d \times N}$ and $\overline{\partial f}(\bm{X}_t)=\left[ \overline{\nabla f}(\bm{X}_t), \overline{\nabla f}(\bm{X}_t),\cdots, \overline{\nabla f}(\bm{X}_t)\right] \in \mathbbm{R}^{d \times N}$ where $\overline{\nabla f}(\bm{X}_t) = \frac{1}{N}\sum _{i=1}^N\nabla f_i(\bm{x}_{t}^{(i)})$;
    \item We use lowercase letters with apostrophe to denote the iterations in which the communication is performed, i.e., $t^\prime \in \mathcal{T}:= \{t| t=0,\cdots, T \ {\rm and} \ {\rm mod}(t, \tau)=0\}$;
    \item To simplify the proof, we set ${\rm mod}(T,\tau)=0$;
    \item For $\bm{A}=[\bm{a}_1, \bm{a}_2, \cdots, \bm{a}_N]\in\mathbbm{R}^{d \times N}$, $\bm{B}=[\bm{b}_1, \bm{b}_2, \cdots, \bm{b}_N]\in\mathbbm{R}^{d \times N}$, we set $\langle\bm{A},\bm{B}\rangle=\sum_{i=1}^{N}\langle\bm{a}_i,\bm{b}_i\rangle = {\rm Tr}(\bm{A}^T\bm{B})=\|\bm{A}^T\bm{B}\|_F^2$.
\end{itemize}

\subsubsection{Facts}
\begin{fact}
    \label{FC1:}
    For $\bm{x}_1, \bm{x}_2, \cdots, \bm{x}_N \in \mathbbm{R}^d$,
         {\rm (1)}  we have
         \begin{align*}
             \label{FC1_1:}
             \left\|\sum_{i=1}^{N}\bm{x}_i\right\|^2 \leq N\sum_{i=1}^{N}\|\bm{x}_i\|^2, \left\|\frac{1}{N}\sum_{i=1}^{N}\bm{x}_i\right\|^2 \leq \frac{1}{N}\sum_{i=1}^{N}\|\bm{x}_i\|^2.
             \tag{i}
         \end{align*}
         {\rm (2)} If $\bm{x}_i\left(i \in [N]\right)$ are independent with $\bm{0}$ means, i.e. $\mathbbm{E}[\left\langle\bm{x}_i,\bm{x}_j \right\rangle]=0$,$i \neq j \in [N]$ , then we have
        \begin{align*}
            \label{FC1_2:}
            \left\|\sum_{i=1}^{N}\bm{x}_i\right\|^2 = \sum_{i=1}^{N}\|\bm{x}_i\|^2, \left\|\frac{1}{N}\sum_{i=1}^{N}\bm{x}_i\right\|^2 = \frac{1}{N^2}\sum_{i=1}^{N}\|\bm{x}_i\|^2.
            \tag{ii}
        \end{align*}
\end{fact}

\begin{fact}
    \label{FC2:}
     For $\bm{x}, \bm{y}  \in \mathbbm{R}^{d}$, 
     we have
     \begin{align*}
         \label{FC2_1:}
         \langle\bm{x},\bm{y}\rangle = \frac{1}{2} (\|\bm{x}\|^2 + \|\bm{y}\|^2-\|\bm{x}-\bm{y}\|^2),
          \tag{i}
     \end{align*}
     \begin{align*}
         \label{FC2_2:}
         \pm \langle\bm{x},\bm{y}\rangle \leq \frac{1}{2\eta} \|\bm{x}\|^2 + 2\eta \|\bm{y}\|^2,
         \tag{ii}
     \end{align*}
     \begin{align*}
         \label{FC2_3:}
         \|\bm{x} + \bm{y}\|^2  \leq (1+\eta)\|\bm{x}\|^2+(1+\frac{1}{\eta})\|\bm{y}\|^2,
         \tag{iii}
     \end{align*}
     where $\eta > 0$ is some constant.
\end{fact}

\begin{fact}
    \label{FC3:}
     For $\bm{A}, \bm{B}  \in \mathbbm{R}^{d\times N}$, 
      we have
     \begin{align*}
         \label{FC3_1:}
          \langle\bm{A},\bm{B}\rangle = \frac{1}{2} (\|\bm{A}\|_F^2 + \|\bm{B}\|_F^2-\|\bm{A}-\bm{B}\|_F^2),
          \tag{i}
     \end{align*}
     \begin{align*}
         \label{FC3_2:}
         \pm \langle\bm{A},\bm{B}\rangle \leq \frac{1}{2\eta} \|\bm{A}\|_F^2 + 2\eta \|\bm{B}\|_F^2,
         \tag{ii}
     \end{align*}
     \begin{align*}
         \label{F3_3:}
         \|\bm{A} + \bm{B}\|_F^2  \leq (1+\eta)\|\bm{A}\|_F^2+(1+\frac{1}{\eta})\|\bm{B}\|_F^2,
         \tag{iii}
     \end{align*}
     where $\eta > 0$ is some constant.
\end{fact}

\begin{fact}
    \label{FC4:}
     {\rm (1)} For any independent matrix random variables $\bm{A} \in \mathbbm{R}^{d\times N}$, we have
     \begin{align*}
         \label{F4_1:} \mathbbm{E}\|\bm{A}-\mathbbm{E}[\bm{A}]\|_F^2 \leq \mathbbm{E}\|\bm{A}\|_F^2.
         \tag{i}
     \end{align*}
     {\rm (2)} For any independent vector random variables $\bm{a} \in \mathbbm{R}^{d}$, we have
     \begin{align*}
         \label{F4_2:} \mathbbm{E}\|\bm{a}-\mathbbm{E}[\bm{a}]\|^2 \leq \mathbbm{E}\|\bm{a}\|^2.
         \tag{ii}
     \end{align*}
     {\rm (3)} For $\bm{A}=[\bm{a}_1, \bm{a}_2, \cdots, \bm{a}_N]\in\mathbbm{R}^{d \times N}$, we have
     \begin{align*} \|\bm{A}-\overline{\bm{A}}\|_F^2 =\sum_{i=1}^{N} \|\bm{a}_i-\overline{\bm{a}}\| \leq \sum_{i=1}^{N} \|\bm{a}_i\|^2 \leq \|\bm{A}\|_F^2,
         \tag{iii}
     \end{align*}
     where $\overline{\bm{a}} = \sum_{i=1}^{N}\bm{a}_i$ and $\overline{\bm{A}}=\left[\overline{\bm{a}}, \overline{\bm{a}}, \cdots, \overline{\bm{a}}\right]$.
\end{fact}

\subsection{Convergence Proof of Algorithm \ref{alg:}}
\label{con-sgd:}
In this and the next section, we present complete theoretical proofs of our algorithms DSE-SGD and DSE-MVR.

Now, we provide the theoretical analysis on the convergence of DSE-SGD.
Before formally proceeding with the analysis, we recall the key steps of Algorithm \ref{alg:} and note the following statements about Algorithm \ref{alg:}.

\textbf{First}, if the iteration $t$ satisfies ${\rm mod}(t+1, \tau)=0$, it is easy to check that the following relations hold for lines 7 to 9 of Algorithm \ref{alg:} (given $\bm{y}_0 = \bm{h}_0 = \bm{0}$):
\begin{align}
    &\overline{\bm{h}}_{t+1} = \overline{\bm{x}}_{\tau(t)}  - \overline{\bm{x}}_{t+\frac{1}{2}} = \sum_{j=\tau(t)}^{t} \gamma_j \overline{\bm{g}}_{j}, \label{al1_avrag_rule_h:}\\
    & \overline{\bm{y}}_{t+1} = \overline{\bm{y}}_{\tau(t)} +\overline{\bm{h}}_{t+1} - \overline{\bm{h}}_{\tau(t)} = \overline{\bm{h}}_{t+1}, \label{al1_avrag_rule_y:}\\
    & \overline{\bm{x}}_{t+1} = \overline{\bm{x}}_{\tau(t)}  - \overline{\bm{y}}_{t+1}. \label{al1_avrag_rule_x:}
\end{align}

If ${\rm mod}(t+1, \tau) \neq 0$ (lines 5 and 11), we have: 
\begin{equation}
    \label{al1_avrag_rule_x_1:}
    \begin{split}
        \overline{\bm{x}}_{t+1} =\overline{\bm{x}}_{\tau(t)}  - \gamma_t \overline{\bm{g}}_{t},
    \end{split}
\end{equation}
where $\overline{\bm{x}}_{\tau(t)}$ denotes the average model parameter of all nodes in the previous communication round, i.e., $\tau(t)=\max \left\{l:l\leq t \ {\rm and} \ {\rm mod} {(l,\tau)} = 0 \right\}$.

Then, noting that $\overline{\bm{y}}_{t+1} = \overline{\bm{h}}_{t+1}$ (lines 7-8), we have: for all $t \in [0, \cdots, T-1]$,
\begin{equation}
    \label{al1_avrag_rule_x_2:}
    \begin{split}  \overline{\bm{x}}_{t+1} = \overline{\bm{x}}_{t} - \gamma_t \overline{\bm{g}}_{t},
    \end{split}
\end{equation}
where $\overline{\bm{g}}_{t} = \frac{1}{N}\sum_{i=1}^{N}\bm{g}_{t}^{(i)} = \frac{1}{Nb}\sum_{i=1}^{N}\sum_{r=1}^{b}\nabla f_i(\bm{x}_{t}^{(i)}; \bm{\xi}_{r}^{(i)})$, $\bm{\xi}_{{r}}^{(i)} \sim \mathcal{D}_i$.

\textbf{Second},  we represent the update rules of the model parameters in matrix form as follows. 

If ${\rm mod}(t+1, \tau)=0$,  then each node communicates with its neighbor nodes and the communication among nodes is controlled by mixing matrix $\bm{W}$. Particularly, we have:
\begin{align}  
    &\bm{H}_{t+1} = \left(\bm{X}_{\tau(t)} - \bm{X}_{t+\frac{1}{2}}\right) = \sum_{j=\tau(t)}^{t} \gamma_j \bm{G}_{j}, \label{al1_matrix_rule_H:}\\
    &\bm{Y}_{t+1} = \left(\bm{Y}_{\tau(t)} + \bm{H}_{t+1} - \bm{H}_{\tau(t)}\right)\bm{W}, \label{al1_matrix_rule_Y:}\\
    &\bm{X}_{t+1} =\left(\bm{X}_{\tau(t)} - \bm{Y}_{t+1}\right)\bm{W}. \label{al1_matrix_rule_X1:}
\end{align}
If ${\rm mod}(t+1, \tau) \neq 0$,  then  each node performs local update steps. And we have:
\begin{align}  
    &\bm{X}_{t+1} = \bm{X}_{t} - \gamma_t \bm{G}_{t}. \label{al1_matrix_rule_X2:}
\end{align}

Further, throughout the section, we assume that assumptions \ref{Global_Function_Below_Bounds:} to \ref{CN:} hold.

\subsubsection{Preliminary Lemmas}
\begin{lemm} 
    \label{griadient_consensus:}
    For $\partial f(\bm{X}_t)=[ \nabla f_1(\bm{x}_t^{(1)}), \nabla f_2(\bm{x}_t^{(2)}),\cdots, \nabla f_N(\bm{x}_t^{(N)})]$ where $\nabla f_i(\bm{x}_t^{(i)})$ for all $i\in [N]$ and $t\in[0,\cdots, T-1]$ is generated by Algorithm~\ref{alg:}, we have:
    \begin{align*}
        \mathbbm{E}\left\|\partial f(\bm{X}_t)-\overline{\partial f}(\bm{X}_t)\right\|_F^2  \overset{}{\leq}  8L^2 \mathbbm{E}\left\|\bm{X}_t-\overline{\bm{X}}_t\right\|_F^2 + 2N\varsigma^2,
    \end{align*}
    where the expectation $\mathbbm{E}[\cdot]$ is w.r.t the stochasticity of the algorithm, $\left\|\partial f(\bm{X}_t)-\overline{\partial f}(\bm{X}_t)\right\|_F^2 =  \sum\limits_{i=1}^{N}\left\|\nabla f_i(\bm{x}_t^{(i)})-\overline{\nabla f}(\bm{X}_t)\right\|^2$ and $\left\|\bm{X}_t-\overline{\bm{X}}_t\right\|_F^2 =  \sum\limits_{i=1}^{N}\|\bm{x}_t^{(i)}-\overline{\bm{x}}_t\|^2$.
\end{lemm}
\begin{proof}
    \begin{equation}
        \begin{split}
            \left\|\partial f(\bm{X}_t)-\overline{\partial f}(\bm{X}_t)\right\|_F^2 & =  \sum_{i=1}^{N}\left\|\nabla f_i(\bm{x}_t^{(i)})-\overline{\nabla f}(\bm{X}_t)\right\|^2\\
            & =  \sum_{i=1}^{N}\left\|\nabla f_i(\bm{x}_t^{(i)})\mp\nabla f_i(\overline{\bm{x}}_t)\mp\nabla F(\overline{\bm{x}}_t)-\overline{\nabla f}(\bm{X}_t)\right\|^2\\
            & \overset{(a)}{\leq} 4 \sum_{i=1}^{N}\left\|\nabla f_i(\bm{x}_t^{(i)})-\nabla f_i(\overline{\bm{x}}_t)\right\|^2+2 \sum_{i=1}^{N}\left\|\nabla f_i(\overline{\bm{x}}_t)-\nabla F(\overline{\bm{x}}_t)\right\|^2\\
            & \quad \quad \quad \quad \quad \quad \quad \quad \quad \quad \quad \quad \quad \quad \quad \quad \quad  +4 \sum_{i=1}^{N}\left\|\nabla F(\overline{\bm{x}}_t)-\overline{\nabla f}(\bm{X}_t)\right\|^2\\
            & \overset{(b)}{\leq} 4L^2 \sum_{i=1}^{N}\left\|\bm{x}_t^{(i)}-\overline{\bm{x}}_t\right\|^2+2N\varsigma^2+4 N\left\|\frac{1}{N}\sum_{i=1}^{N}(\nabla f_i(\overline{\bm{x}}_t)-\nabla f_i(\bm{x}_t^{(i)}))\right\|^2\\
            & \overset{(c)}{\leq} 8L^2 \sum_{i=1}^{N}\left\|\bm{x}_t^{(i)}-\overline{\bm{x}}_t\right\|^2+2N\varsigma^2\\
            & \overset{}{=} 8L^2 \left\|\bm{X}_t-\overline{\bm{X}}_t\right\|_F^2 +2N\varsigma^2,
        \end{split}
    \end{equation}
    where ($a$) and ($c$) result from the inequality (i) from Fact \ref{FC1:} and the inequality (\ref{LG}) from Assumption \ref{L_smooth:}, and ($b$) holds by using the bound (\ref{inter-node_bounded_variance}) from Assumption \ref{UG_BV:}. 
    
    We obtain the statement of the lemma by considering the expectation of both sides of the inequality.
\end{proof}

\begin{lemm}
\label{al1_di_con_di:}
For $\bm{G}_{t}=[ \bm{g}_{t}^{(1)}, \bm{g}_{t}^{(2)},\cdots, \bm{g}_{t}^{(N)}]$ where $\bm{g}_{t}^{(i)}$ for all $i\in[N]$ and $t\in[t^\prime+1,\cdots,t^\prime+\tau-1]$ with $t^\prime \in \mathcal{T}$ is generated by Algorithm~\ref{alg:}, we have:
\begin{align*}
    \mathbbm{E}\left\|\sum_{j=t^\prime}^{t-1}(\bm{G}_{j} -\overline{\bm{G}}_{j})\right\|_F^2  \overset{}{\leq} \frac{2N\sigma^2(t-t^\prime)}{b}+2\tau\sum_{j=t^\prime}^{t-1}\mathbbm{E}\left\|\partial f(\bm{X}_j)-\overline{\partial f}(\bm{X}_j)\right\|_F^2,
\end{align*}
where the expectation $\mathbbm{E}[\cdot]$ is w.r.t the stochasticity of the algorithm.
\end{lemm}

\begin{proof}
\begin{align*}
\mathbbm{E}\left\|\sum_{j=t^\prime}^{t-1}(\bm{G}_{j} -\overline{\bm{G}}_{j})\right\|_F^2 
& \overset{}{=} 
    \mathbbm{E}\left\| \sum_{j=t^\prime}^{t-1}\left( \bm{G}_{j}-\partial f(\bm{X}_j)-\left(\overline{\bm{G}}_{j}-\overline{\partial f}(\bm{X}_j)\right)+\partial f(\bm{X}_j)-\overline{\partial f}(\bm{X}_j)\right)\right\|_F^2\\
& \overset{(a)}{\leq} 
    2\mathbbm{E}\left\| \sum_{j=t^\prime}^{t-1}\left(\bm{G}_{j}-\partial f(\bm{X}_j)-\left(\overline{\bm{G}}_{j}-\overline{\partial f}(\bm{X}_j)\right)\right)\right\|_F^2\\
    & \quad \quad \quad \quad \quad \quad \quad \quad \quad \quad \quad \quad \quad \quad \quad  +2\mathbbm{E}\left\|\sum_{j=t^\prime}^{t-1}\left(\partial f(\bm{X}_j)-\overline{\partial f}(\bm{X}_j)\right)\right\|_F^2\\
& \overset{(b)}{\leq} 
    2\mathbbm{E}\left\| \sum_{j=t^\prime}^{t-1}\left(\bm{G}_{j}-\partial f(\bm{X}_j)\right)\right\|_F^2+2\mathbbm{E}\left\|\sum_{j=t^\prime}^{t-1}\left(\partial f(\bm{X}_j)-\overline{\partial f}(\bm{X}_j)\right)\right\|_F^2\\
& \overset{(c)}{\leq} 
    2\mathbbm{E}\left\| \sum_{j=t^\prime}^{t-1}\left(\bm{G}_{j}-\partial f(\bm{X}_j)\right)\right\|_F^2+2\tau\sum_{j=t^\prime}^{t-1}\mathbbm{E}\left\|\partial f(\bm{X}_j)-\overline{\partial f}(\bm{X}_j)\right\|_F^2\\
& \overset{(d)}{=} 
    2\sum_{j=t^\prime}^{t-1}\mathbbm{E}\left\| \bm{G}_{j}-\partial f(\bm{X}_j)\right\|_F^2+2\tau\sum_{j=t^\prime}^{t-1}\mathbbm{E}\left\|\partial f(\bm{X}_j)-\overline{\partial f}(\bm{X}_j)\right\|_F^2\\
& \overset{}{=} 
    2\sum_{j=t^\prime}^{t-1}\sum_{i=1}^{N}\mathbbm{E}\left\| \frac{1}{b}\sum_{r=1}^{b}\left(\nabla f_i(\bm{x}_j^{(i)};\bm{\xi}_r^{(i)})-\nabla f_i(\bm{x}_j^{(i)})\right)\right\|^2\\
    & \quad \quad \quad \quad \quad \quad \quad \quad \quad \quad \quad \quad \quad \quad \quad  +2\tau\sum_{j=t^\prime}^{t-1}\mathbbm{E}\left\|\partial f(\bm{X}_j)-\overline{\partial f}(\bm{X}_j)\right\|_F^2\\
& \overset{(e)}{\leq}
    \frac{2N\sigma^2(t-t^\prime)}{b}+2\tau\sum_{j=t^\prime}^{t-1}\mathbbm{E}\left\|\partial f(\bm{X}_j)-\overline{\partial f}(\bm{X}_j)\right\|_F^2,
\end{align*}
where ($a$) results from the inequality (iii) from Fact \ref{FC3:} with $\eta = 1$, ($b$) follows by using the inequality (iii) from Fact \ref{FC4:}, ($d$) holds by using the fact that $\mathbbm{E}[ \bm{G}_{j}-\partial f(\bm{X}_j)]=\bm{0}$ for any $t^\prime \leq j$, and ($c$) and ($e$) use the inequalities (i) and (ii) from Fact \ref{FC1:}, respectively.  

Hence, the lemma is proved.
\end{proof}

\subsubsection{Descent Lemma and Consensus Distance}
\begin{lemm}
\label{al_1_Descent_Lemma:}
{\rm (Descent Lemma)}
 For $t \in [0,\cdots, T-1]$, the averages  $\overline{\bm{x}}_{t}=\frac{1}{N}\sum\limits_{i=1}^{N}\bm{x}_{t}^{(i)}$ of the iterates generated by Algorithm~\ref{alg:} with $\gamma_t = \gamma$ satisfy that
\begin{align*}
    \mathbbm{E}[F(\bm{\overline{x}}_{t+1})] 
    & \overset{}{\leq}
        \mathbbm{E}[F(\bm{\overline{x}}_{t})]  -\frac{\gamma}{2}\mathbbm{E}\|\nabla F(\overline{\bm{x}}_{t})\|^{2} -\frac{\gamma}{2}(1-L\gamma)\mathbbm{E}\left\|\overline{\nabla f}(\bm{X}_t)\right\|^{2}   \\
        & \quad \quad \quad \quad \quad \quad \quad \quad \quad \quad \quad \quad \quad \quad \quad \quad \quad \quad +\frac{L^2\gamma}{2N} \mathbbm{E}\|\bm{X}_t-\overline{\bm{X}}_{t}\|_F^{2} +  \frac{L\gamma^2\sigma^2}{2Nb},
\end{align*}
where the expectation $\mathbbm{E}[\cdot]$ is w.r.t the stochasticity of the algorithm.
\end{lemm}

\begin{proof}
    By the smoothness of $F(\cdot)$ (Assumption \ref{L_smooth:}), we have:
    \begin{equation}
        \begin{split}
            \label{al_1_Descent_Lemma_1:}
            \mathbbm{E}[F(\bm{\overline{x}}_{t+1})] & \overset{}{\leq} \mathbbm{E}[F(\bm{\overline{x}}_{t})] + \mathbbm{E}[\langle\nabla F(\bm{\overline{x}}_{t}), \bm{\overline{x}}_{t+1}-\bm{\overline{x}}_{t}\rangle] + \frac{ L}{2} \mathbbm{E}\|\bm{\overline{x}}_{t+1}-\bm{\overline{x}}_{t}\|^{2} \\
             & \overset{(a)}{=} \mathbbm{E}[F(\bm{\overline{x}}_{t})]  \underbrace{-\gamma \mathbbm{E}[\langle\nabla F(\bm{\overline{x}}_{t}),  \bm{\overline{g}}_t\rangle]}_{T_1} + \frac{L\gamma^2 }{2} \underbrace{\mathbbm{E}\|\bm{\overline{g}}_t\|^{2}}_{T_2
             },
        \end{split}
    \end{equation}
    where ($a$) holds  because of (\ref{al1_avrag_rule_x_1:}).

    We note that
    \begin{equation}
        \begin{split}
            T_1 & = -\gamma \mathbbm{E}[\langle\nabla F(\overline{\bm{x}}_{t}),  \bm{\overline{g}}_t\rangle]\\
            & \overset{}{=} 
                -\gamma \mathbbm{E}\left\langle\nabla F(\overline{\bm{x}}_{t}),  \frac{1}{N}\sum_{i=1}^{N}\nabla f_i(\bm{x}_t^{(i)})\right\rangle\\
            & \overset{(a)}{=} 
                -\frac{\gamma}{2}\mathbbm{E}\|\nabla F(\overline{\bm{x}}_{t})\|^{2} -\frac{\gamma}{2}\mathbbm{E}\left\|\frac{1}{N}\sum_{i=1}^{N}\nabla f_i(\bm{x}_t^{(i)})\right\|^{2} +\frac{\gamma}{2} \mathbbm{E}\left\|\nabla F(\bm{\overline{x}}_{t})-\frac{1}{N}\sum_{i=1}^{N}\nabla f_i(\bm{x}_t^{(i)})\right\|^{2}\\
            & \overset{(b)}{\leq}
                -\frac{\gamma}{2}\mathbbm{E}\|\nabla F(\overline{\bm{x}}_{t})\|^{2} -\frac{\gamma}{2}\mathbbm{E}\left\|\overline{\nabla f}(\bm{X}_t)\right\|^{2}  +\frac{L^2\gamma}{2N} \mathbbm{E}\|\bm{X}_t-\overline{\bm{X}}_{t}\|_F^{2},
        \end{split}
    \end{equation}
    where ($a$) holds because of  the equality (i) from Fact \ref{FC2:}, and the last inequality ($b$) uses the inequality (ii) from Fact \ref{FC1:}. Note that $\mathbbm{E}\|\bm{X}_t-\overline{\bm{X}}_{t}\|_F^{2}=\sum_{i=1}^{N}\|\bm{x}_t^{(i)}-\overline{\bm{x}}_{t}\|^{2}$.
    And we observe that
    
    \begin{align*}
        T_2 & =  \mathbbm{E}\|\bm{\overline{g}}_t\|^{2}\\
        & \overset{}{=}
            \mathbbm{E}\|\bm{\overline{g}}_t-\overline{\nabla f}(\bm{X}_t)+\overline{\nabla f}(\bm{X}_t)\|^{2}\\
        & \overset{(a)}{\leq}
            \mathbbm{E}\|\bm{\overline{g}}_t-\overline{\nabla f}(\bm{X}_t)\|^{2}+\|\overline{\nabla f}(\bm{X}_t)\|^{2} \\
        & \overset{}{=}
            \mathbbm{E}\left\|\frac{1}{Nb}\sum_{i=1}^{N}\sum_{r=1}^{b}\left(\nabla f_i(\bm{x}_{t}^{(i)};\bm{\xi}_{{t,r}}^{(i)})-\nabla f_i(\bm{x}_{t}^{(i)})\right)\right\|^{2} +\left\|\overline{\nabla f}(\bm{X}_t)\right\|^{2}\\
        & \overset{(b)}{\leq} 
            \frac{\sigma^2}{Nb}+\left\|\overline{\nabla f}(\bm{X}_t)\right\|^{2},
    \end{align*}
    where ($a$) uses the inequality (ii) from Fact \ref{FC1:} and ($b$) results from the inequality (\ref{intra-node_bounded_variance}) from Assumption~\ref{UG_BV:}.
    
    Substituting the upper bounds of $T_1$ and $T_2$ into (\ref{al_1_Descent_Lemma_1:}), the lemma is proved.
\end{proof}
Next, we present the upper bound of the expectation of the consensus distance for the estimation of global average accumulated direction, i.e., $\mathbbm{E}\|\bm{Y}_{t^{\prime}}- \bm{\overline{Y}}_{t^{\prime}}\|_F^2$ and $t^\prime \in \mathcal{T}:= \{t| t=0,\cdots, T \ {\rm and} \ {\rm mod}(t, \tau)=0\}$. 

\begin{lemm}
\label{al1_G_A_A_D_CD:}
For $\bm{Y}_{\tau}=[\bm{y}_{\tau}^{(1)}, \bm{y}_{\tau}^{(2)},\cdots, \bm{y}_{\tau}^{(N)}]$ where $\bm{y}_{\tau}^{(i)}$ for any $i \in [N]$ is
generated by Algorithm~\ref{alg:}, we have:
\begin{align*}
    \mathbbm{E}\|\bm{Y}_{\tau}-\bm{\overline{Y}}_{\tau}\|_F^2
     \overset{}{=} \frac{2N\lambda^2 \gamma^2\tau\sigma^2}{b} +  \frac{2\lambda^2\gamma^2\tau}{1-\lambda^2} \sum_{t=0}^{\tau-1}\mathbbm{E}\left\|\partial f(\bm{X}_t)-\overline{\partial f}(\bm{X}_t))\right\|_F^2,
\end{align*}
where the expectation $\mathbbm{E}[\cdot]$ is w.r.t the stochasticity of the algorithm.
\end{lemm}

\begin{proof}
We recall the initialization of Algorithm \ref{alg:} that $\bm{Y}_{0}=[\bm{y}_{0}^{(1)},\cdots,\bm{y}_{0}^{(N)}]=\bm{0}^{d \times N}$ and $\bm{H}_{0}=[\bm{h}_{0}^{(1)},\cdots,\bm{h}_{0}^{(N)}]=\bm{0}^{d \times N}$. Using the update rule (\ref{al1_matrix_rule_Y:}), we get:
\begin{equation}
    \label{al1_G_A_A_D_CD_1:}
    \begin{split}
        \mathbbm{E}\|\bm{Y}_{\tau}-\bm{\overline{Y}}_{\tau}\|_F^2
        & \overset{}{=} \mathbbm{E}\|(\bm{Y}_{0}+\bm{H}_{\tau}-\bm{H}_0)\bm{W}- (\bm{\overline{Y}}_{0}+\bm{\overline{H}}_{\tau}-\bm{\overline{H}}_0)\|_F^2\\
        & \overset{(a)}{=} \mathbbm{E}\|(\bm{Y}_{0}\bm{W}-\bm{\overline{Y}}_{0})+(\bm{H}_{\tau}-\bm{H}_0)(\bm{W}-\bm{Q})\|_F^2\\
        & \overset{(b)}{\leq}  \lambda^2\mathbbm{E}\|\bm{Y}_{0}-\bm{\overline{Y}}_{0}\|_F^2+\underbrace{\mathbbm{E}\|(\bm{H}_{\tau}-\bm{H}_{0})(\bm{W}-\bm{Q})\|_F^2}_{T_3}  \\
        & \quad \quad \quad \quad \quad \quad \quad \quad \quad \quad \quad \quad \quad \quad + \underbrace{2\mathbbm{E}\left[\left\langle\bm{Y}_{0}\bm{W}-\bm{\overline{Y}}_{0},(\bm{H}_{\tau}-\bm{H}_{0})(\bm{W}-\bm{Q})\right\rangle\right]}_{T_4},
    \end{split}
\end{equation}
where ($a$) and ($b$) hold by using the fact that $\bm{WQ}=\bm{Q}$ and the inequality (\ref{consenus_dist:}) from Assumption \ref{CN:}.

We note that
\begin{equation}
    \label{al1_G_A_A_D_CD_2:}
    \begin{split}
        T_3 & = \mathbbm{E}\|(\bm{H}_{\tau}-\bm{H}_{0})(\bm{W}-\bm{Q})\|_F^2\\
        & \overset{(a)}{\leq}
            \lambda^2\mathbbm{E}\left\|\sum_{t=0}^{\tau-1} \gamma(\bm{G}_t-\overline{\bm{G}}_t)\right\|_F^2 \\
        & \overset{(b)}{\leq} 
            \lambda^2 \gamma^2 \left[\frac{2N\tau\sigma^2}{b}+2\tau\sum_{t=0}^{\tau-1}\mathbbm{E}\left\|\partial f(\bm{X}_t)-\overline{\partial f}(\bm{X}_t)\right\|_F^2\right]  \\
        & \overset{}{\leq} 
            \frac{2N\lambda^2 \gamma^2\tau\sigma^2}{b}+2\lambda^2 \gamma^2\tau \sum_{t=0}^{\tau-1}\mathbbm{E}\|\partial f(\bm{X}_t)-\overline{\partial f}(\bm{X}_t)\|_F^2,
     \end{split}
\end{equation}
where ($a$) follows by using the update rule (\ref{al1_matrix_rule_H:}) and the inequality (\ref{consenus_dist:}), and the inequality ($b$) results from the statement of Lemma \ref{al1_di_con_di:}. We also observe that 
\begin{equation}
    \label{al1_G_A_A_D_CD_3:}
    \begin{split}
        T_4 & = 2\mathbbm{E}\left[\left\langle\bm{Y}_{0}\bm{W}-\bm{\overline{Y}}_{0},(\bm{H}_{\tau}-\bm{H}_{0})(\bm{W}-\bm{Q})\right\rangle\right]\\
        & \overset{(a)}{\leq} \eta \mathbbm{E}\|\bm{Y}_{0}\bm{W}-\bm{\overline{Y}}_{0}\|_F^2 + \frac{1}{\eta} \mathbbm{E}\left\|\sum_{t=0}^{\tau-1}\gamma\left(\partial f(\bm{X}_t)\bm{W}-\overline{\partial f}(\bm{X}_t)\right)\right\|_F^2 \\
        & \overset{(b)}{\leq}  \frac{1-\lambda^2}{2} \mathbbm{E}\|\bm{Y}_{0}-\bm{\overline{Y}}_{0}\|_F^2 + \frac{2\lambda^4\gamma^2\tau}{1-\lambda^2} \sum_{t=0}^{\tau-1}\mathbbm{E}\left\|\partial f(\bm{X}_t)-\overline{\partial f}(\bm{X}_t))\right\|_F^2,
    \end{split}
\end{equation}
where ($a$) holds by using the inequality  (ii) from Fact \ref{FC3:} and setting $\eta = \frac{1-\lambda^2}{2\lambda^2}$, and ($b$) uses the inequality (i) from Fact \ref{FC1:} and the inequality (\ref{consenus_dist:}) from Assumption \ref{CN:}.

Substituting the upper bounds of $T_3$ and $T_4$ into (\ref{al1_G_A_A_D_CD_1:}), we have:
\begin{equation}
    \label{al1_G_A_A_D_CD_4:}
    \begin{split}
        \mathbbm{E}\|\bm{Y}_{\tau}-\bm{\overline{Y}}_{\tau}\|_F^2
        & \overset{(a)}{\leq}    
        \frac{1+\lambda^2}{2}\mathbbm{E}\|\bm{Y}_{0}-\bm{\overline{Y}}_{0}\|_F^2+\frac{2N\lambda^2 \gamma^2\tau\sigma^2}{b}   \\
        &\quad \quad \quad \quad \quad \quad \quad \quad \quad \quad \quad \quad  \quad \quad \ +  \frac{2\lambda^2\gamma^2\tau}{1-\lambda^2} \sum_{t=0}^{\tau-1}\mathbbm{E}\left\|\partial f(\bm{X}_t)-\overline{\partial f}(\bm{X}_t))\right\|_F^2\\
        & \overset{}{=}  
        \frac{2N\lambda^2 \gamma^2\tau\sigma^2}{b} +  \frac{2\lambda^2\gamma^2\tau}{1-\lambda^2} \sum_{t=0}^{\tau-1}\mathbbm{E}\left\|\partial f(\bm{X}_t)-\overline{\partial f}(\bm{X}_t))\right\|_F^2,
    \end{split}
\end{equation}
where ($a$) results from the fact that $1+\frac{\lambda^2}{1-\lambda^2}=\frac{1}{1-\lambda^2}$. So far, we complete the proof.
\end{proof}

\begin{lemm}
\label{al1_G_A_A_D_CD_main:}
For $\bm{Y}_{t^\prime}=[\bm{y}_{t^\prime}^{(1)}, \bm{y}_{t^\prime}^{(2)},\cdots, \bm{y}_{t^\prime}^{(N)}]$ where $\bm{y}_{t^\prime}^{(i)}$ for any $i \in [N]$ and $t^\prime \in [\tau, \cdots, T-\tau]$ with $t^\prime \in \mathcal{T}$ is generated by Algorithm~\ref{alg:}, we have:
\begin{align*}
    \sum_{t^\prime=\tau}^{T-\tau}\mathbbm{E}\|\bm{Y}_{t^{\prime}}- \bm{\overline{Y}}_{t^{\prime}}\|_F^2
    &\overset{}{\leq}    \frac{16N\lambda^2 \gamma^2 T\sigma^2}{(1-\lambda^2)b} +  \frac{16\lambda^2\gamma^2\tau}{(1-\lambda^2)^2} \sum_{t=0}^{T-\tau-1} \mathbbm{E}\left\|\partial f(\bm{X}_t)-\overline{\partial f}(\bm{X}_t))\right\|_F^2,
\end{align*}
where the expectation $\mathbbm{E}[\cdot]$ is w.r.t the stochasticity of the algorithm.
\end{lemm}    

\begin{proof}
Using the update rule (\ref{al1_matrix_rule_Y:}), we get:
\begin{equation}
    \label{al1_G_A_A_D_CD_main_1:}
    \begin{split}
        \mathbbm{E}\|\bm{Y}_{t^{\prime}}- \bm{\overline{Y}}_{t^{\prime}}\|_F^2
        & \overset{}{=} \mathbbm{E}\|(\bm{Y}_{t^{\prime}-\tau}+\bm{H}_{t^{\prime}}-\bm{H}_{t^{\prime}-\tau})\bm{W}- (\bm{\overline{Y}}_{t^{\prime}-\tau}+\bm{\overline{H}}_{t^{\prime}}-\bm{\overline{H}}_{t^{\prime}-\tau})\|_F^2\\
        & \overset{(a)}{=} \mathbbm{E}\|(\bm{Y}_{t^{\prime}-\tau}\bm{W}-\bm{\overline{Y}}_{t^{\prime}-\tau})+(\bm{H}_{t^{\prime}}-\bm{H}_{t^{\prime}-\tau})(\bm{W}-\bm{Q})\|_F^2\\
        & \overset{(b)}{\leq} \lambda^2\mathbbm{E}\|\bm{Y}_{t^{\prime}-\tau}-\bm{\overline{Y}}_{t^{\prime}-\tau}\|_F^2+\underbrace{\mathbbm{E}\|(\bm{H}_{t^{\prime}}-\bm{H}_{t^{\prime}-\tau})(\bm{W}-\bm{Q})\|_F^2}_{T_5} \\
        & \quad \quad \quad \quad \quad \quad \quad \quad \quad \quad \quad \ \ + \underbrace{2\mathbbm{E}\left[\left\langle\bm{Y}_{t^{\prime}-\tau}\bm{W}-\bm{\overline{Y}}_{t^{\prime}-\tau},(\bm{H}_{t^{\prime}}-\bm{H}_{t^{\prime}-\tau})(\bm{W}-\bm{Q})\right\rangle\right]}_{T_6},
    \end{split}
\end{equation}
where ($a$) and ($b$) hold by using   $\bm{WQ}=\bm{Q}$ and the inequality (\ref{consenus_dist:}) from Assumption \ref{CN:}, respectively. 
We note that
\begin{equation}
    \label{al1_G_A_A_D_CD_main_2:}
    \begin{split}
        T_5 & = \mathbbm{E}\|(\bm{H}_{t^{\prime}}-\bm{H}_{t^{\prime}-\tau})(\bm{W}-\bm{Q})\|_F^2\\
        &\overset{}{=}
            \mathbbm{E}\left\|\sum_{t=t^{\prime}-\tau}^{t^{\prime}-1}\gamma(\bm{G}_t-\bm{G}_{t-\tau})(\bm{W}-\bm{Q})\right\|_F^2\\
        & \overset{(a)}{\leq}
              2\lambda^2\gamma^2\mathbbm{E}\left\|\sum_{t=t^{\prime}-\tau}^{t^{\prime}-1}(\bm{G}_t-\overline{\bm{G}}_t)\right\|_F^2+2\lambda^2\gamma^2\mathbbm{E}\left\|\sum_{t=t^{\prime}-\tau}^{t^{\prime}-1}(\bm{G}_{t-\tau}-\overline{\bm{G}}_{t-\tau})\right\|_F^2\\
        & \overset{(b)}{\leq} 
            \frac{8N\lambda^2\gamma^2\tau \sigma^2}{b}+4\lambda^2\gamma^2\tau \sum_{t=t^{\prime}-2\tau}^{t^{\prime}-1}\mathbbm{E}\|\partial f(\bm{X}_t)-\overline{\partial f}(\bm{X}_t)\|_F^2,
     \end{split}
\end{equation}
where ($a$) results from the inequality (iii) in Fact \ref{FC3:} and ($b$)  holds by using the statement of Assumption \ref{CN:} and Lemma \ref{al1_di_con_di:}. We also observe that
\begin{equation}
    \label{al1_G_A_A_D_CD_main_3:}
    \begin{split}
        T_6 & = 2\mathbbm{E}\left[\left\langle\bm{Y}_{t^{\prime}-\tau}\bm{W}-\bm{\overline{Y}}_{t^{\prime}-\tau},(\bm{H}_{t^{\prime}}-\bm{H}_{t^{\prime}-\tau})(\bm{W}-\bm{Q})\right\rangle\right]\\
        &\overset{(a)}{=}
            2\mathbbm{E}\left[\left\langle\bm{Y}_{t^{\prime}-\tau}\bm{W}-\bm{\overline{Y}}_{t^{\prime}-\tau},\sum_{t=t^{\prime}-\tau}^{t^{\prime}-1}\gamma\left(\partial f(\bm{X}_t)-\partial f(\bm{X}_{t-\tau})\right)(\bm{W}-\bm{Q})\right\rangle\right]\\
         & \overset{(b)}{\leq}
            \frac{1-\lambda^2}{2\lambda^2} \mathbbm{E}\|\bm{Y}_{t^{\prime}-\tau}\bm{W}-\bm{\overline{Y}}_{t^{\prime}-\tau}\|_F^2 +\\
            & \quad \quad \quad \quad \frac{2\lambda^2\gamma^2\tau }{1-\lambda^2}\sum_{t=t^{\prime}-\tau}^{t^{\prime}-1} \mathbbm{E}\left\|\left(\partial f(\bm{X}_t)\bm{W}-\overline{\partial f}(\bm{X}_t)\right)-\left(\partial f(\bm{X}_{t-\tau})\bm{W}-\overline{\partial f}(\bm{X}_{t-\tau})\right)\right\|_F^2\\
          & \overset{(c)}{\leq}
            \frac{1-\lambda^2}{2} \mathbbm{E}\|\bm{Y}_{t^{\prime}-\tau}-\bm{\overline{Y}}_{t^{\prime}-\tau}\|_F^2 +\\
            & \quad \quad \quad \quad \quad \quad  \frac{4\lambda^4\gamma^2\tau }{1-\lambda^2}\sum_{t=t^{\prime}-\tau}^{t^{\prime}-1} \left[\mathbbm{E}\left\|\partial f(\bm{X}_t)-\overline{\partial f}(\bm{X}_t)\right\|_F^2+\left\|\partial f(\bm{X}_{t-\tau})-\overline{\partial f}(\bm{X}_{t-\tau})\right\|_F^2\right]\\
          & \overset{}{\leq}
            \frac{1-\lambda^2}{2} \mathbbm{E}\|\bm{Y}_{t^{\prime}-\tau}-\bm{\overline{Y}}_{t^{\prime}-\tau}\|_F^2 + \frac{4\lambda^4\gamma^2\tau }{1-\lambda^2}\sum_{t=t^{\prime}-2\tau}^{t^{\prime}-1} \mathbbm{E}\left\|\partial f(\bm{X}_t)-\overline{\partial f}(\bm{X}_t)\right\|_F^2,
      \end{split}
\end{equation}
where ($a$) follows by using the equality $\mathbbm{E}[\bm{G}_t] = \partial f(\bm{X}_t)$ for all $t\in[0,\cdots,T]$,  ($b$) uses the fact that $\pm \langle\bm{A},\bm{B}\rangle \leq \frac{1}{2\eta} \|\bm{A}\|_F^2 + 2\eta \|\bm{B}\|_F^2$ with $\eta = \frac{1-\lambda^2}{2\lambda^2}$ and the inequality (i) from Fact \ref{FC1:}, and ($c$) uses the inequality (\ref{consenus_dist:}) from Assumption~\ref{CN:}.

Substituting the upper bounds of $T_5$ and $T_6$ into (\ref{al1_G_A_A_D_CD_main_1:}), we have:
\begin{equation}
    \label{al1_G_A_A_D_CD_main_4:}
    \begin{split}
        \mathbbm{E}\|\bm{Y}_{t^{\prime}}- \bm{\overline{Y}}_{t^{\prime}}\|_F^2
        & \overset{}{\leq} \frac{1+\lambda^2}{2} \mathbbm{E}\|\bm{Y}_{t^{\prime}-\tau}-\bm{\overline{Y}}_{t^{\prime}-\tau}\|_F^2+\frac{8N\lambda^2\gamma^2\tau\sigma^2}{b}  \\
        & \quad \quad \quad \quad \quad \quad \quad \quad \quad \quad + \frac{4\lambda^2\gamma^2\tau }{1-\lambda^2}\sum_{t=t^{\prime}-2\tau}^{t^{\prime}-1} \mathbbm{E}\left\|\partial f(\bm{X}_t)-\overline{\partial f}(\bm{X}_t)\right\|_F^2.
    \end{split}
\end{equation}
To simplify the description, we set  $\bm{A}_{t}=\frac{2N\lambda^2 \gamma^2\sigma^2}{b} +  \frac{2\lambda^2\gamma^2\tau}{1-\lambda^2} \mathbbm{E}\left\|\partial f(\bm{X}_t)-\overline{\partial f}(\bm{X}_t))\right\|_F^2$ and then recursively apply the (\ref{al1_G_A_A_D_CD_main_4:}) from $t^{\prime}$ to $\tau$ to get:
\begin{equation}
    \label{al1_G_A_A_D_CD_main_5:}
    \begin{split}
        \mathbbm{E}\|\bm{Y}_{t^{\prime}}- \bm{\overline{Y}}_{t^{\prime}}\|_F^2
        & \overset{}{\leq}  \frac{1+\lambda^2}{2}\mathbbm{E}\|\bm{Y}_{t^{\prime}-\tau}-\bm{\overline{Y}}_{t^{\prime}-\tau}\|_F^2 +  2 \sum_{t=t^{\prime}-2\tau}^{t^{\prime}-1} \bm{A}_{t}\\
        & \overset{}{\leq}  \left(\frac{1+\lambda^2}{2}\right)^{\frac{t^\prime-\tau}{\tau}}\mathbbm{E}\|\bm{Y}_{\tau}-\bm{\overline{Y}}_{\tau}\|_F^2 +2 \sum_{j^{\prime}=0}^{t^{\prime}-2\tau} \left(\frac{1+\lambda^2}{2}\right)^{\frac{t^\prime-2\tau-j^\prime}{\tau}} \sum_{t=j^{\prime}}^{j^{\prime}+2\tau-1} \bm{A}_{t}\\
        & \overset{(a)}{\leq}  \left(\frac{1+\lambda^2}{2}\right)^{\frac{t^\prime-\tau}{\tau}}\sum_{t=0}^{\tau-1}\bm{A}_{t} +2 \sum_{j^{\prime}=0}^{t^{\prime}-2\tau} \left(\frac{1+\lambda^2}{2}\right)^{\frac{t^\prime-2\tau-j^\prime}{\tau}} \sum_{t=j^{\prime}}^{j^{\prime}+2\tau-1} \bm{A}_{t},
    \end{split}
\end{equation}
where ($a$) results from the statement of Lemma \ref{al1_G_A_A_D_CD:}.
Finally, Summing over all (\ref{al1_G_A_A_D_CD_main_5:}) for $t^{\prime} \in \{\tau, \cdots, T-\tau\}$,  we have:
\begin{equation}
    \label{al1_G_A_A_D_CD_main_6:}
    \begin{split}
        \sum_{t^\prime=\tau}^{T-\tau}\mathbbm{E}\|\bm{Y}_{t^{\prime}}- \bm{\overline{Y}}_{t^{\prime}}\|_F^2
        & \overset{}{\leq}   \sum_{t^\prime=\tau}^{T-\tau}\left(\frac{1+\lambda^2}{2}\right)^{\frac{t^\prime-\tau}{\tau}}\sum_{t=0}^{\tau-1}\bm{A}_{t} +2 \sum_{t^\prime=\tau}^{T-\tau}\sum_{j^{\prime}=0}^{t^{\prime}-2\tau} \left(\frac{1+\lambda^2}{2}\right)^{\frac{t^\prime-2\tau-j^\prime}{\tau}} \sum_{t=j^{\prime}}^{j^{\prime}+2\tau-1} \bm{A}_{t}\\
        & \overset{}{=}   \sum_{t^\prime=\tau}^{T-\tau}\left(\frac{1+\lambda^2}{2}\right)^{\frac{t^\prime-\tau}{\tau}}\sum_{t=0}^{\tau-1}\bm{A}_{t} +2 \sum_{j^{\prime}=0}^{T-3\tau}\sum_{t^\prime=j^{\prime}+2\tau}^{T-\tau} \left(\frac{1+\lambda^2}{2}\right)^{\frac{t^\prime-2\tau-j^\prime}{\tau}} \sum_{t=j^{\prime}}^{j^{\prime}+2\tau-1} \bm{A}_{t}\\
        & \overset{(a)}{\leq}   \frac{2}{1-\lambda^2}\sum_{t=0}^{\tau-1}\bm{A}_{t} + \frac{4}{1-\lambda^2} \sum_{j^{\prime}=0}^{T-3\tau} \sum_{t=j^{\prime}}^{j^{\prime}+2\tau-1} \bm{A}_{t}\\
        & \overset{}{\leq} \frac{8}{1-\lambda^2} \sum_{t=0}^{T-\tau-1} \bm{A}_{t},
    \end{split}
\end{equation}
where $j^\prime \in \mathcal{T}$, and ($a$) holds by using the inequalities $\sum\limits_{t^\prime=\tau}^{T-\tau}\left(\frac{1+\lambda^2}{2}\right)^{\frac{t^\prime-\tau}{\tau}} \leq \sum\limits_{k=0}^{\infty}\left(\frac{1+\lambda^2}{2}\right)^k \leq \frac{2}{1-\lambda^2}$ and $\sum \limits_{t^\prime=j^{\prime}+2\tau}^{T-\tau} \left(\frac{1+\lambda^2}{2}\right)^{\frac{t^\prime-2\tau-j^\prime}{\tau}} \leq \sum\limits_{k=0}^{\infty}\left(\frac{1+\lambda^2}{2}\right)^k \leq \frac{2}{1-\lambda^2}$. 

By making a simple substitution for (\ref{al1_G_A_A_D_CD_main_6:}) by using $\bm{A}_{t}$, we complete the proof.
\end{proof}

\begin{lemm}
{\rm (Consensus Distance)} 
\label{al_1_CD:}
For $\bm{X}_{t}=[\bm{x}_{t}^{(1)}, \bm{x}_{t}^{(2)},\cdots, \bm{x}_{t}^{(N)}]$ where $\bm{x}_{t}^{(i)}$ for any $i \in [N]$ and $t \in [0, \cdots, T-1]$ is generated by Algorithm \ref{alg:} with $\gamma_t = \gamma \leq \min \left\{ \frac{1}{4\sqrt{2}L\tau}, \frac{(1-\lambda^2)^2}{32\sqrt{6}\lambda^2L\tau}\right\}$, we have:
\begin{align*}
    \sum_{t =0}^{T-1}\mathbbm{E}\|\bm{X}_{t}-\overline{\bm{X}}_{t}\|_F^2 
    & \overset{}{\leq} \frac{768N\lambda^4 \gamma^2 \tau T\sigma^2}{(1-\lambda^2)^3b} +\frac{1536N\lambda^4\gamma^2\tau^2 T\varsigma^2}{(1-\lambda^2)^4} \\
    & \quad \quad \quad \quad \quad \quad \quad \quad \quad \quad +  \frac{12N\gamma^2(\tau-1) T\sigma^2}{b}+24N\gamma^2\tau(\tau-1)T\varsigma^2,
\end{align*}
where the expectation $\mathbbm{E}[\cdot]$ is w.r.t the stochasticity of the algorithm.
\end{lemm}

\begin{proof}
We first transform $\sum\limits_{t =0}^{T-1}\mathbbm{E}\|\bm{X}_{t}-\overline{\bm{X}}_{t}\|_F^2$ to the following form:
\begin{equation}
    \label{al_1_CD1:}
    \sum_{t =0}^{T-1}\mathbbm{E}\|\bm{X}_{t}-\overline{\bm{X}}_{t}\|_F^2 = \sum_{t^\prime =0}^{T-\tau}\sum_{t=t^\prime}^{t^\prime+\tau-1}\mathbbm{E}\|\bm{X}_{t}-\overline{\bm{X}}_{t}\|_F^2,
\end{equation}
where $t^\prime \in \mathcal{T}$. According to the update rule (\ref{al1_matrix_rule_X2:}), we have: for any $t \in [t^\prime+1, \cdots, t^\prime+\tau-1]$,
\begin{align*}
    \mathbbm{E}\|\bm{X}_{t}-\overline{\bm{X}}_{t}\|_F^2 &= \mathbbm{E}\left\|\bm{X}_{t^\prime}-\overline{\bm{X}}_{t^\prime}- \gamma \sum_{j=t^\prime}^{t-1}(\bm{G}_{j} -\overline{\bm{G}}_{j})\right\|_F^2\\
    & \overset{(a)}{\leq}
        2\mathbbm{E}\|\bm{X}_{t^\prime}-\overline{\bm{X}}_{t^\prime}\|_F^2  + 2\gamma^2\mathbbm{E}\left\|\sum_{j=t^\prime}^{t-1}(\bm{G}_{j} -\overline{\bm{G}}_{j})\right\|_F^2\\
    & \overset{(b)}{\leq}
        2\mathbbm{E}\|\bm{X}_{t^\prime}-\overline{\bm{X}}_{t^\prime}\|_F^2  + \frac{4N\gamma^2\sigma^2(t-t^\prime)}{b}+4\gamma^2\tau\sum_{j=t^\prime}^{t-1}\mathbbm{E}\left\|\partial f(\bm{X}_j)-\overline{\partial f}(\bm{X}_j)\right\|_F^2\\
    & \overset{(c)}{\leq}
        2\mathbbm{E}\|\bm{X}_{t^\prime}-\overline{\bm{X}}_{t^\prime}\|_F^2  + \frac{4N\gamma^2\sigma^2(t-t^\prime)}{b}+8N\gamma^2\varsigma^2\tau(t-t^\prime)\\
        & \quad \quad \quad \quad \quad \quad \quad \quad \quad \quad \quad \quad \quad \quad \quad \quad \quad \quad \quad \quad +32L^2\gamma^2\tau\sum_{j=t^\prime}^{t-1} \mathbbm{E}\left\|\bm{X}_j-\overline{\bm{X}}_j\right\|_F^2 \\
    & \overset{(d)}{\leq}
        2\mathbbm{E}\|\bm{X}_{t^\prime}-\overline{\bm{X}}_{t^\prime}\|_F^2  + \frac{4N\gamma^2\sigma^2(t-t^\prime)}{b}+8N\gamma^2\varsigma^2\tau(t-t^\prime)\\
        & \quad \quad \quad \quad \quad \quad \quad \quad \quad \quad \quad \quad \quad \quad \quad \quad \quad \quad \quad \quad \quad \quad \quad  +\frac{1}{\tau}\sum_{j=t^\prime}^{t-1} \mathbbm{E}\left\|\bm{X}_j-\overline{\bm{X}}_j\right\|_F^2 \\
    & \overset{(e)}{\leq}
        2\left(1+\frac{1}{\tau}\right)^{t-t^\prime}\mathbbm{E}\|\bm{X}_{t^\prime}-\overline{\bm{X}}_{t^\prime}\|_F^2  \\
        & \quad \quad \quad \quad \quad \quad \quad \quad \quad  + \left(\frac{4N\gamma^2\sigma^2}{b\tau}+8N\gamma^2\varsigma^2\right)\sum_{j=t^\prime+1}^{t}\left(1+\frac{1}{\tau}\right)^{t-j}(j-t^\prime) \\
    & \overset{(f)}{\leq}
        6\mathbbm{E}\|\bm{X}_{t^\prime}-\overline{\bm{X}}_{t^\prime}\|_F^2  + \left(\frac{12N\gamma^2\sigma^2}{b\tau}+24N\gamma^2\varsigma^2\right)\sum_{j=t^\prime+1}^{t}(j-t^\prime),
\end{align*}
where ($a$) holds by using the inequality (iii) from Fact \ref{FC3:} and setting $\eta = 1$, ($b$) and ($c$) use the statements of lemma \ref{al1_di_con_di:} and lemma \ref{griadient_consensus:}, respectively, ($d$) follows from the fact that $32L^2\gamma^2\tau \leq \frac{1}{\tau}$ holds if $\gamma \leq \frac{1}{4\sqrt{2}L\tau}$, ($e$) results from recursively substituting every $\mathbbm{E}\left\|\bm{X}_j-\overline{\bm{X}}_j\right\|_F^2$ in the last term of ($d$), the inequality ($f$) follows from the fact that $(1+\frac{1}{\tau})^{t-j} \leq (1+\frac{1}{\tau})^{t-t^\prime} \leq (1+\frac{1}{\tau})^{\tau} \leq 3 $. Then, summing over $\mathbbm{E}\|\bm{X}_{t}-\overline{\bm{X}}_{t}\|_F^2$ from $0$ to $
T-1$, we have:
\begin{align}
    \sum_{t =0}^{T-1}\mathbbm{E}\|\bm{X}_{t}-\overline{\bm{X}}_{t}\|_F^2 & = \sum_{t^\prime =0}^{T-\tau}\sum_{t=t^\prime}^{t^\prime+\tau-1}\mathbbm{E}\|\bm{X}_{t}-\overline{\bm{X}}_{t}\|_F^2\notag \\
    &\overset{}{=}
        \sum_{t^\prime =0}^{T-\tau}\bigg[\mathbbm{E}\|\bm{X}_{t^\prime}-\overline{\bm{X}}_{t^\prime}\|_F^2+\sum_{t=t^\prime+1}^{t^\prime+\tau-1}\mathbbm{E}\|\bm{X}_{t}-\overline{\bm{X}}_{t}\|_F^2\bigg]\notag \\
    & \overset{}{\leq}
        6\tau\sum_{t^\prime =0}^{T-\tau}\mathbbm{E}\|\bm{X}_{t^\prime}-\overline{\bm{X}}_{t^\prime}\|_F^2   \notag \\
        & \quad \quad \quad \quad \quad \quad \quad \quad + \left(\frac{12N\gamma^2\sigma^2}{b\tau}+24N\gamma^2\varsigma^2\right)\sum_{t^\prime =0}^{T-\tau}\sum_{t=t^\prime+1}^{t^\prime+\tau-1}\sum_{j=t^\prime+1}^{t}(j-t^\prime)\notag \\
    & \overset{}{\leq}
        6\tau\sum_{t^\prime =0}^{T-\tau}\mathbbm{E}\|\bm{X}_{t^\prime}-\overline{\bm{X}}_{t^\prime}\|_F^2   \notag \\
        & \quad \quad \quad \quad \quad \quad \quad \quad + \left(\frac{12N\gamma^2\sigma^2}{b\tau}+24N\gamma^2\varsigma^2\right)\sum_{t^\prime =0}^{T-\tau}\sum_{t=t^\prime}^{t^\prime+\tau-1}\sum_{j=t^\prime+1}^{t^\prime+\tau-1}(j-t^\prime)\notag \\
    & \overset{}{=}
        6\tau\sum_{t^\prime =0}^{T-\tau}\mathbbm{E}\|\bm{X}_{t^\prime}-\overline{\bm{X}}_{t^\prime}\|_F^2  +  \left(\frac{12N\gamma^2\sigma^2}{b\tau}+24N\gamma^2\varsigma^2\right)\frac{\tau(\tau-1)T}{2} \notag \\
    & \overset{}{=}
        6\tau\sum_{t^\prime =0}^{T-\tau}\mathbbm{E}\|\bm{X}_{t^\prime}-\overline{\bm{X}}_{t^\prime}\|_F^2  +  \frac{6N\gamma^2(\tau-1)T\sigma^2}{b}+12N\gamma^2\tau(\tau-1)T\varsigma^2. \label{al_1_CD3:}
\end{align}

Further, using the update rule (\ref{al1_matrix_rule_X1:}), we get: for $t^\prime>0$,
\begin{align}
    \label{al_1_CD4:}
    \mathbbm{E}\|\bm{X}_{t^{\prime}}-\bm{\overline{X}}_{t^{\prime}}\|_F^2 
    & \overset{}{=} \mathbbm{E}\|(\bm{X}_{t^{\prime}-\tau}\bm{W}- \bm{\overline{X}}_{t^{\prime}-\tau}) -( \bm{Y}_{t^{\prime}}\bm{W}- \bm{\overline{Y}}_{t^{\prime}})\|_F^2 \notag \\
    & \overset{(a)}{\leq} \frac{1+\lambda^2}{2}\mathbbm{E}\|\bm{X}_{t^{\prime}-\tau}- \bm{\overline{X}}_{t^{\prime}-\tau}\|_F^2+\frac{(1+\lambda^2)\lambda^2}{1-\lambda^2}\mathbbm{E}\|\bm{Y}_{t^{\prime}}- \bm{\overline{Y}}_{t^{\prime}}\|_F^2 \notag  \\
    & \overset{(b)}{\leq} \left(\frac{1+\lambda^2}{2}\right)^{\frac{t^\prime}{\tau}} \mathbbm{E}\|\bm{X}_{0}- \bm{\overline{X}}_{0}\|_F^2 + \frac{(1+\lambda^2)\lambda^2}{1-\lambda^2}\sum_{j^\prime=\tau}^{t^\prime}\left(\frac{1+\lambda^2}{2}\right)^{\frac{t^\prime-j^\prime}{\tau}} \mathbbm{E}\|\bm{Y}_{j^{\prime}}- \bm{\overline{Y}}_{j^{\prime}}\|_F^2 \notag \\
    & \overset{(c)}{=} \frac{(1+\lambda^2)\lambda^2}{1-\lambda^2}\sum_{j^\prime=\tau}^{t^\prime}\left(\frac{1+\lambda^2}{2}\right)^{\frac{t^\prime-j^\prime}{\tau}} \mathbbm{E}\|\bm{Y}_{j^{\prime}}- \bm{\overline{Y}}_{j^{\prime}}\|_F^2,
\end{align}
where ($a$) results from the inequality (iii) in Fact \ref{FC3:} with $\eta = \frac{1-\lambda^2}{2\lambda^2}$ and the inequality (\ref{consenus_dist:}) from Assumption \ref{CN:}, ($b$) recursively applies ($a$) from $t^\prime-\tau$ to $0$, and the last inequality ($c$) holds by using the fact that $\bm{X}_0=\overline{\bm{X}}_0$. And now, summing over (\ref{al_1_CD4:}) from $t^\prime = 0$ to $T-\tau$, we have:
\begin{equation}
    \label{al_1_CD5:}
    \begin{split}
        \sum_{t^\prime=0}^{T-\tau}\mathbbm{E}\|\bm{X}_{t^{\prime}}-\bm{\overline{X}}_{t^{\prime}}\|_F^2 
        & \overset{}{=}
            \sum_{t^\prime=\tau}^{T-\tau}\mathbbm{E}\|\bm{X}_{t^{\prime}}-\bm{\overline{X}}_{t^{\prime}}\|_F^2 \\
        & \overset{}{\leq}
            \frac{(1+\lambda^2)\lambda^2}{1-\lambda^2}\sum_{t^\prime=\tau}^{T-\tau}\sum_{j^\prime=\tau}^{t^\prime}\left(\frac{1+\lambda^2}{2}\right)^{\frac{t^\prime-j^\prime}{\tau}} \mathbbm{E}\|\bm{Y}_{j^{\prime}}- \bm{\overline{Y}}_{j^{\prime}}\|_F^2\\
        & \overset{}{=}
            \frac{(1+\lambda^2)\lambda^2}{1-\lambda^2}\sum_{j^\prime=\tau}^{T-\tau}\sum_{t^\prime=j^\prime}^{T-\tau}\left(\frac{1+\lambda^2}{2}\right)^{\frac{t^\prime-j^\prime}{\tau}} \mathbbm{E}\|\bm{Y}_{j^{\prime}}- \bm{\overline{Y}}_{j^{\prime}}\|_F^2\\
        & \overset{(a)}{\leq}
            \frac{4\lambda^2}{(1-\lambda^2)^2}\sum_{t^\prime=\tau}^{T-\tau} \mathbbm{E}\|\bm{Y}_{t^{\prime}}- \bm{\overline{Y}}_{t^{\prime}}\|_F^2\\
        & \overset{(b)}{\leq} 
            \frac{64N\lambda^4 \gamma^2 T\sigma^2}{(1-\lambda^2)^3b} +  \frac{64\lambda^4\gamma^2\tau}{(1-\lambda^2)^4} \sum_{t=0}^{T-\tau-1} \bigg[8L^2 \left\|\bm{X}_t-\overline{\bm{X}}_t\right\|_F^2 +2N\varsigma^2\bigg]\\
        & \overset{}{=} 
            \frac{64N\lambda^4 \gamma^2 T\sigma^2}{(1-\lambda^2)^3b} +\frac{128N\lambda^4\gamma^2\tau T\varsigma^2}{(1-\lambda^2)^4} +  \frac{512\lambda^4L^2\gamma^2\tau}{(1-\lambda^2)^4} \sum_{t=0}^{T-\tau-1}  \left\|\bm{X}_t-\overline{\bm{X}}_t\right\|_F^2,
    \end{split}
\end{equation}
where $j^\prime \in \mathcal{T}$, ($a$) holds from the facts that $\sum \limits_{t^\prime=j^\prime}^{T-\tau}\left(\frac{1+\lambda^2}{2}\right)^{\frac{t^\prime-j^\prime}{\tau}}\leq \sum\limits_{k=0}^{\infty}\left(\frac{1+\lambda^2}{2}\right)^{k} \leq \frac{2}{1-\lambda^2}$ and $1+\lambda^2 \leq 2$, and the third inequality ($b$) uses the statements of lemma \ref{al1_G_A_A_D_CD_main:} and lemma \ref{griadient_consensus:}.

Next, substituting the inequality (\ref{al_1_CD5:}) into (\ref{al_1_CD3:}), we get:
\begin{align*}
    \sum_{t =0}^{T-1}\mathbbm{E}\|\bm{X}_{t}-\overline{\bm{X}}_{t}\|_F^2 
    & \overset{}{\leq}
        6\tau\sum_{t^\prime =0}^{T-\tau}\mathbbm{E}\|\bm{X}_{t^\prime}-\overline{\bm{X}}_{t^\prime}\|_F^2  +  \frac{6N\gamma^2(\tau-1)T\sigma^2}{b}+12N\gamma^2\tau(\tau-1)T\varsigma^2\\
    & \overset{}{\leq}
        \frac{384N\lambda^4 \gamma^2 \tau T\sigma^2}{(1-\lambda^2)^3b} +\frac{768N\lambda^4\gamma^2\tau^2 T\varsigma^2}{(1-\lambda^2)^4} \\
        & \quad \quad \quad \quad \quad \quad \quad \quad \quad \quad \quad \quad \quad  +\frac{6N\gamma^2(\tau-1)T\sigma^2}{b}+12N\gamma^2\tau(\tau-1)T\varsigma^2 \\
        & \quad \quad \quad \quad \quad \quad \quad \quad \quad \quad \quad \quad \quad \quad \quad \quad  +  \frac{3072\lambda^4L^2\gamma^2\tau^2}{(1-\lambda^2)^4} \sum_{t=0}^{T-1}  \left\|\bm{X}_t-\overline{\bm{X}}_t\right\|_F^2.
\end{align*}
Finally, the inequality $1-\frac{3072\lambda^4L^2\gamma^2\tau^2}{(1-\lambda^2)^4} \geq\frac{1}{2}$ holds if we set $\gamma \leq  \frac{(1-\lambda^2)^2}{32\sqrt{6}\lambda^2L\tau}$. Hence, we complete the proof.
\end{proof}

\subsubsection{The Proof of Theorem \ref{alg:}}

In this section, we give the proof of Theorem \ref{alg:} using the statements of several lemmas listed in the previous subsections. 

\begin{lemm}
\label{al_1_PT1:}
For all $t \in [0,\cdots,T-1]$, the averages $\overline{\bm{x}}_{t}=\frac{1}{N}\sum\limits_{i=1}^{N}\bm{x}_{t}^{(i)}$ of the iterates generated by Algorithm \ref{alg:} with $\gamma_t = \gamma \leq \min \left\{ \frac{1}{4\sqrt{2}L\tau}, \frac{(1-\lambda^2)^2}{32\sqrt{6}\lambda^2L\tau}\right\}$ satisfy that
\begin{align*}
    \frac{1}{T}\sum_{t=0}^{T-1}\|\nabla F(\bm{\overline{x}}_{t})\|^{2}
    & \overset{}{\leq}     
        \frac{2(F(\bm{\overline{x}}_{0}) - F^*)}{\gamma T}   + \frac{L\gamma\sigma^2}{Nb} +  \frac{12L^2\gamma^2(\tau-1) \sigma^2}{b}+24L^2\gamma^2\tau(\tau-1)\varsigma^2 \\
        & \quad \quad \quad \quad \quad \quad \quad \quad \quad \quad \quad \quad +  \frac{768\lambda^4 L^2 \gamma^2 \tau \sigma^2}{(1-\lambda^2)^3b} +\frac{1536\lambda^4L^2\gamma^2\tau^2 \varsigma^2}{(1-\lambda^2)^4},   
\end{align*}
where the expectation $\mathbbm{E}[\cdot]$ is w.r.t the stochasticity of the algorithm.
\end{lemm}
\begin{proof}
Using the statement of descent lemma \ref{al_1_Descent_Lemma:}, we get:
\begin{equation}
    \label{al_1_PT1_1:}
    \begin{split}
        \mathbbm{E}[F(\bm{\overline{x}}_{t+1})]
        & \overset{}{\leq} 
            \mathbbm{E}[F(\bm{\overline{x}}_{t})]  -\frac{\gamma}{2}\mathbbm{E}\|\nabla F(\overline{\bm{x}}_{t})\|^{2} -\frac{\gamma}{2}(1-L\gamma)\mathbbm{E}\left\|\overline{\nabla f}(\bm{X}_t)\right\|^{2}   \\
            & \quad \quad \quad \quad \quad \quad \quad \quad \quad \quad \quad \quad \quad \quad \quad \quad \quad \quad +\frac{L^2\gamma}{2N} \mathbbm{E}\|\bm{X}_t-\overline{\bm{X}}_{t}\|_F^{2} +  \frac{L\gamma^2\sigma^2}{2Nb} \\
        & \overset{(a)}{\leq}
            \mathbbm{E}[F(\bm{\overline{x}}_{t})]  -\frac{\gamma}{2}\|\nabla F(\bm{\overline{x}}_{t})\|^{2}  +\frac{L^2\gamma}{2N} \mathbbm{E}\|\bm{X}_t-\overline{\bm{X}}_{t}\|_F^{2} + \frac{L\gamma^2\sigma^2}{2Nb},
    \end{split}
\end{equation}

where ($a$) holds if we set $\gamma \leq \frac{1}{4\sqrt{2}L\tau} \leq \frac{1}{L}$.

Next, summing over all (\ref{al_1_PT1_1:}) for $t \in [0,\cdots,T-1]$ and making a simple arrangement, we have:
\begin{equation}
    \label{al_1_PT1_2:}
    \begin{split}
        \frac{1}{T}\sum_{t=0}^{T-1}\|\nabla F(\bm{\overline{x}}_{t})\|^{2}
        & \overset{}{\leq}  \frac{2}{\gamma T}[\mathbbm{E}[F(\bm{\overline{x}}_{0})]] - \mathbbm{E}[F(\bm{\overline{x}}_{T})]   +\frac{L^2}{NT} \sum_{t=0}^{T-1}\mathbbm{E}\|\bm{X}_t-\overline{\bm{X}}_{t}\|_F^{2} + \frac{L\gamma\sigma^2}{Nb}\\
        & \overset{(a)}{\leq}  \frac{2(F(\bm{\overline{x}}_{0}) - F^*)}{\gamma T}   + \frac{L^2}{NT} \sum_{t=0}^{T-1}\mathbbm{E}\|\bm{X}_t-\overline{\bm{X}}_{t}\|_F^{2} + \frac{L\gamma\sigma^2}{Nb}\\
        & \overset{(b)}{\leq}  \frac{2(F(\bm{\overline{x}}_{0}) - F^*)}{\gamma T}   + \frac{L\gamma\sigma^2}{Nb} +  \frac{12L^2\gamma^2(\tau-1) \sigma^2}{b}+24L^2\gamma^2\tau(\tau-1)\varsigma^2 \\
        & \quad \quad \quad \quad \quad \quad \quad \quad \quad \quad \quad \quad +  \frac{768\lambda^4 L^2 \gamma^2 \tau \sigma^2}{(1-\lambda^2)^3b} +\frac{1536\lambda^4L^2\gamma^2\tau^2 \varsigma^2}{(1-\lambda^2)^4},
    \end{split}
\end{equation}
where ($a$) uses the fact that $F^*=\inf_{\bm{x}\in \mathbbm{R}^d}F(\bm{x})>-\infty$, and ($b$) results from the statement of lemma \ref{al_1_CD:}.
So far, we complete the proof.
\end{proof}

\subsection{Convergence Proof of Algorithm \ref{alg_momen:}}
\label{CP_alg_momen:}
In this section, we provide the theoretical analysis of the proposed algorithm DSE-MVR. Before doing the specific derivation, we briefly review the update rules of Algorithm \ref{alg_momen:} and give the average and matrix expressions of the update process. 

\textbf{First}, we represent the update rules of the model parameters in average form as follows.
        
If the iteration $t$ satisfies ${\rm mod}(t+1, \tau)=0$, it is easy
to check that the following relations hold for lines 7 to 10 of Algorithm \ref{alg_momen:}  (given $\bm{y}_0 = \bm{h}_0 = \bm{0}$):
\begin{align}
    &\overline{\bm{h}}_{t+1} = \overline{\bm{x}}_{\tau(t)}  - \overline{\bm{x}}_{t+\frac{1}{2}} = \sum_{j=\tau(t)}^{t} \gamma_j \overline{\bm{v}}_{j}, \label{al2_avrag_rule_h:}\\
    & \overline{\bm{y}}_{t+1} = \overline{\bm{y}}_{\tau(t)} +\overline{\bm{h}}_{t+1} - \overline{\bm{h}}_{\tau(t)} = \overline{\bm{h}}_{t+1}, \label{al2_avrag_rule_y:} \\
    & \overline{\bm{x}}_{t+1} = \overline{\bm{x}}_{\tau(t)} - \overline{\bm{y}}_{t+1}, \label{al2_avrag_rule_x1:}\\
    &\overline{\bm{v}}_{t+1} = \overline{\nabla f}(\bm{X}_t). \label{al2_avrag_rule_v1:} 
\end{align}

If ${\rm mod}(t+1, \tau) \neq 0$  (lines 5, 12 and 15), we have: 
\begin{align}
    & \overline{\bm{x}}_{t+1} = \overline{\bm{x}}_{t} - \gamma_t\overline{\bm{v}}_{t}, \label{al2_avrag_rule_x2:}\\
    & \overline{\bm{v}}_{t+1} = \overline{\bm{g}}_{t+1} + (1-\alpha_{t+1})(\overline{\bm{v}}_{t}-\overline{\bm{g}}_{t}), \label{al2_avrag_rule_v2:}
\end{align}
where $\overline{\bm{x}}_{\tau(t)}$ denotes the average model parameter of all nodes in the previous communication round. Note that   $\tau(t)=\max \left\{l:l\leq t \ {\rm and} \ {\rm mod} {(l,\tau)} = 0 \right\}$ and $\overline{\bm{g}}_{t} = \frac{1}{N}\sum_{i=1}^{N}\bm{g}_{t}^{(i)} = \frac{1}{Nb}\sum_{i=1}^{N}\sum_{r=1}^{b}\nabla f_i(\bm{x}_{t}^{(i)}; \bm{\xi}_{r}^{(i)})$, $\bm{\xi}_{{r}}^{(i)} \sim \mathcal{D}_i$.

Then, note that $\overline{\bm{y}}_{t+1} = \overline{\bm{h}}_{t+1}$ (lines 7-8),  we get: for all $t \in [0, \cdots, T-1]$,
\begin{equation}
    \label{al2_avrag_rule_x:} \overline{\bm{x}}_{t+1} = \overline{\bm{x}}_{t} - \gamma_t \overline{\bm{v}}_{t}.
\end{equation}
\textbf{Second}, we represent the update rules of the model parameters in matrix form as follows. 

If ${\rm mod}(t+1, \tau)=0$, then each node communicates with its neighbor nodes and the communication among nodes is controlled by mixing matrix $\bm{W}$. And we have:
\begin{align}  
    &\bm{H}_{t+1} = (\bm{X}_{\tau(t)} - \bm{X}_{t+\frac{1}{2}}) = \sum_{j=\tau(t)}^{t} \gamma_j \bm{V}_{j}, \label{al2_matrix_rule_H:}\\
    &\bm{Y}_{t+1} = (\bm{Y}_{\tau(t)} + \bm{H}_{t+1} - \bm{H}_{\tau(t)})\bm{W}, \label{al2_matrix_rule_Y:}\\
    &\bm{X}_{t+1} = (\bm{X}_{\tau(t)} - \bm{Y}_{t+1})\bm{W}, \label{al2_matrix_rule_X1:}\\
    &\bm{V}_{t+1} = \partial f(\bm{X}_t). \label{al2_matrix_rule_V1:}
\end{align}

If ${\rm mod}(t+1, \tau) \neq 0$, then each node performs local update steps. And we have:
\begin{align}  
    &\bm{X}_{t+1} = \bm{X}_{t} - \gamma_t \bm{V}_{t}, \label{al2_matrix_rule_X2:}\\
    &\bm{V}_{t+1} = \bm{G}_{t+1} + (1-\alpha_t)(\bm{V}_{t}-\bm{G}_{t}). \label{al2_matrix_rule_V2:}
\end{align}

Further, we define the gradient error of node $i$ as $\bm{e}_{t}^{(i)}= \bm{v}_{t}^{(i)}-\nabla f_i(\bm{x}_t^{(i)})$ and the average gradient error of all nodes as $\overline{\bm{e}}_{t}^{(i)}=\sum\limits_{i=1}^{N}\bm{e}_{t}^{(i)}$.

Noting that throughout the section, we assume that assumptions \ref{Global_Function_Below_Bounds:} to \ref{CN:} hold.

\subsubsection{Preliminary Lemmas}
\begin{lemm}
\label{al2_expectaion_direction:}
For $\bm{V}_t = [ \bm{v}_t^{(1)}, \bm{v}_t^{(2)},\cdots, \bm{v}_t^{(N)}]$ where $\bm{v}_t^{(i)}$ for any $i\in[N]$ and $t\in[0,\cdots,T]$ is generated according to Algorithm \ref{alg_momen:}, we have:
$$\mathbbm{E}[\bm{V}_t]=\partial f(\bm{X}_{t}),$$
where the expectation $\mathbbm{E}[\cdot]$ is w.r.t the stochasticity of the algorithm.
\end{lemm}
\begin{proof}
The update rule of algorithm DSE-MVR for $\bm{V}_t$ is given in (\ref{al2_matrix_rule_V1:}) and (\ref{al2_matrix_rule_V2:}). Specifically, after every $\tau$ steps, each node locally computes the full gradient as the update direction of the next local step, that is, the local update direction is reset. The local update direction reset is designed to ensure that $\bm{V}_t$ is unbiased. 

For $\mod(t,\tau) = 0$, we obviously have $\mathbbm{E}[\bm{V}_t]=\partial f(\bm{X}_{t})$. For $\mod(t,\tau) \neq 0$, i.e. $t\in [t^\prime+1, t^\prime+\tau-1]$ with $t^\prime \in \mathcal{T}$,  we have: 
\begin{align*}
    \mathbbm{E}[\bm{V}_t] &= \mathbbm{E}[\bm{G}_{t}+(1-\alpha_{t-1})(\bm{V}_{t-1}-\bm{G}_{t-1})]\\
    &= \partial f(\bm{X}_t)+\prod_{j=t-2}^{t-1} (1-\alpha_{j})(\mathbbm{E}[\bm{V}_{t-2}]-\partial f(\bm{X}_{t-2}))\\
    &\cdots\\
    &= \partial f(\bm{X}_t) + \prod_{j=t-(t^{\prime}-\tau)}^{t-1} (1-\alpha_{j})(\mathbbm{E}[\bm{V}_{t^{\prime}-\tau}] - \partial f(\bm{X}_{t^{\prime}-\tau}))\\
    & \overset{(a)}{\leq} \partial f(\bm{X}_t),
\end{align*}
where ($a$) follows from the fact that $\mathbbm{E}[\bm{V}_t]=\partial f(\bm{X}_{t})$ holds if $t \in \mathcal{T}$. Hence, the lemma is proved.
\end{proof}

\begin{lemm}
\label{al2_di_con_di:}
 For $\bm{V}_t = [ \bm{v}_t^{(1)}, \bm{v}_t^{(2)},\cdots, \bm{v}_t^{(N)}]$ where $\bm{v}_t^{(i)}$ for any $i\in[N]$ and $t\in[0,\cdots,T]$ is generated according to Algorithm \ref{alg_momen:}, we have:
\begin{align*}
    \mathbbm{E}\| \bm{V}_{t}-\overline{\bm{V}}_{t}\|_F^2   \overset{}{\leq} 2\sum_{i=1}^{N} \mathbbm{E}\|\bm{e}_{t}^{(i)}\|^2 +2\mathbbm{E}\left\|\partial f(\bm{X}_{t})-\overline{\partial f}(\bm{X}_{t})\right\|_F^2,
\end{align*}
where the expectation $\mathbbm{E}[\cdot]$ is w.r.t the stochasticity of the algorithm.
\end{lemm}

\begin{proof}
At the communication round, as stated in Lemma \ref{al2_expectaion_direction:}, for $t \in \mathcal{T}$, we directly get:
\begin{equation}
    \begin{split}
        \mathbbm{E}\| \bm{V}_{t}-\overline{\bm{V}}_{t}\|_F^2 \overset{}{=} \mathbbm{E}\left\|\partial f(\bm{X}_{t})-\overline{\partial f}(\bm{X}_{t})\right\|_F^2.
    \end{split}
\end{equation}

And for $t \notin \mathcal{T}$, we have:
\begin{equation}
    \begin{split}
        \mathbbm{E}\| \bm{V}_{t}-\overline{\bm{V}}_{t}\|_F^2 & = \sum_{i=1}^{N}\mathbbm{E}\| \bm{v}_{t}^{(i)}-\overline{\bm{v}}_{t}\|^2\\
        & \overset{}{=} \sum_{i=1}^{N} \mathbbm{E}\left\|\bm{v}_{t}^{(i)}-\nabla f_i(\bm{x}_{t}^{(i)})-\left(\overline{\bm{v}}_{t}-\overline{\nabla f}(\bm{X}_{t})\right) +\left(\nabla f_i(\bm{x}_{t}^{(i)})-\overline{\nabla f}(\bm{X}_{t})\right)\right\|^2\\
        & \overset{(a)}{\leq} 2\sum_{i=1}^{N} \left[\mathbbm{E}\left\|\bm{v}_{t}^{(i)}-\nabla f_i(\bm{x}_{t}^{(i)})-\left(\overline{\bm{v}}_{t}-\overline{\nabla f}(\bm{X}_{t})\right)\right\|^2 +\mathbbm{E}\left\|\nabla f_i(\bm{x}_{t}^{(i)})-\overline{\nabla f}(\bm{X}_{t})\right\|^2\right]\\
        & \overset{(b)}{\leq} 2\sum_{i=1}^{N} \left[\mathbbm{E}\left\|\bm{v}_{t}^{(i)}-\nabla f_i(\bm{x}_{t}^{(i)})\right\|^2 +\mathbbm{E}\left\|\nabla f_i(\bm{x}_{t}^{(i)})-\overline{\nabla f}(\bm{X}_{t})\right\|^2\right]\\
        & \overset{}{=} 2\sum_{i=1}^{N} \mathbbm{E}\|\bm{e}_{t}^{(i)}\|^2 +2\mathbbm{E}\left\|\partial f(\bm{X}_{t})-\overline{\partial f}(\bm{X}_{t})\right\|_F^2,
    \end{split}
\end{equation}
where ($a$) and ($b$) result from the inequalities (i) from Fact \ref{FC1:} and (iii) from Fact \ref{FC4:}, respectively. 
Hence, the lemma is proved.
\end{proof}

\subsubsection{Descent Lemma, Gradient Error Contraction and Consensus Distance}

In this section, we present a series of lemmas and their corresponding proof procedures. These lemmas are the key elements used to derive Theorem \ref{thm:DSE_MVR}.
\begin{lemm}
\label{al2_Descent_Lemma:}
{\rm (Descent Lemma)} 
For any $t \in [0,\cdots, T-1]$,
then the averages $\overline{\bm{x}}_{t}=\frac{1}{N}\sum\limits_{i=1}^{N}\bm{x}_{t}^{(i)}$ of the iterates generated by algorithm \ref{alg_momen:} satisfy that 
\begin{align*}
    \mathbbm{E}[F(\bm{\overline{x}}_{t+1})] 
    & \leq \mathbbm{E}[F(\bm{\overline{x}}_{t})]  -\frac{\gamma}{2}\mathbbm{E}\|\nabla F(\bm{\overline{x}}_{t})\|^{2}-\frac{\gamma}{2}(1-\gamma L)\mathbbm{E}\|\bm{\overline{v}}_t\|^{2}+\gamma\mathbbm{E}\|\bm{\overline{e}}_t\|^{2} \\
    & \quad \quad \quad \quad \quad \quad \quad \quad \quad \quad \quad \quad \quad \quad \quad \quad \quad \quad \quad \quad  +\frac{\gamma L^2}{N} \mathbbm{E}\|\bm{X}_t-\overline{\bm{X}}_{t}\|_F^{2},
\end{align*}
where the expectation $\mathbbm{E}[\cdot]$ is w.r.t the stochasticity of the algorithm. 
\end{lemm}

\begin{proof}
    By the smoothness of $F(\cdot)$ (Assumption \ref{L_smooth:}), we have:
    \begin{equation}
        \begin{split}
            \label{al2_Descent_Lemma_1:}
            \mathbbm{E}[F(\bm{\overline{x}}_{t+1})] & \overset{}{\leq} \mathbbm{E}[F(\bm{\overline{x}}_{t})] + \mathbbm{E}[\langle\nabla F(\bm{\overline{x}}_{t}), \bm{\overline{x}}_{t+1}-\bm{\overline{x}}_{t}\rangle] + \frac{ L}{2} \mathbbm{E}\|\bm{\overline{x}}_{t+1}-\bm{\overline{x}}_{t}\|^{2} \\
             & \overset{(a)}{=} \mathbbm{E}[F(\bm{\overline{x}}_{t})]  \underbrace{-\gamma \mathbbm{E}[\langle\nabla F(\bm{\overline{x}}_{t}),  \bm{\overline{v}}_t\rangle]}_{T_7} + \frac{\gamma^2 L}{2} \mathbbm{E}\|\bm{\overline{v}}_t\|^{2},
        \end{split}
    \end{equation}
    where ($a$) holds  because of (\ref{al2_avrag_rule_x:}).

    We note that
    \begin{equation}
        \begin{split}
            T_7 & = -\gamma \mathbbm{E}[\langle\nabla F(\bm{\overline{x}}_{t}),  \bm{\overline{v}}_t\rangle]\\
            & \overset{(a)}{=} -\frac{\gamma}{2}\mathbbm{E}\|\nabla F(\bm{\overline{x}}_{t})\|^{2}-\frac{\gamma}{2}\mathbbm{E}\|\bm{\overline{v}}_t\|^{2}+\frac{\gamma}{2}\mathbbm{E}\|\bm{\overline{v}}_t-\nabla F(\bm{\overline{x}}_{t})\|^{2}\\
            & \overset{(b)}{\leq} -\frac{\gamma}{2}\mathbbm{E}\|\nabla F(\bm{\overline{x}}_{t})\|^{2}-\frac{\gamma}{2}\mathbbm{E}\|\bm{\overline{v}}_t\|^{2}+\gamma\mathbbm{E}\|\bm{\overline{v}}_t-\overline{\nabla f}(\bm{X}_t)\|^{2}+\gamma \mathbbm{E}\|\overline{\nabla f}(\bm{X}_t)-\nabla F(\bm{\overline{x}}_{t})\|^{2}\\
            & \overset{}{=} -\frac{\gamma}{2}\mathbbm{E}\|\nabla F(\bm{\overline{x}}_{t})\|^{2}-\frac{\gamma}{2}\mathbbm{E}\|\bm{\overline{v}}_t\|^{2}+\gamma\mathbbm{E}\|\bm{\overline{v}}_t-\overline{\nabla f}(\bm{X}_t)\|^{2}+\gamma \mathbbm{E}\left\|\frac{1}{N}\sum_{i=1}^{N}\left(\nabla f_i(\bm{x}_t^{(i)})-\nabla f_i(\overline{\bm{x}}_t)\right)\right\|^{2}\\
            & \overset{(c)}{\leq} -\frac{\gamma}{2}\mathbbm{E}\|\nabla F(\bm{\overline{x}}_{t})\|^{2}-\frac{\gamma}{2}\mathbbm{E}\|\bm{\overline{v}}_t\|^{2}+\gamma\mathbbm{E}\|\bm{\overline{v}}_t-\overline{\nabla f}(\bm{X}_t)\|^{2}+\frac{\gamma}{N}\sum_{i=1}^{N} \mathbbm{E}\left\|\nabla f_i(\bm{x}_t^{(i)})-\nabla f_i(\overline{\bm{x}}_t)\right\|^{2}\\
            & \overset{(d)}{\leq} -\frac{\gamma}{2}\mathbbm{E}\|\nabla F(\bm{\overline{x}}_{t})\|^{2}-\frac{\gamma}{2}\mathbbm{E}\|\bm{\overline{v}}_t\|^{2}+\gamma\mathbbm{E}\|\bm{\overline{v}}_t-\overline{\nabla f}(\bm{X}_t)\|^{2}+\frac{\gamma L^2}{N} \mathbbm{E}\|\bm{X}_t-\overline{\bm{X}}_{t}\|_F^{2},
        \end{split}
    \end{equation}
    where ($a$) follows by using the equality (i) from Fact \ref{FC2:},  ($b$) holds because of  adding and subtracting $\overline{\nabla f}(\bm{X}_t)$ in the last term of ($a$) and the inequality (i) of Fact \ref{FC1:},  ($c$) results from the inequality (i) of Fact \ref{FC1:}, and last inequality ($d$) uses the statement of Assumption \ref{L_smooth:}.
    
    Substituting the upper bounds of $T_7$ into (\ref{al2_Descent_Lemma_1:}), we get:
    \begin{equation}
        \begin{split}        
            \mathbbm{E}[F(\bm{\overline{x}}_{t+1})]
            & \leq \mathbbm{E}[F(\bm{\overline{x}}_{t})]  -\frac{\gamma}{2}\mathbbm{E}\|\nabla F(\bm{\overline{x}}_{t})\|^{2}-\frac{\gamma}{2}(1-\gamma L)\mathbbm{E}\|\bm{\overline{v}}_t\|^{2}+\gamma\mathbbm{E}\|\bm{\overline{e}}_t\|^{2}\\
            & \quad \quad \quad \quad \quad \quad \quad \quad \quad \quad \quad \quad \quad \quad \quad \quad \quad \quad \quad \quad \quad \quad \quad  +\frac{\gamma L^2}{N} \mathbbm{E}\|\bm{X}_t-\overline{\bm{X}}_{t}\|_F^{2}.
        \end{split}
    \end{equation}
    Hence, the lemma is proved.
\end{proof}

\begin{lemm}{\rm (Gradient Error Contraction)} 
\label{al2_lemm_GEC:}
 For $t \in [0, \cdots,T -1]$, then the iterates generated by Algorithm \ref{alg_momen:} with $\gamma_t = \gamma \leq \frac{1}{8L\tau}$ and $\alpha_t=\alpha=\frac{32L^2\gamma^2}{Nb}$ satisfy that
\begin{align*}
    & \sum_{t=0}^{T-1}\sum_{i=1}^{N}\mathbbm{E}\|\bm{e}_{t}^{(i)}\|^2 
     \overset{}{\leq}\frac{8 L^2  \gamma^2 (\tau-1) }{b}\sum_{t=0}^{T-1}\|\partial f(\bm{X}_{t})-\overline{\partial f}(\bm{X}_{t})\|_F^2  + \frac{4N L^2  \gamma^2(\tau-1)}{b}\sum_{t=0}^{T-1}  \mathbbm{E}\|\overline{\bm{v}}_{t}\|^2  \\
    & \quad \quad \quad \quad \quad \quad \quad \quad \quad \quad \quad \quad \quad \quad \quad \quad \quad \quad \quad \quad \quad \quad \quad \quad \quad \quad \quad \quad + \frac{2N\alpha^2(\tau-1) T \sigma^2}{b},\\
    & \sum_{t=0}^{T-1}\mathbbm{E}\|\bm{\overline{e}}_{t}\|^2 
    \overset{}{\leq}\frac{16  L^2  \gamma^2 (\tau-1)}{N^2b} \sum_{t=0}^{T-1}\mathbbm{E}\|\partial f(\bm{X}_{t})-\overline{\partial f}(\bm{X}_{t})\|_F^2  + \frac{8  L^2  \gamma^2 (\tau-1)}{Nb}  \sum_{t=0}^{T-1}\mathbbm{E}\|\overline{\bm{v}}_{t}\|^2 \\
    & \quad \quad \quad \quad \quad \quad \quad \quad \quad \quad \quad \quad \quad \quad \quad \quad \quad \quad \quad \quad \quad \quad \quad \quad \quad \quad \quad \quad +\frac{4\alpha^2 (\tau-1) T \sigma^2}{Nb},
\end{align*}
where the expectation $\mathbbm{E}[\cdot]$ is w.r.t the stochasticity of the algorithm. 
\end{lemm}

\begin{proof}
    We first recall that the update rule of each local stochastic gradient estimator $\bm{v}_t^{(i)}$, $t \in [t^\prime+1, \cdots, t^\prime + \tau -1]$. The line 15 of Algorithm \ref{alg_momen:} can be equivalently written  as follows:
    \begin{equation}
    \label{al2_lemm_GEC:1}
        \begin{split}
            \bm{v}_{t}^{(i)} & =   \bm{g}_{t}^{(i)} + (1-\alpha)(\bm{v}_{t-1}^{(i)}-\bm{g}_{t-1}^{(i)})\\
            & \overset{}{=}   \alpha \bm{g}_{t}^{(i)} + (1-\alpha)(\bm{v}_{t-1}^{(i)}+\bm{g}_{t}^{(i)}-\bm{g}_{t-1}^{(i)}).
        \end{split}
    \end{equation}
    
    By subtracting $\nabla f_i(\bm{x}_t^{(i)})$ from both sides of (\ref{al2_lemm_GEC:1}), we get:
    \begin{equation}
        \label{al2_lemm_GEC:2}
        \begin{split}
            \bm{e}_{t}^{(i)} 
            & \overset{}{=} 
                \bm{v}_{t}^{(i)} - \nabla f_i(\bm{x}_t^{(i)}) \\
            & \overset{}{=} 
                \alpha (\bm{g}_{t}^{(i)}-\nabla f_i(\bm{x}_t^{(i)})) + (1-\alpha)(\bm{v}_{t-1}^{(i)}-\nabla f_i(\bm{x}_t^{(i)})+\bm{g}_{t}^{(i)}-\bm{g}_{t-1}^{(i)})\\
            & \overset{}{=}   
                \alpha (\bm{g}_{t}^{(i)}-\nabla f_i(\bm{x}_t^{(i)})) + (1-\alpha)(\bm{v}_{t-1}^{(i)}\pm \nabla f_i(\bm{x}_{t-1}^{(i)})-\nabla f_i(\bm{x}_t^{(i)})+\bm{g}_{t}^{(i)}-\bm{g}_{t-1}^{(i)}) \\
            & \overset{}{=}  
                (1-\alpha)(\bm{v}_{t-1}^{(i)}- \nabla f_i(\bm{x}_{t-1}^{(i)}))+\alpha (\bm{g}_{t}^{(i)}-\nabla f_i(\bm{x}_t^{(i)})) \\
                & \quad \quad \quad \quad \quad \quad \quad \quad \quad \quad \quad \quad \quad \quad \quad \quad + (1-\alpha)(\bm{g}_{t}^{(i)}-\bm{g}_{t-1}^{(i)}+\nabla f_i(\bm{x}_{t-1}^{(i)})-\nabla f_i(\bm{x}_t^{(i)}))\\
            & \overset{(a)}{=} 
                (1-\alpha)\bm{e}_{t-1}^{(i)}+ \frac{\alpha}{b} \sum \limits_{r=1}^b \underbrace{\left(\nabla f_i(\bm{x}_{t}^{(i)};\bm{\xi}_{{r}}^{(i)})-\nabla f_i(\bm{x}_t^{(i)})\right)}_{=:\bm{a}_{t,r}^{(i)}} \\
                & \quad  \quad \quad \quad \quad \quad  + \frac{1-\alpha}{b} \sum \limits_{r=1}^b \underbrace{\left(\nabla f_i(\bm{x}_{t}^{(i)};\bm{\xi}_{{r}}^{(i)})-\nabla f_i(\bm{x}_{t-1}^{(i)};\bm{\xi}_{{r}}^{(i)})+\nabla f_i(\bm{x}_{t-1}^{(i)})-\nabla f_i(\bm{x}_t^{(i)})\right)}_{=:\bm{b}_{t,r}^{(i)}},
        \end{split}
    \end{equation}
    where ($a$) follows from $\bm{g}_{t}^{(i)} = \frac{1}{b} \sum \limits_{r=1}^b \nabla f_i(\bm{x}_{t}^{(i)};\bm{\xi}_{{r}}^{(i)})$ and $\bm{g}_{t-1}^{(i)} = \frac{1}{b} \sum \limits_{r=1}^b \nabla f_i(\bm{x}_{t-1}^{(i)};\bm{\xi}_{{r}}^{(i)})$,  $\bm{\xi}_{{r}}^{(i)} \sim \mathcal{D}_i$.
    
    Note that for $\bm{a}_{t,r}^{(i)}$, we directly have the following results from Assumption \ref{UG_BV:}:
    \begin{align}
        & \mathbbm{E}[\bm{a}_{t,r}^{(i)}] = \mathbbm{E}[\nabla f_i(\bm{x}_{t}^{(i)};\bm{\xi}_{{r}}^{(i)})-\nabla f_i(\bm{x}_t^{(i)})]=\bm{0}, \label{al2_lemm_GEC:3} \\
        & \mathbbm{E}\|\bm{a}_{t,r}^{(i)}\|^2 
         =  \mathbbm{E}\|\nabla f_i(\bm{x}_{t}^{(i)};\bm{\xi}_{{r}}^{(i)})-\nabla f_i(\bm{x}_t^{(i)})\|^2 \leq \sigma^2. \label{al2_lemm_GEC:4}
    \end{align}
    
    For $\bm{b}_{t,r}^{(i)}$, we observe that
    
    \begin{equation}
        \label{al2_lemm_GEC:5}
        \begin{split}
            \mathbbm{E}[\bm{b}_{t,r}^{(i)}] & = \mathbbm{E}[\nabla f_i(\bm{x}_{t}^{(i)};\bm{\xi}_{{r}}^{(i)})-\nabla f_i(\bm{x}_{t-1}^{(i)};\bm{\xi}_{{r}}^{(i)})+\nabla f_i(\bm{x}_{t-1}^{(i)})-\nabla f_i(\bm{x}_t^{(i)})]\\
            & =\mathbbm{E}[\nabla f_i(\bm{x}_{t}^{(i)};\bm{\xi}_{{r}}^{(i)})-\nabla f_i(\bm{x}_t^{(i)})] - \mathbbm{E}[\nabla f_i(\bm{x}_{t-1}^{(i)};\bm{\xi}_{{r}}^{(i)})-\nabla f_i(\bm{x}_{t-1}^{(i)})]\\
            & \overset{(a)}{=}\bm{0},
        \end{split}
    \end{equation}
    where ($a$) follows from the equality (\ref{Unbiased_Gradient}) of Assumption \ref{UG_BV:}, and
    \begin{equation}
        \label{al2_lemm_GEC:6}
        \begin{split}
            \mathbbm{E}\|\bm{b}_{t,r}^{(i)}\|^2 
             & =  \mathbbm{E}\left\|\nabla f_i(\bm{x}_{t}^{(i)};\bm{\xi}_{{r}}^{(i)})-\nabla f_i(\bm{x}_{t-1}^{(i)};\bm{\xi}_{{r}}^{(i)})+\nabla f_i(\bm{x}_{t-1}^{(i)})-\nabla f_i(\bm{x}_t^{(i)})\right\|^2 \\
             & \overset{(a)}{\leq} \mathbbm{E}\left\|\nabla f_i(\bm{x}_{t}^{(i)};\bm{\xi}_{{r}}^{(i)})-\nabla f_i(\bm{x}_{t-1}^{(i)};\bm{\xi}_{{r}}^{(i)})\right\|^2\\
             & \overset{(b)}{\leq} L^2\mathbbm{E}\|\bm{x}_t^{(i)}-\bm{x}_{t-1}^{(i)}\|^2,
        \end{split}
    \end{equation}
    where $(a)$ uses the inequality (ii) from Fact \ref{FC4:}, and $(b)$ results from the inequality (\ref{LSG}) of Assumption \ref{L_smooth:}.
    
    Now, let's consider $\mathbbm{E}\|\bm{e}_{t}^{(i)}\|^2$ of $N$ nodes simultaneously, i.e., $\sum_{i=1}^{N}\mathbbm{E}\|\bm{e}_{t}^{(i)}\|^2$, as follows:
    \begin{align}
        &\sum_{i=1}^{N}\mathbbm{E}\|\bm{e}_{t}^{(i)}\|^2 \notag \\
        & \overset{}{=}
            \sum_{i=1}^{N}\mathbbm{E}\left\|(1-\alpha)\bm{e}_{t-1}^{(i)}+ \frac{\alpha}{b} \sum \limits_{r=1}^b \bm{a}_{t,r}^{(i)} + \frac{1-\alpha}{b} \sum \limits_{r=1}^b\bm{b}_{t,r}^{(i)}\right\|^2\notag \\
        & \overset{(a)}{\leq}
            (1-\alpha)^2\sum_{i=1}^{N}\mathbbm{E}\|\bm{e}_{t-1}^{(i)}\|^2 + \frac{(1-\alpha)^2}{b^2} \sum_{i=1}^{N}\sum_{r=1}^{b} \mathbbm{E}\|   \bm{b}_{t,r}^{(i)}\|^2 +\frac{\alpha^2}{b^2}\sum_{i=1}^{N}\sum_{r=1}^{b}\mathbbm{E}\|\bm{a}_{t,r}^{(i)}\|^2 \notag \\
        & \overset{(b)}{\leq}
            (1-\alpha)^2\sum_{i=1}^{N}\mathbbm{E}\|\bm{e}_{t-1}^{(i)}\|^2 + \frac{(1-\alpha)^2 L^2}{b} \sum_{i=1}^{N}\mathbbm{E}\|\bm{x}_t^{(i)}-\bm{x}_{t-1}^{(i)}\|^2 +\frac{N\alpha^2 \sigma^2}{b}\notag \\
        & \overset{(c)}{=}
            (1-\alpha)^2\sum_{i=1}^{N}\mathbbm{E}\|\bm{e}_{t-1}^{(i)}\|^2+ \frac{(1-\alpha)^2 L^2 \gamma^2}{ b}\sum_{i=1}^{N}\mathbbm{E}\|\bm{v}_{t-1}^{(i)}\|^2 +\frac{N\alpha^2 \sigma^2}{b}\notag \\
        & \overset{}{\leq}
            (1-\alpha)^2\sum_{i=1}^{N}\mathbbm{E}\|\bm{e}_{t-1}^{(i)}\|^2 + \frac{2(1-\alpha)^2 L^2 \gamma^2}{b}  \sum_{i=1}^{N}\mathbbm{E}\bigg\|\bm{v}_{t-1}^{(i)}-\nabla f_i(\bm{x}_{t-1}^{(i)})-\left(\overline{\bm{v}}_{t-1}-\overline{\nabla f}(\bm{X}_{t-1})\right) \notag \\
            & \quad \quad \quad \quad \quad \quad \quad \ +\left(\nabla f_i(\bm{x}_{t-1}^{(i)})-\overline{\nabla f}(\bm{X}_{t-1})\right)\bigg\|^2 + \frac{2N (1-\alpha)^2 L^2 \gamma^2}{b} \mathbbm{E}\|\overline{\bm{v}}_{t-1}\|^2+ \frac{N\alpha^2 \sigma^2}{b} \notag \\
        & \overset{}{\leq}
            (1-\alpha)^2\sum_{i=1}^{N}\mathbbm{E}\|\bm{e}_{t-1}^{(i)}\|^2+ \frac{4(1-\alpha)^2 L^2 \gamma^2}{b}  \sum_{i=1}^{N}\bigg[\mathbbm{E}\left\|\bm{v}_{t-1}^{(i)}-\nabla f_i(\bm{x}_{t-1}^{(i)})-\left(\overline{\bm{v}}_{t-1}-\overline{\nabla f}(\bm{X}_{t-1})\right)\right\|^2 \notag \\
            & \quad \quad \quad \quad \quad \quad \quad +  \mathbbm{E}\left\|\nabla f_i(\bm{x}_{t-1}^{(i)})-\overline{\nabla f}(\bm{X}_{t-1})\right\|^2\bigg] + \frac{2 N(1-\alpha)^2 L^2 \gamma^2}{b} \mathbbm{E}\|\overline{\bm{v}}_{t-1}\|^2 +\frac{N\alpha^2 \sigma^2}{b}\notag \\
        & \overset{(d)}{\leq}
            (1-\alpha)^2\left(1+\frac{4 L^2 \gamma^2}{b} \right)\sum_{i=1}^{N}\mathbbm{E}\|\bm{e}_{t-1}^{(i)}\|^2 +\frac{4(1-\alpha)^2 L^2 \gamma^2}{b}\mathbbm{E}\|\partial f(\bm{X}_{t-1})-\overline{\partial f}(\bm{X}_{t-1})\|_F^2 \notag \\
            & \quad \quad \quad \quad \quad \quad \quad \quad \quad \quad \quad \quad \quad \quad \quad \quad \quad \quad \quad \quad \quad \quad   + \frac{2N(1-\alpha)^2 L^2 \gamma^2}{b} \mathbbm{E}\|\overline{\bm{v}}_{t-1}\|^2 +\frac{N\alpha^2 \sigma^2}{b}\notag \\
        & \overset{(e)}{\leq}
            \left(1+\frac{4 L^2 \gamma^2}{b} \right)\sum_{i=1}^{N}\mathbbm{E}\|\bm{e}_{t-1}^{(i)}\|^2 +\frac{4 L^2 \gamma^2}{b}\mathbbm{E}\|\partial f(\bm{X}_{t-1})-\overline{\partial f}(\bm{X}_{t-1})\|_F^2 \notag \\
            & \quad \quad \quad \quad \quad \quad \quad \quad \quad \quad \quad \quad \quad \quad \quad \quad \quad \quad \quad \quad \quad \quad \quad \quad \quad \quad + \frac{2N L^2 \gamma^2}{b} \mathbbm{E}\|\overline{\bm{v}}_{t-1}\|^2  +\frac{N\alpha^2 \sigma^2}{b}\notag \\
        & \overset{(f)}{\leq}
            \left(1+\frac{1}{16b\tau} \right)\sum_{i=1}^{N}\mathbbm{E}\|\bm{e}_{t-1}^{(i)}\|^2 +\frac{4 L^2 \gamma^2}{b}\mathbbm{E}\|\partial f(\bm{X}_{t-1})-\overline{\partial f}(\bm{X}_{t-1})\|_F^2 \notag \\
            & \quad \quad \quad \quad \quad \quad \quad \quad \quad \quad \quad \quad \quad \quad \quad \quad \quad \quad \quad \quad \quad \quad \quad \quad \quad \quad + \frac{2N L^2 \gamma^2}{b} \mathbbm{E}\|\overline{\bm{v}}_{t-1}\|^2  +\frac{N\alpha^2 \sigma^2}{b}\notag \\
        & \overset{(g)}{\leq}
            \left(1+\frac{1}{16b\tau} \right)^{t-t^\prime}\sum_{i=1}^{N}\mathbbm{E}\|\bm{e}_{t^\prime}^{(i)}\|^2 +\frac{4 L^2 \gamma^2 }{b}\sum_{j=t^\prime}^{t-1}\left(1+\frac{1}{16b\tau} \right)^{t-1-j}\mathbbm{E}\|\partial f(\bm{X}_{t-1})-\overline{\partial f}(\bm{X}_{t-1})\|_F^2 \notag \\
            & \quad \quad \quad \quad \quad \quad \  + \frac{2N L^2\gamma^2}{b} \sum_{j=t^\prime}^{t-1}\left(1+\frac{1}{16b\tau} \right)^{t-1-j} \mathbbm{E}\|\overline{\bm{v}}_{j}\|^2  + \frac{N\alpha^2\sigma^2}{b} \sum_{j=t^\prime}^{t-1}\left(1+\frac{1}{16b\tau} \right)^{t-1-j} \notag \\
        & \overset{(h)}{\leq}
            \sum_{j=t^\prime}^{t-1}\left[\frac{8 L^2 \gamma^2 }{b}\mathbbm{E}\|\partial f(\bm{X}_{j})-\overline{\partial f}(\bm{X}_{j})\|_F^2   + \frac{4N L^2 \gamma^2}{b}  \mathbbm{E}\|\overline{\bm{v}}_{j}\|^2  + \frac{2N\alpha^2\sigma^2}{b} \right], \label{al2_lemm_GEC:7}
    \end{align}
    
    where ($a$) results from expanding the norm using inner product (i.e. the inequality (ii) from Fact \ref{FC1:}) and noting that the cross terms are zero in expectation from (\ref{al2_lemm_GEC:3}) and (\ref{al2_lemm_GEC:5}), ($b$) holds because of the inequalities (\ref{al2_lemm_GEC:4}) and (\ref{al2_lemm_GEC:6}), ($c$) follows from the update rule (\ref{al2_avrag_rule_x:}), ($d$) uses the inequality (iii) from Fact \ref{FC4:}, ($e$) and ($f$) follow from the facts that $\alpha =\frac{32L^2\gamma^2}{Nb} \leq \frac{1}{2Nb\tau^2} < 1$ and $1+\frac{4 L^2 \gamma^2}{b}  \leq 1+\frac{1}{16b\tau} $ hold if $\gamma \leq \frac{1}{8L\tau}$,  ($g$) results from recursively applying ($f$) from $t-1$ to $t^\prime$, and the last inequality ($h$) holds because $\mathbbm{E}\|\bm{e}_{t^\prime}^{(i)}\|^2=0$ for $t^\prime \in \mathcal{T}$ and $\left(1+\frac{1}{16b\tau} \right)^{t-1-j} \leq \left(1+\frac{1}{16b\tau} \right)^{\tau} \leq e^{\frac{1}{16b}} \leq e^{\frac{1}{16}} \leq 2$.
   
   Then, summing over $\sum_{i=1}^{N}\mathbbm{E}\|\bm{e}_{t}^{(i)}\|^2$ from $t=0$ to $t=T-1$, we have:
   \begin{align} 
            \sum_{t=0}^{T-1}\sum_{i=1}^{N}&\mathbbm{E}\|\bm{e}_{t}^{(i)}\|^2 
         \overset{}{=}    
            \sum_{t^\prime=0}^{T-\tau}\sum_{t=t^\prime}^{t^\prime+\tau-1}\sum_{i=1}^{N}\mathbbm{E}\|\bm{e}_{t}^{(i)}\|^2=\sum_{t^\prime=0}^{T-\tau}\sum_{t=t^\prime+1}^{t^\prime+\tau-1}\sum_{i=1}^{N}\mathbbm{E}\|\bm{e}_{t}^{(i)}\|^2\notag\\
        & \overset{}{\leq}    
            \sum_{t^\prime=0}^{T-\tau}\sum_{t=t^\prime+1}^{t^\prime+\tau-1}\sum_{j=t^\prime}^{t-1}\bigg[\frac{8 L^2 \gamma^2 }{b}\mathbbm{E}\|\partial f(\bm{X}_{j})-\overline{\partial f}(\bm{X}_{j})\|_F^2   + \frac{4N L^2 \gamma^2}{b}  \mathbbm{E}\|\overline{\bm{v}}_{j}\|^2  + \frac{2N\alpha^2\sigma^2}{b} \bigg] \notag \\
        & \overset{}{\leq}    
            (\tau-1)\sum_{t^\prime=0}^{T-\tau}\sum_{t=t^\prime}^{t^\prime+\tau-1}\bigg[\frac{8 L^2 \gamma^2 }{b}\mathbbm{E}\|\partial f(\bm{X}_{t})-\overline{\partial f}(\bm{X}_{t})\|_F^2   + \frac{4N L^2 \gamma^2}{b}  \mathbbm{E}\|\overline{\bm{v}}_{t}\|^2+ \frac{2N\alpha^2 \sigma^2}{b}\bigg]  \notag \\
        & \overset{}{=}
            \frac{8 L^2  \gamma^2 (\tau-1) }{b}\sum_{t=0}^{T-1}\|\partial f(\bm{X}_{t})-\overline{\partial f}(\bm{X}_{t})\|_F^2  + \frac{4N L^2  \gamma^2(\tau-1)}{b}\sum_{t=0}^{T-1}  \mathbbm{E}\|\overline{\bm{v}}_{t}\|^2 \notag \\
            & \quad \quad \quad \quad \quad \quad \quad \quad \quad \quad \quad \quad \quad \quad \quad \quad \quad \quad \quad \quad \quad \quad \quad \quad \quad \quad \quad \quad + \frac{2N\alpha^2(\tau-1) T \sigma^2}{b}. \label{al2_lemm_GEC:8}
    \end{align}

    Next, we derive the upper bound of $\mathbbm{E}\|\overline{\bm{e}}_t\|^2$, and the derivation process is similar to that of $\sum_{i=1}^{N}\mathbbm{E}\|\bm{e}_{t}^{(i)}\|^2$, as follows:
    
    \begin{align*}
        \mathbbm{E}&\|\bm{\overline{e}}_{t}\|^2 \\
        & \overset{(a)}{=}
            \mathbbm{E}\left\|(1-\alpha)\bm{\overline{e}}_{t-1} + \frac{\alpha}{Nb} \sum_{i=1}^{N} \sum_{r=1}^{b} \bm{a}_{t,r}^{(i)} + \frac{1-\alpha}{Nb} \sum_{i=1}^{N} \sum_{r=1}^{b} \bm{b}_{t,r}^{(i)} \right\|^2\\
        & \overset{}{\leq}
            (1-\alpha)^2\mathbbm{E}\|\bm{\overline{e}}_{t-1}\|^2  + \frac{(1-\alpha)^2}{N^2b^2} \sum_{i=1}^{N} \sum_{r=1}^{b} \mathbbm{E}\|\bm{b}_{t,r}^{(i)}\|^2 +\frac{\alpha^2}{N^2b^2}\sum_{i=1}^{N} \sum_{r=1}^{b}\mathbbm{E}\|\bm{a}_{t,r}^{(i)}\|^2\\
        & \overset{}{\leq}
            (1-\alpha)^2\mathbbm{E}\|\bm{\overline{e}}_{t-1}\|^2 + \frac{(1-\alpha)^2 L^2}{N^2 b}\sum_{i=1}^{N}\mathbbm{E}\|\bm{x}_t^{(i)}-\bm{x}_{t-1}^{(i)}\|^2+\frac{\alpha^2 \sigma^2}{Nb}\\
        & \overset{}{=}
            (1-\alpha)^2\mathbbm{E}\|\bm{\overline{e}}_{t-1}\|^2 + \frac{(1-\alpha)^2 L^2 \gamma^2}{N^2 b}\sum_{i=1}^{N}\mathbbm{E}\|\bm{v}_{t-1}^{(i)}\|^2 +\frac{\alpha^2 \sigma^2}{Nb}\\
        & \overset{}{\leq}
            (1-\alpha)^2\mathbbm{E}\|\bm{\overline{e}}_{t-1}\|^2+ \frac{4(1-\alpha)^2 L^2 \gamma^2}{N^2b}  \sum_{i=1}^{N}\bigg[\mathbbm{E}\left\|\bm{v}_{t-1}^{(i)}-\nabla f_i(\bm{x}_{t-1}^{(i)})-\left(\overline{\bm{v}}_{t-1}-\overline{\nabla f}(\bm{X}_{t-1})\right)\right\|^2 \\
            & \quad \quad \quad \quad \quad \quad \quad +  \mathbbm{E}\left\|\nabla f_i(\bm{x}_{t-1}^{(i)})-\overline{\nabla f}(\bm{X}_{t-1})\right\|^2\bigg] + \frac{2 (1-\alpha)^2 L^2 \gamma^2}{Nb} \mathbbm{E}\|\overline{\bm{v}}_{t-1}\|^2 +\frac{\alpha^2 \sigma^2}{Nb}\\
        & \overset{}{\leq}
            (1-\alpha)^2\left(1+ \frac{8  L^2 \gamma^2}{Nb} \right)\mathbbm{E}\|\bm{\overline{e}}_{t-1}\|^2  + \frac{4 (1-\alpha)^2 L^2 \gamma^2}{N^2b} \mathbbm{E}\|\partial f(\bm{X}_{t-1})-\overline{\partial f}(\bm{X}_{t-1})\|_F^2  \\
            & \quad \quad \quad \quad \quad  \quad \quad \quad + \frac{8 (1-\alpha)^2 L^2 \gamma^2}{N^2b}  \sum_{i=1}^{N}\mathbbm{E}\|\bm{e}_{t-1}^{(i)}\|^2 + \frac{2 (1-\alpha)^2 L^2 \gamma^2}{Nb} \mathbbm{E}\|\overline{\bm{v}}_{t-1}\|^2 +\frac{\alpha^2 \sigma^2}{Nb}\\
        & \overset{}{\leq}
            \left(1+ \frac{8  L^2 \gamma^2}{Nb} \right)\mathbbm{E}\|\bm{\overline{e}}_{t-1}\|^2  + \frac{4  L^2 \gamma^2}{N^2b} \mathbbm{E}\|\partial f(\bm{X}_{t-1})-\overline{\partial f}(\bm{X}_{t-1})\|_F^2  \\
            & \quad \quad \quad \quad \quad \quad \quad \quad \quad \quad \quad \quad \quad \quad \quad \quad + \frac{8  L^2 \gamma^2}{N^2b}  \sum_{i=1}^{N}\mathbbm{E}\|\bm{e}_{t-1}^{(i)}\|^2 + \frac{2 L^2 \gamma^2}{Nb} \mathbbm{E}\|\overline{\bm{v}}_{t-1}\|^2 +\frac{\alpha^2 \sigma^2}{Nb}\\
        &\overset{\gamma\leq\frac{1}{8L\tau}}{\leq}
            \left(1+ \frac{1}{8Nb\tau} \right)\mathbbm{E}\|\bm{\overline{e}}_{t-1}\|^2  + \frac{4  L^2 \gamma^2}{N^2b} \mathbbm{E}\|\partial f(\bm{X}_{t-1})-\overline{\partial f}(\bm{X}_{t-1})\|_F^2  \\
            & \quad \quad \quad \quad \quad \quad \quad \quad \quad \quad \quad \quad \quad \quad \quad \quad  + \frac{8  L^2 \gamma^2}{N^2b}  \sum_{i=1}^{N}\mathbbm{E}\|\bm{e}_{t-1}^{(i)}\|^2 + \frac{2 L^2 \gamma^2}{Nb} \mathbbm{E}\|\overline{\bm{v}}_{t-1}\|^2 +\frac{\alpha^2 \sigma^2}{Nb}\\
        &\overset{}{\leq}
            \left(1+ \frac{1}{8Nb\tau} \right)^{t-t^\prime}\mathbbm{E}\|\bm{\overline{e}}_{t^\prime}\|^2  + \frac{4L^2 \gamma^2}{N^2b} \sum_{j=t^\prime}^{t-1}\left(1+\frac{1}{8Nb\tau} \right)^{t-1-j}\mathbbm{E}\|\partial f(\bm{X}_{j})-\overline{\partial f}(\bm{X}_{j})\|_F^2  \\
            & \quad  + \frac{8  L^2 \gamma^2}{N^2b}  \sum_{j=t^\prime}^{t-1}\left(1+\frac{1}{8Nb\tau} \right)^{t-1-j}\sum_{i=1}^{N}\mathbbm{E}\|\bm{e}_{j}^{(i)}\|^2 + \frac{2 L^2 \gamma^2}{Nb} \sum_{j=t^\prime}^{t-1}\left(1+\frac{1}{8Nb\tau} \right)^{t-1-j}\mathbbm{E}\|\overline{\bm{v}}_{j}\|^2 \\
            & \quad \quad \quad \quad \quad \quad \quad \quad \quad \quad \quad \quad \quad \quad \quad \quad \quad \quad \quad \quad \quad \quad \quad \quad  +\frac{\alpha^2 \sigma^2}{Nb}\sum_{j=t^\prime}^{t-1}\left(1+\frac{1}{8Nb\tau} \right)^{t-1-j}\\
        &\overset{(b)}{\leq} 
            \sum_{j=t^\prime}^{t-1}\bigg[\frac{8 L^2 \gamma^2}{N^2b} \mathbbm{E}\|\partial f(\bm{X}_{j})-\overline{\partial f}(\bm{X}_{j})\|_F^2   + \frac{16  L^2 \gamma^2}{N^2b}  \sum_{i=1}^{N}\mathbbm{E}\|\bm{e}_{j}^{(i)}\|^2 + \frac{4  L^2 \gamma^2}{Nb} \mathbbm{E}\|\overline{\bm{v}}_{j}\|^2+\frac{2\alpha^2 \sigma^2}{Nb}\bigg],
    \end{align*}
    where ($a$) holds because $\bm{\overline{e}}_t =\frac{1}{N}\sum_{i=1}^{N}\bm{e}_{t}^{(i)}$, and ($b$) results from the facts that  $\mathbbm{E}\|\bm{e}_{t^\prime}^{(i)}\|^2=0$ for $t^\prime \in \mathcal{T}$ and $\left(1+\frac{1}{8Nb\tau} \right)^{t-1-j} \leq \left(1+\frac{1}{8Nb\tau} \right)^{\tau} \leq e^{\frac{1}{8Nb}} \leq e^{\frac{1}{8}} \leq 2$. The rest of the derivation process is similar to that of $\mathbbm{E}\|\bm{e}_{t}^{(i)}\|^2$, then we do not explain the conditions under which the equations and inequalities hold.
    
    Finally, summing over $\mathbbm{E}\|\overline{\bm{e}}_t\|^2$ from $t=0$ to $t=T-1$, we have:
    \begin{align*}
        &\sum_{t=0}^{T-1}\mathbbm{E}\|\bm{\overline{e}}_{t}\|^2 
        \overset{}{=}
            \sum_{t^\prime=0}^{T-\tau}\sum_{t=t^\prime}^{t^\prime+\tau-1}\mathbbm{E}\|\bm{\overline{e}}_{t}\|^2 = \sum_{t^\prime=0}^{T-\tau}\sum_{t=t^\prime+1}^{t^\prime+\tau-1}\mathbbm{E}\|\bm{\overline{e}}_{t}\|^2\\
        &\overset{}{=}
            \sum_{t^\prime=0}^{T-\tau}\sum_{t=t^\prime+1}^{t^\prime+\tau-1}\sum_{j=t^\prime}^{t-1}\bigg[\frac{8  L^2 \gamma^2}{N^2b} \mathbbm{E}\|\partial f(\bm{X}_{j})-\overline{\partial f}(\bm{X}_{j})\|_F^2   + \frac{16  L^2 \gamma^2}{N^2b}  \sum_{i=1}^{N}\mathbbm{E}\|\bm{e}_{j}^{(i)}\|^2 \\ 
            & \quad \quad \quad \quad \quad \quad \quad \quad \quad \quad \quad \quad \quad \quad \quad \quad \quad \quad \quad \quad \quad \quad \quad \quad \quad \quad \quad + \frac{4  L^2 \gamma^2}{Nb} \mathbbm{E}\|\overline{\bm{v}}_{j}\|^2+\frac{2\alpha^2 \sigma^2}{Nb}\bigg] \\
        &\overset{}{\leq}
            (\tau-1)\sum_{t^\prime=0}^{T-\tau}\sum_{t=t^\prime}^{t^\prime+\tau-1}\bigg[\frac{8 L^2 \gamma^2}{N^2b} \mathbbm{E}\|\partial f(\bm{X}_{t})-\overline{\partial f}(\bm{X}_{t})\|_F^2   + \frac{16  L^2 \gamma^2}{N^2b}  \sum_{i=1}^{N}\mathbbm{E}\|\bm{e}_{t}^{(i)}\|^2 \\ 
            & \quad \quad \quad \quad \quad \quad \quad \quad \quad \quad \quad \quad \quad \quad \quad \quad \quad \quad \quad \quad \quad \quad \quad \quad \quad \quad \quad  + \frac{4L^2 \gamma^2}{Nb} \mathbbm{E}\|\overline{\bm{v}}_{t}\|^2+\frac{2\alpha^2 \sigma^2}{Nb}\bigg] \\
        &\overset{(a)}{\leq}
            \left(1+\frac{16  L^2  \gamma^2 (\tau-1)}{b}\right)\bigg[\frac{8  L^2  \gamma^2 (\tau-1)}{N^2b} \sum_{t=0}^{T -1}\mathbbm{E}\|\partial f(\bm{X}_{t})-\overline{\partial f}(\bm{X}_{t})\|_F^2   \\
            & \quad \quad \quad \quad \quad \quad \quad \quad \quad \quad \quad \quad \quad \quad \quad \quad \quad \quad \quad  + \frac{4  L^2  \gamma^2 (\tau-1)}{Nb}  \sum_{t=0}^{T -1}\mathbbm{E}\|\overline{\bm{v}}_{t}\|^2  +\frac{2\alpha^2 (\tau-1) T \sigma^2}{Nb}\bigg]\\
        &\overset{}{\leq} 
            \frac{16  L^2  \gamma^2 (\tau-1)}{N^2b} \sum_{t=0}^{T-1}\mathbbm{E}\|\partial f(\bm{X}_{t})-\overline{\partial f}(\bm{X}_{t})\|_F^2  + \frac{8  L^2  \gamma^2 (\tau-1)}{Nb}  \sum_{t=0}^{T-1}\mathbbm{E}\|\overline{\bm{v}}_{t}\|^2 +\frac{4\alpha^2 (\tau-1) T \sigma^2}{Nb},
    \end{align*}
    where ($a$) results from substituting (\ref{al2_lemm_GEC:8}) into ($a$). Hence, we complete the proof.
\end{proof}

Next, we analysis the upper bound of the expectation of the consensus distance for the estimation of global average accumulated direction, i.e., $\mathbbm{E}\|\bm{Y}_{t^{\prime}}- \bm{\overline{Y}}_{t^{\prime}}\|_F^2$, $t^\prime \in \mathcal{T}:= \{t| t=0,\cdots, T, {\rm mod}(t, \tau)=0\}$.

\begin{lemm}
\label{al2_G_A_A_D_CD:}
For $\bm{Y}_{\tau}=[\bm{y}_{\tau}^{(1)}, \bm{y}_{\tau}^{(2)},\cdots, \bm{y}_{\tau}^{(N)}]$ where $\bm{y}_{\tau}^{(i)}$ for any $i \in [N]$ is
generated by Algorithm \ref{alg_momen:}, we have:
\begin{align*}
    \mathbbm{E}\|\bm{Y}_{\tau}-\bm{\overline{Y}}_{\tau}\|_F^2
     \overset{}{=}  
    2\lambda^2\gamma^2(\tau-1)\sum_{t=0}^{\tau-1}\sum_{i=1}^{N}\mathbbm{E}\|\bm{e}_{t}^{(i)}\|^2  +  \frac{2\lambda^2\gamma^2\tau}{1-\lambda^2} \sum_{t=0}^{\tau-1}\mathbbm{E}\left\|\partial f(\bm{X}_t)-\overline{\partial f}(\bm{X}_t))\right\|_F^2,
\end{align*}
where the expectation $\mathbbm{E}[\cdot]$ is w.r.t the stochasticity of the algorithm.
\end{lemm}

\begin{proof}
We first recall the initialization of Algorithm \ref{alg_momen:} that $\bm{Y}_{0}=[\bm{y}_{0}^{(1)},\cdots,\bm{y}_{0}^{(N)}]=\bm{0}^{d \times N}$ and $\bm{H}_{0}=[\bm{h}_{0}^{(1)},\cdots,\bm{h}_{0}^{(N)}]=\bm{0}^{d \times N}$. Using the update rule (\ref{al2_matrix_rule_Y:}), we get:
\begin{equation}
    \label{al2_G_A_A_D_CD:1}
    \begin{split}
        \mathbbm{E}\|\bm{Y}_{\tau}-\bm{\overline{Y}}_{\tau}\|_F^2
        & \overset{}{=} \mathbbm{E}\|(\bm{Y}_{0}+\bm{H}_{\tau}-\bm{H}_0)\bm{W}- (\bm{\overline{Y}}_{0}+\bm{\overline{H}}_{\tau}-\bm{\overline{H}}_0)\|_F^2\\
        & \overset{(a)}{=} \mathbbm{E}\|(\bm{Y}_{0}\bm{W}-\bm{\overline{Y}}_{0})+(\bm{H}_{\tau}-\bm{H}_0)(\bm{W}-\bm{Q})\|_F^2\\
        & \overset{(b)}{\leq}  \lambda^2\mathbbm{E}\|\bm{Y}_{0}-\bm{\overline{Y}}_{0}\|_F^2+\underbrace{\mathbbm{E}\|(\bm{H}_{\tau}-\bm{H}_{0})(\bm{W}-\bm{Q})\|_F^2}_{T_8}  \\
        & \quad \quad \quad \quad \quad \quad \quad \quad \quad \quad \quad \quad \quad \quad + \underbrace{2\mathbbm{E}\left[\left\langle\bm{Y}_{0}\bm{W}-\bm{\overline{Y}}_{0},(\bm{H}_{\tau}-\bm{H}_{0})(\bm{W}-\bm{Q})\right\rangle\right]}_{T_9},
    \end{split}
\end{equation}
where ($a$) and ($b$) hold by using $\bm{WQ}=\bm{Q}$ and the inequality (\ref{consenus_dist:}) from Assumption \ref{CN:}, respectively.

We note that
\begin{equation}
    \label{al2_G_A_A_D_CD:2}
    \begin{split}
        T_8 & = \mathbbm{E}\|(\bm{H}_{\tau}-\bm{H}_{0})(\bm{W}-\bm{Q})\|_F^2  \overset{}{\leq} \lambda^2 \mathbbm{E}\|\bm{H}_{\tau}-\overline{\bm{H}}_{\tau}\|_F^2\\
        & \overset{(a)}{=} \lambda^2\gamma^2\mathbbm{E}\left\|\sum_{t=0}^{\tau-1} (\bm{V}_t-\overline{\bm{V}}_t)\right\|_F^2 \\
        & \overset{(b)}{\leq}  2\lambda^2\gamma^2(\tau-1)\sum_{t=0}^{\tau-1}\sum_{i=1}^{N}\mathbbm{E}\|\bm{e}_{t}^{(i)}\|^2 + 2\lambda^2 \gamma^2 \tau \sum_{t=0}^{\tau-1}  \mathbbm{E}\|\partial f(\bm{X}_{t})-\overline{\partial f}(\bm{X}_{t})\|_F^2,
     \end{split}
\end{equation}
where ($a$) follows from the update rule (\ref{al2_matrix_rule_H:}),  ($b$) holds by using the inequality (i) from Fact \ref{FC1:} and the statement of lemma \ref{al2_di_con_di:}. We also observe that 
\begin{equation}
    \label{al2_G_A_A_D_CD:3}
    \begin{split}
        T_9 & = 2\mathbbm{E}\left[\left\langle\bm{Y}_{0}\bm{W}-\bm{\overline{Y}}_{0},(\bm{H}_{\tau}-\bm{H}_{0})(\bm{W}-\bm{Q})\right\rangle\right]\\
        & \overset{(a)}{\leq} \eta \mathbbm{E}\|\bm{Y}_{0}\bm{W}-\bm{\overline{Y}}_{0}\|_F^2 + \frac{1}{\eta} \mathbbm{E}\left\|\sum_{t=0}^{\tau-1}\gamma\left(\partial f(\bm{X}_t)\bm{W}-\overline{\partial f}(\bm{X}_t)\right)\right\|_F^2 \\
        & \overset{(b)}{\leq}  \frac{1-\lambda^2}{2} \mathbbm{E}\|\bm{Y}_{0}-\bm{\overline{Y}}_{0}\|_F^2 + \frac{2\lambda^4\gamma^2\tau}{1-\lambda^2} \sum_{t=0}^{\tau-1}\mathbbm{E}\left\|\partial f(\bm{X}_t)-\overline{\partial f}(\bm{X}_t))\right\|_F^2,
    \end{split}
\end{equation}
where ($a$) holds by using the inequality (ii) from Fact \ref{FC3:} and setting $\eta = \frac{1-\lambda^2}{2\lambda^2}$, and the statement of Lemma \ref{al2_expectaion_direction:}, ($b$) results from the inequality (i) from Fact \ref{FC1:} and the inequality  (\ref{consenus_dist:}) from Assumption \ref{CN:}.

Substituting the upper bounds of $T_8$ and $T_9$ into (\ref{al2_G_A_A_D_CD:1}), we have:
\begin{equation}
    \label{al2_G_A_A_D_CD:4}
    \begin{split}
        \mathbbm{E}\|\bm{Y}_{\tau}-\bm{\overline{Y}}_{\tau}\|_F^2
        & \overset{(a)}{\leq}    
        \frac{1+\lambda^2}{2}\mathbbm{E}\|\bm{Y}_{0}-\bm{\overline{Y}}_{0}\|_F^2+2\lambda^2\gamma^2(\tau-1)\sum_{t=0}^{\tau-1}\sum_{i=1}^{N}\mathbbm{E}\|\bm{e}_{t}^{(i)}\|^2   \\
        &\quad \quad \quad \quad \quad \quad \quad \quad \quad \quad \quad \quad  \quad \quad \ +  \frac{2\lambda^2\gamma^2\tau}{1-\lambda^2} \sum_{t=0}^{\tau-1}\mathbbm{E}\left\|\partial f(\bm{X}_t)-\overline{\partial f}(\bm{X}_t))\right\|_F^2\\
        & \overset{}{=}  
        2\lambda^2\gamma^2(\tau-1)\sum_{t=0}^{\tau-1}\sum_{i=1}^{N}\mathbbm{E}\|\bm{e}_{t}^{(i)}\|^2  +  \frac{2\lambda^2\gamma^2\tau}{1-\lambda^2} \sum_{t=0}^{\tau-1}\mathbbm{E}\left\|\partial f(\bm{X}_t)-\overline{\partial f}(\bm{X}_t))\right\|_F^2,
    \end{split}
\end{equation}
where ($a$) results from the fact that $1+\frac{\lambda^2}{1-\lambda^2}=\frac{1}{1-\lambda^2}$. So far, we complete the proof.
\end{proof}

\begin{lemm}
\label{al2_G_A_A_D_CD_main:}
For $\bm{Y}_{t^\prime}=[\bm{y}_{t^\prime}^{(1)}, \bm{y}_{t^\prime}^{(2)},\cdots, \bm{y}_{t^\prime}^{(N)}]$ where $\bm{y}_{t^\prime}^{(i)}$ for any $i \in [N]$ and $t^\prime \in [\tau, \cdots, T-\tau]$ with $t^\prime \in \mathcal{T}$ is generated by Algorithm \ref{alg_momen:}, we have:
\begin{align*}
    \sum_{t^\prime=\tau}^{T-\tau}\mathbbm{E}\|\bm{Y}_{t^{\prime}}- \bm{\overline{Y}}_{t^{\prime}}\|_F^2
    & \overset{}{\leq} \frac{16\lambda^2\gamma^2(\tau-1)}{1-\lambda^2}  \sum_{t=0}^{T-\tau-1} \sum_{i=1}^{N}\mathbbm{E}\|\bm{e}_{t}^{(i)}\|^2\\
    & \quad \quad \quad \quad \quad \quad \quad \quad +\frac{16\lambda^2\gamma^2\tau}{(1-\lambda^2)^2} \sum_{t=0}^{T-\tau-1} \mathbbm{E}\left\|\partial f(\bm{X}_t)-\overline{\partial f}(\bm{X}_t)\right\|_F^2,
\end{align*}
where the expectation $\mathbbm{E}[\cdot]$ is w.r.t the stochasticity of the algorithm.
\end{lemm}    

\begin{proof}
Using the update rule (\ref{al2_matrix_rule_Y:}), we get: for $t^\prime > \tau$,
\begin{equation}
    \label{al2_G_A_A_D_CD_main:1}
    \begin{split}
        \mathbbm{E}\|\bm{Y}_{t^{\prime}}- \bm{\overline{Y}}_{t^{\prime}}\|_F^2
        & \overset{}{=} \mathbbm{E}\|(\bm{Y}_{t^{\prime}-\tau}+\bm{H}_{t^{\prime}}-\bm{H}_{t^{\prime}-\tau})\bm{W}- (\bm{\overline{Y}}_{t^{\prime}-\tau}+\bm{\overline{H}}_{t^{\prime}}-\bm{\overline{H}}_{t^{\prime}-\tau})\|_F^2\\
        & \overset{(a)}{=} \mathbbm{E}\|(\bm{Y}_{t^{\prime}-\tau}\bm{W}-\bm{\overline{Y}}_{t^{\prime}-\tau})+(\bm{H}_{t^{\prime}}-\bm{H}_{t^{\prime}-\tau})(\bm{W}-\bm{Q})\|_F^2\\
        & \overset{(b)}{\leq} \lambda^2\mathbbm{E}\|\bm{Y}_{t^{\prime}-\tau}-\bm{\overline{Y}}_{t^{\prime}-\tau}\|_F^2+\underbrace{\mathbbm{E}\|(\bm{H}_{t^{\prime}}-\bm{H}_{t^{\prime}-\tau})(\bm{W}-\bm{Q})\|_F^2}_{T_{10}} \\
        & \quad \quad \quad \quad \quad \quad \quad \quad \quad \quad   + \underbrace{2\mathbbm{E}\left[\left\langle\bm{Y}_{t^{\prime}-\tau}\bm{W}-\bm{\overline{Y}}_{t^{\prime}-\tau},(\bm{H}_{t^{\prime}}-\bm{H}_{t^{\prime}-\tau})(\bm{W}-\bm{Q})\right\rangle\right]}_{T_{11}},
    \end{split}
\end{equation}
where ($a$) uses the fact that $\bm{WQ}=\bm{Q}$, and ($b$) follows from the inequality (i) from Fact \ref{FC4:} and the inequality (\ref{consenus_dist:}) from Assumption \ref{CN:}.

We note that
\begin{equation}
    \label{al2_G_A_A_D_CD_main:2}
    \begin{split}
        T_{10} & = \mathbbm{E}\|(\bm{H}_{t^{\prime}}-\bm{H}_{t^{\prime}-\tau})(\bm{W}-\bm{Q})\|_F^2\\
        &\overset{}{=}
            \gamma^2\mathbbm{E}\left\|\sum_{t=t^{\prime}-\tau}^{t^{\prime}-1}(\bm{V}_t-\bm{V}_{t-\tau})(\bm{W}-\bm{Q})\right\|_F^2\\
        & \overset{(a)}{\leq}
            2\gamma^2  \mathbbm{E}\left\|\sum_{t=t^{\prime}-\tau}^{t^{\prime}-1}(\bm{V}_t\bm{W}-\overline{\bm{V}}_t)\right\|_F^2+2\gamma^2  \mathbbm{E}\left\|\sum_{t=t^{\prime}-\tau}^{t^{\prime}-1}(\bm{V}_{t-\tau}\bm{W}-\overline{\bm{V}}_{t-\tau})\right\|_F^2\\
         & \overset{(b)}{\leq}
            4\lambda^2\gamma^2(\tau-1)\sum_{t=t^{\prime}-2\tau}^{t^{\prime}-1}\sum_{i=1}^{N}\mathbbm{E}\|\bm{e}_{t}^{(i)}\|^2 + 4\lambda^2\gamma^2\tau\sum_{t=t^{\prime}-2\tau}^{t^{\prime}-1}  \mathbbm{E}\|\partial f(\bm{X}_{t})-\overline{\partial f}(\bm{X}_{t})\|_F^2,
     \end{split}
\end{equation}
where ($a$) and ($b$) hold by using the inequality (i) from Fact \ref{FC1:}, the inequality (\ref{consenus_dist:}) from Assumption \ref{CN:} and the statement of lemma \ref{al2_di_con_di:}. We also observe that
\begin{equation}
    \label{al2_G_A_A_D_CD_main:3}
    \begin{split}
        T_{11} & =
            2\mathbbm{E}\left[\left\langle\bm{Y}_{t^{\prime}-\tau}\bm{W}-\bm{\overline{Y}}_{t^{\prime}-\tau},(\bm{H}_{t^{\prime}}-\bm{H}_{t^{\prime}-\tau})(\bm{W}-\bm{Q})\right\rangle\right]\\
        &\overset{(a)}{=}
            2\mathbbm{E}\left[\left\langle\bm{Y}_{t^{\prime}-\tau}\bm{W}-\bm{\overline{Y}}_{t^{\prime}-\tau},\sum_{t=t^{\prime}-\tau}^{t^{\prime}-1}\gamma\left(\partial f(\bm{X}_t)-\partial f(\bm{X}_{t-\tau})\right)(\bm{W}-\bm{Q})\right\rangle\right]\\
         & \overset{(b)}{\leq}
            \frac{1-\lambda^2}{2\lambda^2} \mathbbm{E}\|\bm{Y}_{t^{\prime}-\tau}\bm{W}-\bm{\overline{Y}}_{t^{\prime}-\tau}\|_F^2 +\\
            & \quad \quad \quad \quad \frac{2\lambda^2\gamma^2\tau}{1-\lambda^2}\sum_{t=t^{\prime}-\tau}^{t^{\prime}-1} \mathbbm{E}\left\|\left(\partial f(\bm{X}_t)\bm{W}-\overline{\partial f}(\bm{X}_t)\right)-\left(\partial f(\bm{X}_{t-\tau})\bm{W}-\overline{\partial f}(\bm{X}_{t-\tau})\right)\right\|_F^2\\
          & \overset{(c)}{\leq}
            \frac{1-\lambda^2}{2} \mathbbm{E}\|\bm{Y}_{t^{\prime}-\tau}-\bm{\overline{Y}}_{t^{\prime}-\tau}\|_F^2 +\\
            & \quad \quad \quad \quad \quad \quad  \frac{4\lambda^4\gamma^2\tau}{1-\lambda^2}\sum_{t=t^{\prime}-\tau}^{t^{\prime}-1} \left[\mathbbm{E}\left\|\partial f(\bm{X}_t)-\overline{\partial f}(\bm{X}_t)\right\|_F^2+\left\|\partial f(\bm{X}_{t-\tau})-\overline{\partial f}(\bm{X}_{t-\tau})\right\|_F^2\right]\\
          & \overset{}{\leq}
            \frac{1-\lambda^2}{2} \mathbbm{E}\|\bm{Y}_{t^{\prime}-\tau}-\bm{\overline{Y}}_{t^{\prime}-\tau}\|_F^2 + \frac{4\lambda^4\gamma^2\tau}{1-\lambda^2}\sum_{t=t^{\prime}-2\tau}^{t^{\prime}-1} \mathbbm{E}\left\|\partial f(\bm{X}_t)-\overline{\partial f}(\bm{X}_t)\right\|_F^2,
      \end{split}
\end{equation}
where ($a$) follows by using $\mathbbm{E}[\bm{V}_t] = \partial f(\bm{X}_t)$ for all $t\in[0,\cdots,T]$ from Lemma \ref{al2_expectaion_direction:},  ($b$) results from the fact that $\pm \langle\bm{A},\bm{B}\rangle \leq \frac{1}{2\eta} \|\bm{A}\|_F^2 + 2\eta \|\bm{B}\|_F^2$ with $\eta = \frac{1-\lambda^2}{2\lambda^2}$ and the inequality (i) from Fact \ref{FC2:}, and ($c$) uses the inequality (\ref{consenus_dist:}) from Assumption \ref{CN:} and the inequality (iii) with $\eta = 1$ from Fact \ref{FC3:}.

Substituting the upper bounds of $T_{10}$ and $T_{11}$ into (\ref{al2_G_A_A_D_CD_main:1}), we have:
\begin{equation}
    \label{al2_G_A_A_D_CD_main:4}
    \begin{split}
        \mathbbm{E}\|\bm{Y}_{t^{\prime}}- \bm{\overline{Y}}_{t^{\prime}}\|_F^2
        & \overset{}{\leq}
            \frac{1+\lambda^2}{2} \mathbbm{E}\|\bm{Y}_{t^{\prime}-\tau}-\bm{\overline{Y}}_{t^{\prime}-\tau}\|_F^2+4\lambda^2\gamma^2(\tau-1)\sum_{t=t^{\prime}-2\tau}^{t^{\prime}-1}\sum_{i=1}^{N}\mathbbm{E}\|\bm{e}_{t}^{(i)}\|^2  \\
            & \quad \quad \quad \quad \quad \quad \quad \quad \quad \quad \quad \quad \quad \quad \quad +\frac{4\lambda^2\gamma^2\tau}{1-\lambda^2}\sum_{t=t^{\prime}-2\tau}^{t^{\prime}-1} \mathbbm{E}\left\|\partial f(\bm{X}_t)-\overline{\partial f}(\bm{X}_t)\right\|_F^2.
    \end{split}
\end{equation}
To simplify the description, we set  $\bm{A}_{t}=2(\tau-1)\lambda^2\gamma^2\sum_{i=1}^{N}\mathbbm{E}\|\bm{e}_{t}^{(i)}\|^2  +\frac{2\lambda^2\tau\gamma^2}{1-\lambda^2} \mathbbm{E}\left\|\partial f(\bm{X}_t)-\overline{\partial f}(\bm{X}_t)\right\|_F^2$.

Finally, the statement of the lemma can be obtained directly by applying (\ref{al1_G_A_A_D_CD_main_5:}) and (\ref{al1_G_A_A_D_CD_main_6:}) in Lemma \ref{al1_G_A_A_D_CD_main:}.
\end{proof}

\begin{lemm}
{\rm (Consensus Distance)} 
\label{al_2_CD:}
For $\bm{X}_{t}=[\bm{x}_{t}^{(1)}, \bm{x}_{t}^{(2)},\cdots, \bm{x}_{t}^{(N)}]$ where $\bm{x}_{t}^{(i)}$ for any $i \in [N]$ and $t \in [0, \cdots, T-1]$ is generated by Algorithm \ref{alg_momen:} with $\gamma_t = \gamma \leq \min \left\{ \frac{1}{8L\tau}, \frac{(1-\lambda^2)^2}{64\sqrt{6}\lambda^2L\tau}\right\}$, we have:

\begin{align*}
    \sum_{t =0}^{T-1}\mathbbm{E}\|\bm{X}_{t}-\overline{\bm{X}}_{t}\|_F^2 
    & \overset{}{\leq} \left(\frac{32N L^2 \gamma^4(\tau-1)^2}{b}+ \frac{2048N\lambda^4 L^2  \gamma^4\tau(\tau-1)^2}{(1-\lambda^2)^3b}\right)\sum_{t=0}^{T-1} \mathbbm{E}\|\overline{\bm{v}}_{t}\|^2  \\ 
    & \quad \quad \quad \quad \quad \quad \quad \quad + \frac{1024N\lambda^4\alpha^2\gamma^2\tau(\tau-1)^2 T \sigma^2}{(1-\lambda^2)^3b} +\frac{2048N\lambda^4\gamma^2\tau^2 T\varsigma^2}{(1-\lambda^2)^4} \\
    & \quad \quad \quad \quad \quad \quad \quad \quad \quad \quad+ \frac{8N\alpha^2\gamma^2\tau(\tau-1)^2T\sigma^2}{b} + 24N\gamma^2\tau(\tau-1)T\varsigma^2, 
\end{align*}
where the expectation $\mathbbm{E}[\cdot]$ is w.r.t the stochasticity of the algorithm.
\end{lemm}

\begin{proof}
Firstly, we transform $\sum\limits_{t =0}^{T-1}\mathbbm{E}\|\bm{X}_{t}-\overline{\bm{X}}_{t}\|_F^2$ to the following form:
\begin{equation}
    \label{al_2_CD1:}
    \sum_{t =0}^{T-1}\mathbbm{E}\|\bm{X}_{t}-\overline{\bm{X}}_{t}\|_F^2 = \sum_{t^\prime =0}^{T-\tau}\sum_{t=t^\prime}^{t^\prime+\tau-1}\mathbbm{E}\|\bm{X}_{t}-\overline{\bm{X}}_{t}\|_F^2,
\end{equation}
where $t^\prime \in \mathcal{T}-\{T\} = \{t| t=0,\cdots, T-\tau \ {\rm and} \ {\rm mod}(t, \tau)=0\}$.

Note that according to the update rule (\ref{al2_matrix_rule_X2:}), we have: for any $t \in [t^\prime+1, \cdots, t^\prime+\tau-1]$,

\begin{align*}
    \mathbbm{E}\|\bm{X}_{t}&-\overline{\bm{X}}_{t}\|_F^2 = \mathbbm{E}\left\|\bm{X}_{t^\prime}-\overline{\bm{X}}_{t^\prime}- \gamma \sum_{j=t^\prime}^{t-1}(\bm{V}_{j} -\overline{\bm{V}}_{j})\right\|_F^2\\
    & \overset{(a)}{\leq}
        2\mathbbm{E}\|\bm{X}_{t^\prime}-\overline{\bm{X}}_{t^\prime}\|_F^2  + 2\gamma^2(t-t^\prime)\sum_{j=t^\prime}^{t-1}\mathbbm{E}\left\|\bm{V}_{j} -\overline{\bm{V}}_{j}\right\|_F^2\\
    & \overset{(b)}{\leq}
        2\mathbbm{E}\|\bm{X}_{t^\prime}-\overline{\bm{X}}_{t^\prime}\|_F^2   +4\gamma^2(t-t^\prime)\sum_{j=t^\prime}^{t-1}\mathbbm{E}\left\|\partial f(\bm{X}_{t})-\overline{\partial f}(\bm{X}_{t})\right\|_F^2\\
        & \quad \quad \quad \quad   + 4\gamma^2(t-t^\prime)\sum_{j=t^\prime}^{t-1}\sum_{k=t^\prime}^{j-1}\bigg[\frac{8 L^2 \gamma^2 }{b}\mathbbm{E}\|\partial f(\bm{X}_{k})-\overline{\partial f}(\bm{X}_{k})\|_F^2   + \frac{4N L^2 \gamma^2}{b}  \mathbbm{E}\|\overline{\bm{v}}_{k}\|^2  \\
        & \quad \quad \quad \quad \quad \quad \quad \quad \quad \quad \quad \quad \quad \quad \quad \quad \quad \quad \quad \quad \quad \quad \quad \quad \quad \quad \quad \quad \quad \quad + \frac{2N\alpha^2\sigma^2}{b} \bigg]\\
    & \overset{}{\leq}
        2\mathbbm{E}\|\bm{X}_{t^\prime}-\overline{\bm{X}}_{t^\prime}\|_F^2   +4\gamma^2(t-t^\prime)\sum_{j=t^\prime}^{t-1}\mathbbm{E}\left\|\partial f(\bm{X}_{t})-\overline{\partial f}(\bm{X}_{t})\right\|_F^2\\
        & \quad \quad \quad \quad \quad \quad + 4\gamma^2(t-t^\prime)^2\sum_{j=t^\prime}^{t-1}\bigg[\frac{8 L^2 \gamma^2 }{b}\mathbbm{E}\|\partial f(\bm{X}_{j})-\overline{\partial f}(\bm{X}_{j})\|_F^2   + \frac{4N L^2 \gamma^2}{b}  \mathbbm{E}\|\overline{\bm{v}}_{j}\|^2  \\
        & \quad \quad \quad \quad \quad \quad \quad \quad \quad \quad \quad \quad \quad \quad \quad \quad \quad \quad \quad \quad \quad \quad \quad \quad \quad \quad \quad \quad \quad \quad + \frac{2N\alpha^2\sigma^2}{b} \bigg]\\
    & \overset{(c)}{\leq}
        2\mathbbm{E}\|\bm{X}_{t^\prime}-\overline{\bm{X}}_{t^\prime}\|_F^2   +4\gamma^2(t-t^\prime)\left(1+\frac{8 L^2 \gamma^2(t-t^\prime) }{b}\right)\sum_{j=t^\prime}^{t-1}\bigg[8L^2 \mathbbm{E}\left\|\bm{X}_t-\overline{\bm{X}}_t\right\|_F^2 \\
        & \quad \quad \quad \quad \quad \quad \quad \quad  \quad+2N\varsigma^2\bigg] +  \frac{16N L^2 \gamma^4}{b} (t-t^\prime)^2\sum_{j=t^\prime}^{t-1}  \mathbbm{E}\|\overline{\bm{v}}_{j}\|^2  + \frac{8N\alpha^2\gamma^2\sigma^2}{b}(t-t^\prime)^3 \\
    & \overset{(d)}{\leq}
        2\mathbbm{E}\|\bm{X}_{t^\prime}-\overline{\bm{X}}_{t^\prime}\|_F^2   +64L^2\gamma^2(t-t^\prime)\sum_{j=t^\prime}^{t-1} \left\|\bm{X}_t-\overline{\bm{X}}_t\right\|_F^2 \\
        & \quad \quad \quad \quad \quad  + \frac{16N L^2 \gamma^4}{b} (t-t^\prime)^2\sum_{j=t^\prime}^{t-1}  \mathbbm{E}\|\overline{\bm{v}}_{j}\|^2  + \frac{8N\alpha^2\gamma^2\sigma^2}{b}(t-t^\prime)^3 +16N\gamma^2\varsigma^2(t-t^\prime)^2  \\
    & \overset{(e)}{\leq}
        2\mathbbm{E}\|\bm{X}_{t^\prime}-\overline{\bm{X}}_{t^\prime}\|_F^2   +\frac{1}{\tau}\sum_{j=t^\prime}^{t-1} \left\|\bm{X}_j-\overline{\bm{X}}_j\right\|_F^2\\
        & \quad \quad \quad \quad \quad + \frac{16N L^2 \gamma^4}{b} (t-t^\prime)^2\sum_{j=t^\prime}^{t-1}  \mathbbm{E}\|\overline{\bm{v}}_{j}\|^2   + \frac{8N\alpha^2\gamma^2\sigma^2}{b}(t-t^\prime)^3 +16N\gamma^2\varsigma^2(t-t^\prime)^2\\
    & \overset{(f)}{\leq}
        2\left(1+\frac{1}{\tau}\right)^{t-t^\prime}\mathbbm{E}\|\bm{X}_{t^\prime}-\overline{\bm{X}}_{t^\prime}\|_F^2 + \frac{16N L^2 \gamma^4}{b\tau}\sum_{j=t^\prime+1}^{t}\left(1+\frac{1}{\tau}\right)^{t-j}(j-t^\prime)\sum_{k=t^\prime}^{j-1}   \mathbbm{E}\|\overline{\bm{v}}_{k}\|^2  \\
        & \quad \quad + \frac{8N\alpha^2\gamma^2\sigma^2}{b\tau} \sum_{j=t^\prime+1}^{t}\left(1+\frac{1}{\tau}\right)^{t-j}(j-t^\prime)^3+ \frac{16N\gamma^2\varsigma^2}{\tau}\sum_{j=t^\prime+1}^{t}\left(1+\frac{1}{\tau}\right)^{t-j}(j-t^\prime)^2\\
    & \overset{(g)}{\leq}
        6\mathbbm{E}\|\bm{X}_{t^\prime}-\overline{\bm{X}}_{t^\prime}\|_F^2 + \frac{48N L^2 \gamma^4}{b\tau}\sum_{j=t^\prime+1}^{t}(j-t^\prime)\sum_{k=t^\prime}^{j-1}   \mathbbm{E}\|\overline{\bm{v}}_{k}\|^2  \\
        & \quad \quad \quad \quad \quad \quad \quad \quad \quad \quad \quad \quad \quad + \frac{24N\alpha^2\gamma^2\sigma^2}{b\tau} \sum_{j=t^\prime+1}^{t}(j-t^\prime)^3+ \frac{48N\gamma^2\varsigma^2}{\tau}\sum_{j=t^\prime+1}^{t}(j-t^\prime)^2,
\end{align*}
where ($a$) holds by using the inequalities (i) from Fact \ref{FC1:} and (iii) from Fact \ref{FC3:} with $\eta=1$, ($b$) uses the statement of lemma \ref{al2_di_con_di:} and the inequality (\ref{al2_lemm_GEC:7}) from lemma \ref{al2_lemm_GEC:}, ($c$) uses the statement of lemma \ref{griadient_consensus:}, ($d$) and ($e$) follow from the facts that $1+\frac{8 L^2 \gamma^2(t-t^\prime) }{b} \leq 1+\frac{8 L^2 \gamma^2\tau}{b} \leq 2$ and   $64L^2\gamma^2(t-t^\prime)\leq64L^2\gamma^2\tau \leq \frac{1}{\tau}$ hold if $\gamma \leq \frac{1}{8L\tau}$, ($f$) results from recursively substituting every $\left\|\bm{X}_j-\overline{\bm{X}}_j\right\|_F^2$ in the 
second term of ($e$), and the last inequality ($g$) follows from the fact that $(1+\frac{1}{\tau})^{t-j} \leq (1+\frac{1}{\tau})^{t-t^\prime} \leq (1+\frac{1}{\tau})^{\tau} \leq 3$. Then, summing over $\mathbbm{E}\|\bm{X}_{t}-\overline{\bm{X}}_{t}\|_F^2$ from $0$ to $
T-1$, we get:
\begin{align} 
    \sum_{t =0}^{T-1}\mathbbm{E}&\|\bm{X}_{t}-\overline{\bm{X}}_{t}\|_F^2  = \sum_{t^\prime =0}^{T-\tau}\sum_{t=t^\prime}^{t^\prime+\tau-1}\mathbbm{E}\|\bm{X}_{t}-\overline{\bm{X}}_{t}\|_F^2\notag\\
    &\overset{}{=}
        \sum_{t^\prime =0}^{T-\tau}\bigg[\mathbbm{E}\|\bm{X}_{t^\prime}-\overline{\bm{X}}_{t^\prime}\|_F^2+\sum_{t=t^\prime+1}^{t^\prime+\tau-1}\mathbbm{E}\|\bm{X}_{t}-\overline{\bm{X}}_{t}\|_F^2\bigg]\notag\\
    & \overset{}{\leq}
        6\tau\sum_{t^\prime =0}^{T-\tau}\mathbbm{E}\|\bm{X}_{t^\prime}-\overline{\bm{X}}_{t^\prime}\|_F^2  + \frac{48N L^2 \gamma^4}{b\tau}\sum_{t^\prime =0}^{T-\tau}\sum_{t=t^\prime+1}^{t^\prime+\tau-1}\sum_{j=t^\prime+1}^{t}(j-t^\prime)\sum_{k=t^\prime}^{j-1}   \mathbbm{E}\|\overline{\bm{v}}_{k}\|^2  \notag\\
        & \quad \quad \quad \quad \quad \quad + \sum_{t^\prime =0}^{T-\tau}\sum_{t=t^\prime+1}^{t^\prime+\tau-1}\sum_{j=t^\prime+1}^{t}\left(\frac{24N\alpha^2\gamma^2\sigma^2}{b\tau} (j-t^\prime)^3+ \frac{48N\gamma^2\varsigma^2}{\tau}(j-t^\prime)^2\right)\notag\\
    & \overset{}{\leq}
        6\tau\sum_{t^\prime =0}^{T-\tau}\mathbbm{E}\|\bm{X}_{t^\prime}-\overline{\bm{X}}_{t^\prime}\|_F^2  + \frac{48N L^2 \gamma^4}{b\tau}\sum_{t^\prime =0}^{T-\tau}\sum_{t=t^\prime+1}^{t^\prime+\tau-1}\sum_{j=t^\prime+1}^{t^\prime+\tau-1}(j-t^\prime)\sum_{k=t^\prime}^{t-1}   \mathbbm{E}\|\overline{\bm{v}}_{k}\|^2  \notag\\
        &\quad \quad \quad \quad \quad \quad + \sum_{t^\prime =0}^{T-\tau}\sum_{t=t^\prime}^{t^\prime+\tau-1}\sum_{j=t^\prime+1}^{t^\prime+\tau-1}\left(\frac{24N\alpha^2\gamma^2\sigma^2}{b\tau} (j-t^\prime)^3+ \frac{48N\gamma^2\varsigma^2}{\tau}(j-t^\prime)^2\right) \notag\\
    & \overset{(a)}{\leq}
        6\tau\sum_{t^\prime =0}^{T-\tau}\mathbbm{E}\|\bm{X}_{t^\prime}-\overline{\bm{X}}_{t^\prime}\|_F^2 + \frac{24N L^2 \gamma^4(\tau-1)^2}{b}\sum_{t=0}^{T-1} \mathbbm{E}\|\overline{\bm{v}}_{t}\|^2  \notag\\
        & \quad \quad \quad \quad \quad \quad \quad \quad \quad \quad \quad \quad \quad \quad + \frac{6N\alpha^2\gamma^2\tau(\tau-1)^2T\sigma^2}{b}+ 16N\gamma^2\tau(\tau-1)T\varsigma^2, \label{al_2_CD3:}
\end{align}
where ($a$) holds by using the facts that
$$\sum_{k=1}^{\tau-1}k \leq \frac{\tau(\tau-1)}{2}, \sum_{k=1}^{\tau-1}k^2 \leq \frac{\tau^2(\tau-1)}{3}, \sum_{k=1}^{\tau-1}k^3 \leq \frac{\tau^2(\tau-1)^2}{4}.$$

Further, using the inequalities (\ref{al_1_CD4:}) and (\ref{al_1_CD5:}) from lemma \ref{al_1_CD:}, we get: for $t^\prime>0$,
\begin{equation}
    \label{al_2_CD4:}
    \begin{split}
        \sum_{t^\prime=0}^{T-\tau}\mathbbm{E}&\|\bm{X}_{t^{\prime}}-\bm{\overline{X}}_{t^{\prime}}\|_F^2 
         \overset{}{\leq}
            \frac{4\lambda^2}{(1-\lambda^2)^2}\sum_{t^\prime=\tau}^{T-\tau} \mathbbm{E}\|\bm{Y}_{t^{\prime}}- \bm{\overline{Y}}_{t^{\prime}}\|_F^2\\
        & \overset{(a)}{\leq}
            \frac{64\lambda^4\gamma^2\tau}{(1-\lambda^2)^4}\left(1+\frac{8 L^2  \gamma^2 (\tau-1)}{b}\right) \sum_{t=0}^{T-1} \bigg[8L^2 \mathbbm{E}\left\|\bm{X}_t-\overline{\bm{X}}_t\right\|_F^2 +2N\varsigma^2\bigg]\\
            & \quad \quad \quad \quad \quad \quad  + \frac{256N\lambda^4 L^2  \gamma^4(\tau-1)^2}{(1-\lambda^2)^3b}\sum_{t=0}^{T-1}  \mathbbm{E}\|\overline{\bm{v}}_{t}\|^2 + \frac{128N\lambda^4\alpha^2\gamma^2(\tau-1)^2 T \sigma^2}{(1-\lambda^2)^3b}\\
        & \overset{(b)}{\leq}
            \frac{1024\lambda^4L^2\gamma^2\tau}{(1-\lambda^2)^4} \sum_{t=0}^{T-1}  \mathbbm{E}\left\|\bm{X}_t-\overline{\bm{X}}_t\right\|_F^2 +\frac{256N\lambda^4\gamma^2\tau T\varsigma^2}{(1-\lambda^2)^4} \\
            & \quad \quad \quad \quad \quad \quad + \frac{256N\lambda^4 L^2  \gamma^4(\tau-1)^2}{(1-\lambda^2)^3b}\sum_{t=0}^{T-1}  \mathbbm{E}\|\overline{\bm{v}}_{t}\|^2 + \frac{128N\lambda^4\alpha^2\gamma^2(\tau-1)^2 T \sigma^2}{(1-\lambda^2)^3b},    
    \end{split}
\end{equation}
where ($a$) follows from the statements of lemma \ref{al2_G_A_A_D_CD_main:}, lemma \ref{al2_lemm_GEC:} and lemma \ref{griadient_consensus:}, and the fact that $\frac{1}{(1-\lambda^2)^3}\leq\frac{1}{(1-\lambda^2)^4}$ holds if $\lambda \in (0,1)$, and ($b$) holds because $1+\frac{8 L^2  \gamma^2 (\tau-1)}{b} \leq 2$ holds if $\gamma \leq \frac{1}{8L\tau}$.

Substituting (\ref{al_2_CD4:}) into (\ref{al_2_CD3:}), we have:
\begin{equation}
    \label{al_2_CD5:}
    \begin{split}
        \sum_{t =0}^{T-1}\mathbbm{E}\|\bm{X}_{t}-\overline{\bm{X}}_{t}\|_F^2 
        & \overset{}{\leq}
             \left(\frac{24N L^2 \gamma^4(\tau-1)^2}{b}+ \frac{1536N\lambda^4 L^2  \gamma^4\tau(\tau-1)^2}{(1-\lambda^2)^3b}\right)\sum_{t=0}^{T-1} \mathbbm{E}\|\overline{\bm{v}}_{t}\|^2  \\ 
            & \quad \quad \quad \quad \quad \quad \quad \quad + \frac{768N\lambda^4\alpha^2\gamma^2\tau(\tau-1)^2 T \sigma^2}{(1-\lambda^2)^3b} +\frac{1536N\lambda^4\gamma^2\tau^2 T\varsigma^2}{(1-\lambda^2)^4} \\
            & \quad \quad \quad \quad \quad \quad \quad \quad \quad \quad+ \frac{6N\alpha^2\gamma^2\tau(\tau-1)^2T\sigma^2}{b} + 16N\gamma^2\tau(\tau-1)T\varsigma^2\\
            & \quad \quad \quad \quad \quad \quad \quad \quad \quad \quad \quad \quad \quad \quad +\frac{6144\lambda^4L^2\gamma^2\tau^2}{(1-\lambda^2)^4} \sum_{t=0}^{T-1}  \mathbbm{E}\left\|\bm{X}_t-\overline{\bm{X}}_t\right\|_F^2.
    \end{split}
\end{equation}
Next, if we set $\gamma \leq \frac{(1-\lambda^2)^2}{64\sqrt{6}\lambda^2L\tau}$, then the inequality $1-\frac{6144\lambda^4L^2\gamma^2\tau^2}{(1-\lambda^2)^4}\geq\frac{3}{4}$ holds. Finally, by a simple arrangement, we complete the proof.
\end{proof}

\subsubsection{D.3 \quad The Proof of Theorem \ref{alg_momen:}}
In this section, we give the proof of Theorem \ref{alg_momen:} using the statements of several lemmas listed in the previous subsections. 

\begin{lemm}
    \label{al_2_PT2:}
    For all $t \in [0,\cdots,T-1]$, the averages $\overline{\bm{x}}_{t}=\frac{1}{N}\sum\limits_{i=1}^{N}\bm{x}_{t}^{(i)}$ of the iterates generated by algorithm \ref{alg_momen:} with $\gamma_t = \gamma \leq \min \left\{ \frac{1}{8L\tau}, \frac{(1-\lambda^2)^2}{64\sqrt{6}\lambda^2L\tau}\right\}$ and $\alpha_t=\alpha=\frac{32L^2\gamma^2}{Nb}$ satisfy that
    \begin{align*}
        \frac{1}{T}\sum_{t=0}^{T-1}\mathbbm{E}\|\nabla F(\bm{\overline{x}}_{t})\|^{2} 
        & \overset{}{\leq} 
            \frac{2(F(\bm{\overline{x}}_{0})-F^\star)}{\gamma T} +\frac{8\alpha^2 (\tau-1)  \sigma^2}{Nb} +\frac{64 L^2 \gamma^2(\tau-1)  \varsigma^2}{Nb} \notag \\
            & \quad \quad \quad \quad   + \frac{32\alpha^2L^2\gamma^2\tau(\tau-1)^2\sigma^2}{b} + 96L^2\gamma^2\tau(\tau-1)\varsigma^2 \notag \\
            & \quad \quad  + \frac{4096\lambda^4\alpha^2L^2\gamma^2\tau(\tau-1)^2  \sigma^2}{(1-\lambda^2)^3b} +\frac{8192\lambda^4L^2\gamma^2\tau^2 \varsigma^2}{(1-\lambda^2)^4},
    \end{align*}
    where the expectation $\mathbbm{E}[\cdot]$ is w.r.t the stochasticity of the algorithm.
\end{lemm}

\begin{proof}
    Using the statement of Lemma \ref{al2_Descent_Lemma:}, making a simple arrangement and doing the summation operation from $t=0$ to $T-1$, we get:
    \begin{align}      
        \sum_{t=0}^{T-1}\mathbbm{E}\|\nabla F(\bm{\overline{x}}_{t})\|^{2} 
        & \overset{}{\leq}     
            \frac{2(\mathbbm{E}[F(\bm{\overline{x}}_{0})]-\mathbbm{E}[F(\bm{\overline{x}}_{T})])}{\gamma}-(1-\gamma L)\sum_{t=0}^{T-1}\mathbbm{E}\|\bm{\overline{v}}_t\|^{2}+2\sum_{t=0}^{T-1}\mathbbm{E}\|\bm{\overline{e}}_t\|^{2}\notag\\
            & \quad \quad \quad \quad \quad \quad \quad \quad \quad \quad \quad \quad \quad \quad \quad \quad \quad \quad \quad \quad  +\frac{2 L^2}{N} \sum_{t=0}^{T-1}\mathbbm{E}\|\bm{X}_t-\overline{\bm{X}}_{t}\|_F^{2}\notag \\
        & \overset{(a)}{\leq}     
            \frac{2(F(\bm{\overline{x}}_{0})-F^\star)}{\gamma}-(1-\gamma L)\sum_{t=0}^{T-1}\mathbbm{E}\|\bm{\overline{v}}_t\|^{2}+2\sum_{t=0}^{T-1}\mathbbm{E}\|\bm{\overline{e}}_t\|^{2}\notag \\
            & \quad \quad \quad \quad \quad \quad \quad \quad \quad \quad \quad \quad \quad \quad \quad \quad \quad \quad \quad \quad   +\frac{2 L^2}{N} \sum_{t=0}^{T-1}\mathbbm{E}\|\bm{X}_t-\overline{\bm{X}}_{t}\|_F^{2}\notag \\
        & \overset{(b)}{\leq}    
            \frac{2(F(\bm{\overline{x}}_{0})-F^\star)}{\gamma}-\left(1-\gamma L-\frac{16 L^2 \gamma^2(\tau-1)}{Nb}\right)\sum_{t=0}^{T-1}\mathbbm{E}\|\bm{\overline{v}}_t\|^{2}\notag \\
            & \quad \quad \quad \quad \quad \quad \quad \quad \quad   + \frac{2L^2}{N}\left(1+\frac{128 L^2 \gamma^2(\tau-1)}{Nb}\right) \sum_{t=0}^{T-1} \left\|\bm{X}_t-\overline{\bm{X}}_t\right\|_F^2 \notag \\
            & \quad \quad \quad \quad \quad \quad \quad \quad \quad \quad \quad \quad \quad \quad  +\frac{8\alpha^2 (\tau-1) T \sigma^2}{Nb} +\frac{64 L^2 \gamma^2(\tau-1) T\varsigma^2}{Nb} \notag \\
        & \overset{(c)}{\leq}     
            \frac{2(F(\bm{\overline{x}}_{0})-F^\star)}{\gamma}-\left(1-2\gamma L\right)\sum_{t=0}^{T-1}\mathbbm{E}\|\bm{\overline{v}}_t\|^{2} \notag \\
            & \quad \quad \quad   + \frac{4L^2}{N} \sum_{t=0}^{T-1} \left\|\bm{X}_t-\overline{\bm{X}}_t\right\|_F^2 +\frac{8\alpha^2 (\tau-1) T \sigma^2}{Nb} +\frac{64 L^2 \gamma^2(\tau-1) T\varsigma^2}{Nb}\notag \\
        & \overset{(d)}{\leq}     
            \frac{2(F(\bm{\overline{x}}_{0})-F^\star)}{\gamma}-\bigg(1-2\gamma L-\frac{128 L^4 \gamma^4(\tau-1)^2}{b} \notag \\
            & \quad \quad \quad \quad \quad \quad \quad \quad \quad \quad \quad \quad \quad \quad  - \frac{8192\lambda^4 L^4  \gamma^4\tau(\tau-1)^2}{(1-\lambda^2)^3b}\bigg)\sum_{t=0}^{T-1}\mathbbm{E}\|\bm{\overline{v}}_t\|^{2} \notag \\
            & \quad \quad \quad \quad \quad \quad \quad \quad \quad \quad \quad \quad \quad \quad +\frac{8\alpha^2 (\tau-1) T \sigma^2}{Nb} +\frac{64 L^2 \gamma^2(\tau-1)  T\varsigma^2}{Nb} \notag \\
            & \quad \quad \quad \quad \quad \quad \quad \quad \quad \quad   + \frac{32\alpha^2L^2\gamma^2\tau(\tau-1)^2T\sigma^2}{b} + 96L^2\gamma^2\tau(\tau-1)T\varsigma^2 \notag \\
            & \quad \quad \quad \quad \quad \quad \quad \quad  + \frac{4096\lambda^4\alpha^2L^2\gamma^2\tau(\tau-1)^2 T \sigma^2}{(1-\lambda^2)^3b} +\frac{8192\lambda^4L^2\gamma^2\tau^2 T\varsigma^2}{(1-\lambda^2)^4},\label{al_2_PT2:1}
    \end{align}
    
    where ($a$) holds by using the fact that $F^*=\inf_{\bm{x}\in \mathbbm{R}^d}F(\bm{x})>-\infty$, the inequality ($b$) results from the statements of lemma \ref{al2_lemm_GEC:} and lemma \ref{griadient_consensus:}, ($c$) follows from the facts that $\frac{16 L^2 \gamma^2(\tau-1)}{Nb} \leq\frac{2 L \gamma}{Nb} \leq  L \gamma$ and $\frac{128 L^2 \gamma^2(\tau-1)}{Nb} \leq\frac{2 }{Nb\tau} \leq  1 $ hold if $\gamma \leq \frac{1}{8L\tau}$, $N\geq 2$, $b\geq1$ and $\tau\geq1$, and ($d$) holds by using the statement of lemma \ref{al_2_CD:}.
    
    By a simple calculation, we can get the fact that if $\gamma \leq \min \left\{ \frac{1}{8L\tau}, \frac{(1-\lambda^2)^2}{64\sqrt{6}\lambda^2L\tau}\right\}$, then 
    $$1-2\gamma L-\frac{128 L^4 \gamma^4(\tau-1)^2}{b} - \frac{8192\lambda^4 L^4  \gamma^4\tau(\tau-1)^2}{(1-\lambda^2)^3b}\geq0$$ 
    holds. Finally, by dividing both sides of (\ref{al_2_PT2:1}) by $T$ , we complete the proof.
\end{proof}

\subsection{E \quad Full Experiments}
\label{Ex_Ex:}
In this section, we report the full numerical results.
For the convenience of the reader, we give full the experimental setup and the detail of computing devices and platforms for all the implemented algorithms on top of the MNIST and CIFAR-10 datasets.

\textbf{Experimental Setup:} We conduct 10-class image classification on MNIST~\cite{lecun1998gradient} and CIFAR-10~\cite{krizhevsky2009learning} datasets.
For both datasets, we consider common network topology, i.e., {\it ring graph} and use the Metropolis-Hasting mixing matrix $W$~\cite{koloskova2020unified}, i.e., $w_{ij}=w_{ji}=\frac{1}{deg(i)+1}=\frac{1}{deg(j)+1}$ for any edge $(i,j) \in \mathcal{E}$, to parameterize the communication.  
For MNIST, 
a convolutional neural network (CNN) with two convolutional hidden layers plus two linear layers is implemented for each node. 
We set $T = 400$
and 
fine-tune the learning rate from $\{0.1, 0.2, 0.3, 0.4, 0.5\}$,
the batch size  
$b$ from $\{64, 128, 256\}$ and 
the partial average interval 
$\tau$ from $\{3, 7, 20\}$. 
Note that
we divide the learning rate 
by $2$ at iterations $0.5\cdot T$ and $0.75\cdot T$.  
Moreover,  
the control parameter $\alpha$ is tuned from $\{0.01,  0.05\}$,
which 
is decayed with a decay weight $0.99$.
For CIFAR-10, 
each node implements a Resnet-20-BN~\cite{he2016deep} architecture. 
We fix $T$ to $10000$ and 
schedule 
the learning rate (the control parameter $\alpha$) setting as 
$0.01 (0.002)$, $0.1 (0.02)$, $0.01 (0.002)$, and $0.001 (0.0002)$ at iterations $0\cdot T$, $0.1\cdot T$, $0.75\cdot T$, and $0.9\cdot T$, respectively.
Further, we fine-tune the batch size  
$b$ from $\{256, 512, 1024\}$ and 
the partial average interval 
$\tau$ from $\{4, 8, 20\}$. 
We use  Dirichlet process $Dp(\omega)$~\cite{vogels2021relaysum, lin2021quasi} to strictly partition training data across 20 (40) nodes for MNIST (CIFAR-10), where the scaling parameter $\omega$ controls the data heterogeneity
across nodes. 
For both datasets,
we set $\omega = 0.5$ and $\omega = 10$ to generate the non-iid and iid settings, respectively.
For fairness, 
we compare all the methods 
under uniform data heterogeneity settings w.r.t. the best training loss and test accuracy.

\textbf{Computing devices and platforms:}
\begin{itemize}
    \item OS: Ubuntu 18.04.3 LTS
    \item CPU: Intel(R) Xeon(R) Gold 6126 CPU @ 2.60GHz
    \item CPU Memory: 256 GB.
    \item GPU: 4 * NVIDIA Tesla V100 PCIe
    \item GPU Memory: 4 * 32GB
    \item Programming platform: Python 3.7.4
    \item Deep learning platform: Pytorch 1.11.0
\end{itemize}

\begin{table}
  \centering
  \caption{Top-1 test accuracy($\%$) and training loss overview given difference  settings.}
    \resizebox{1.0\columnwidth}{!}{
    \begin{tabular}{llcccccccccc}
    \toprule[2pt]
    \multirow{2}[2]{*}{Datasets} & \multirow{2}[2]{*}{Settings} & \multicolumn{2}{c}{DLSGD} & \multicolumn{2}{c}{SLOWMo-D} & \multicolumn{2}{c}{PD-SGDM} & \multicolumn{2}{c}{DSE-SGD} & \multicolumn{2}{c}{DSE-MVR} \\
          &       & test accuracy & training loss & test accuracy & training loss & test accuracy & training loss & test accuracy & training loss & test accuracy & training loss \\
    \midrule
    \multirow{3}[2]{*}{MNIST,$\omega=0.5$} & $b=64$  & 97.34$\pm$0.23 & 0.050$\pm$0.008 & \textbf{97.89$\pm$0.19} & 0.029$\pm$0.006 & 97.84$\pm$0.31 & 0.029$\pm$0.011 & 97.76$\pm$0.24 & 0.032$\pm$0.012 & 97.89$\pm$0.55 & \textbf{0.019$\pm$0.012} \\
          & $b=128$ & 97.36$\pm$0.31 & 0.045$\pm$0.007 & 98.09$\pm$0.41 & 0.019$\pm$0.006 & 97.94$\pm$0.43 & 0.035$\pm$0.014 & 97.89$\pm$0.28 & 0.031$\pm$0.011 & \textbf{98.49$\pm$0.46} & \textbf{0.016$\pm$0.009} \\
          & $b=256$ & 97.47$\pm$0.17 & 0.041$\pm$0.007 & 98.23$\pm$0.22 & 0.016$\pm$0.005 & 98.05$\pm$0.46 & 0.025$\pm$0.014 & 97.94$\pm$0.41 & 0.026$\pm$0.012 & \textbf{98.53$\pm$0.36} & \textbf{0.009$\pm$0.006} \\
    \midrule
    \multirow{3}[2]{*}{MNIST,$\omega=10$} & $b=64$  & 97.82$\pm$0.12 & 0.034$\pm$0.004 & \textbf{98.19$\pm$0.31} & 0.029$\pm$0.012 & 97.91$\pm$0.43 & \multicolumn{1}{c}{0.034$\pm$0.023} & 97.75$\pm$0.49 & \multicolumn{1}{c}{0.036$\pm$0.012} & 98.17$\pm$0.69 & \multicolumn{1}{c}{\textbf{0.023$\pm$0.016}} \\
          & $b=128$ & 97.89$\pm$0.11 & 0.029$\pm$0.006 & 98.26$\pm$0.27 & 0.016$\pm$0.012 & 98.24$\pm$0.39 & 0.017$\pm$0.012 & 98.14$\pm$0.23 & 0.020$\pm$0.011 & \textbf{98.45$\pm$0.54} & \textbf{0.012$\pm$0.006} \\
          & $b=256$ & 97.91$\pm$0.09 & 0.024$\pm$0.004 & 98.31$\pm$0.27 & 0.014$\pm$0.012 & 98.50$\pm$0.37 & 0.012$\pm$.0.009 & 98.32$\pm$0.27 & 0.019$\pm$0.011 & \textbf{98.67$\pm$0.38} & \textbf{0.011$\pm$0.009} \\
    \midrule
    \multirow{3}[2]{*}{MNIST, $\omega=0.5$} & $\tau=3$ & 97.59$\pm$0.04 & 0.038$\pm$0.004 & 98.23$\pm$0.19 & 0.018$\pm$0.005 & 98.24$\pm$0.24 & 0.017$\pm$0.007 & 98.15$\pm$0.23 & 0.018$\pm$0.005 & \textbf{98.71$\pm$0.22} & \textbf{0.009$\pm$0.004} \\
          & $\tau=7$ & 97.42$\pm$0.05 & 0.044$\pm$0.004 & 98.19$\pm$0.29 & 0.021$\pm$0.011 & 98.00$\pm$0.34 & 0.039$\pm$0.011 & 97.89$\pm$0.05 & 0.029$\pm$0.002 & \textbf{98.22$\pm$0.75} & \textbf{0.016$\pm$0.014} \\
          & $\tau=20$ & 97.13$\pm$0.15 & 0.053$\pm$0.006 & 97.79$\pm$0.22 & 0.026$\pm$0.005 & 97.58$\pm$0.11 & 0.034$\pm$0.003 & 97.55$\pm$0.05 & 0.041$\pm$0.004 & \textbf{97.98$\pm$0.13} & \textbf{0.019$\pm$0.006} \\
    \midrule
    \multirow{3}[2]{*}{MNIST, $\omega=10$} & $\tau=3$ & 97.96$\pm$0.03 & 0.026$\pm$0.005 & 98.46$\pm$0.08 & 0.015$\pm$0.005 & 98.60$\pm$0.18 & 0.009$\pm$0.004 & 98.38$\pm$0.19 & 0.016$\pm$0.008 & \textbf{99.02$\pm$0.09} & \textbf{0.005$\pm$0.002} \\
          & $\tau=7$ & 97.90$\pm$0.05 & 0.027$\pm$0.005 & \textbf{98.37$\pm$0.01} & 0.018$\pm$0.022 & 98.16$\pm$0.54 & 0.023$\pm$0.028 & 97.95$\pm$0.58 & 0.026$\pm$0.018 & 98.29$\pm$0.27 & \textbf{0.016$\pm$0.009} \\
          & $\tau=20$ & 97.75$\pm$0.06 & 0.035$\pm$0.005 & 97.93$\pm$0.09 & 0.027$\pm$0.002 & 97.89$\pm$0.17 & 0.032$\pm$0.006 & 97.88$\pm$0.19 & 0.033$\pm$0.006 & \textbf{97.98$\pm$0.42} & \textbf{0.026$\pm$0.009} \\
    \midrule
    \multirow{3}[2]{*}{CIFAR-10, $\omega=0.5$} & $b=256$ & 79.59$\pm$1.22 & 0.419$\pm$0.056 & 84.21$\pm$0.49 & 0.192$\pm$0.042 & 84.31$\pm$0.83 & 0.209$\pm$0.113 & 82.19$\pm$0.98 & 0.309$\pm$0.052 & \textbf{84.65$\pm$0.57} & \textbf{0.185$\pm$0.052} \\
          & $b=512$ & 80.46$\pm$1.03 & 0.367$\pm$0.021 & 85.28$\pm$0.39 & 0.134$\pm$0.026 & 84.96$\pm$0.44 & 0.129$\pm$0.046 & 83.15$\pm$0.59 & 0.294$\pm$0.066 & \textbf{85.30$\pm$0.41} & \textbf{0.107$\pm$0.047} \\
          & $b=1024$ & 81.69$\pm$0.21 & 0.339$\pm$0.022 & \textbf{86.03$\pm$0.54} & 0.105$\pm$0.031 & 85.33$\pm$0.37 & 0.160$\pm$0.033 & 83.42$\pm$0.43 & 0.279$\pm$0.069 & 85.83$\pm$0.56 & \textbf{0.099$\pm$0.023} \\
    \midrule
    \multirow{3}[2]{*}{CIFAR-10, $\omega=10$} & $b=256$ & 84.47$\pm$0.31 & 0.194$\pm$0.021 & 87.22$\pm$1.21 & 0.066$\pm$0.063 & 87.04$\pm$0.77 & 0.079$\pm$0.095 & 86.12$\pm$0.86 & 0.130$\pm$0.087 & \textbf{87.50$\pm$1.04} & \textbf{0.054$\pm$0.052} \\
          & $b=512$ & 84.69$\pm$0.32 & 0.277$\pm$0.041 & 87.61$\pm$0.98 & 0.042$\pm$0.045 & 87.65$\pm$0.76 & 0.039$\pm$0.018 & 86.63$\pm$0.76 & 0.099$\pm$0.081 & \textbf{87.87$\pm$1.08} & \textbf{0.013$\pm$0.014} \\
          & $b=1024$ & 85.16$\pm$0.24 & 0.177$\pm$0.064 & 88.11$\pm$1.04 & 0.029$\pm$0.006 & 87.95$\pm$0.83 & 0.035$\pm$0.015 & 87.19$\pm$0.56 & 0.071$\pm$0.058 & \textbf{88.41$\pm$0.75} & \textbf{0.027$\pm$0.018} \\
    \midrule
    \multirow{3}[2]{*}{CIFAR-10, $\omega=0.5$} & $\tau=4$ & 81.44$\pm$0.59 & 0.346$\pm$0.027 & 85.53$\pm$1.03 & 0.164$\pm$0.035 & 85.29$\pm$0.31 & 0.166$\pm$0.042 & 83.54$\pm$0.37 & 0.227$\pm$0.031 & \textbf{85.81$\pm$0.52} & \textbf{0.094$\pm$0.029} \\
          & $\tau=8$ & 80.47$\pm$1.03 & 0.375$\pm$0.036 & \textbf{85.32$\pm$0.77} & 0.174$\pm$0.047 & 85.07$\pm$0.45 & 0.141$\pm$0.038 & 83.02$\pm$0.62 & 0.313$\pm$0.011 & 85.13$\pm$0.77 & \textbf{0.158$\pm$0.047} \\
          & $\tau=20$ & 79.83$\pm$1.59 & 0.404$\pm$0.067 & 84.67$\pm$0.96 & \textbf{0.113$\pm$0.029} & 84.25$\pm$0.79 & 0.192$\pm$0.126 & 82.21$\pm$0.94 & 0.343$\pm$0.024 & \textbf{84.85$\pm$0.50} & 0.140$\pm$0.078 \\
    \midrule
    \multirow{3}[2]{*}{CIFAR-10, $\omega=10$} & $\tau=4$ & 85.05$\pm$0.36 & 0.220$\pm$0.043 & 88.29$\pm$0.41 & 0.017$\pm$0.008 & 88.01$\pm$0.62 & 0.017$\pm$0.006 & 87.22$\pm$0.59 & 0.026$\pm$0.004 & \textbf{88.54$\pm$0.39} & \textbf{0.011$\pm$0.017} \\
          & $\tau=8$ & 84.78$\pm$0.23 & 0.210$\pm$0.030 & 88.25$\pm$0.41 & 0.031$\pm$0.016 & 87.98$\pm$0.33 & 0.038$\pm$0.001 & 86.85$\pm$0.23 & 0.105$\pm$0.059 & \textbf{88.41$\pm$0.35} & \textbf{0.013$\pm$0.012} \\
          & $\tau=20$ & 84.49$\pm$0.45 & 0.214$\pm$0.109 & 86.41$\pm$0.54 & 0.088$\pm$0.050 & 86.65$\pm$0.43 & 0.097$\pm$0.078 & 85.87$\pm$0.78 & 0.169$\pm$0.029 & \textbf{86.83$\pm$0.64} & \textbf{0.059$\pm$0.046} \\
    \bottomrule
    \bottomrule
    \end{tabular}}%
  \label{tab:full_test_train_ov}%
\end{table}%

\begin{figure}[htbp]
  \small
  \centering
  \subfigure[$\tau=3, \omega=0.5$]{
  \includegraphics[scale=0.29]{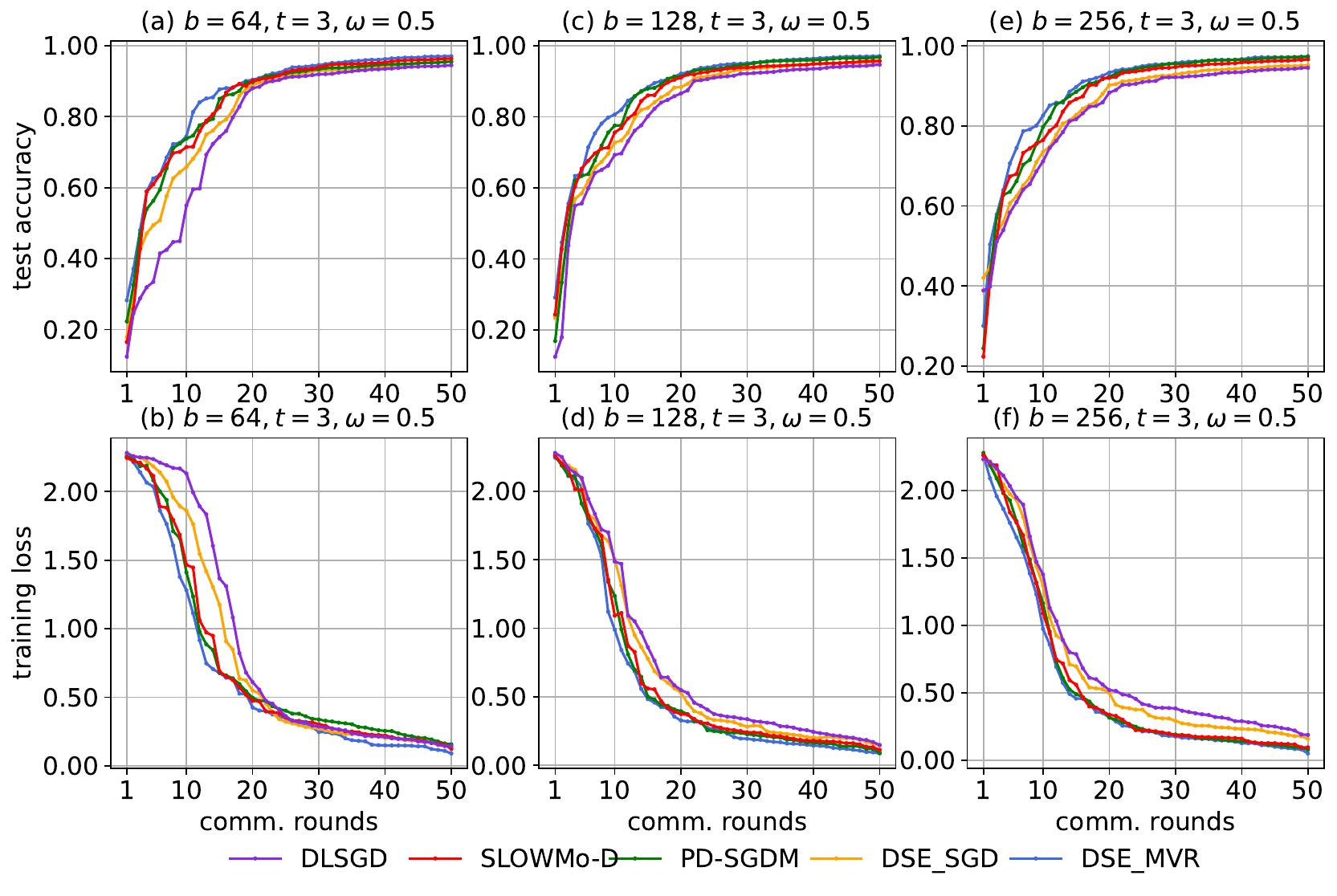}
  }
  \
  \subfigure[$\tau=7, \omega=0.5$]{
  \includegraphics[scale=0.29]{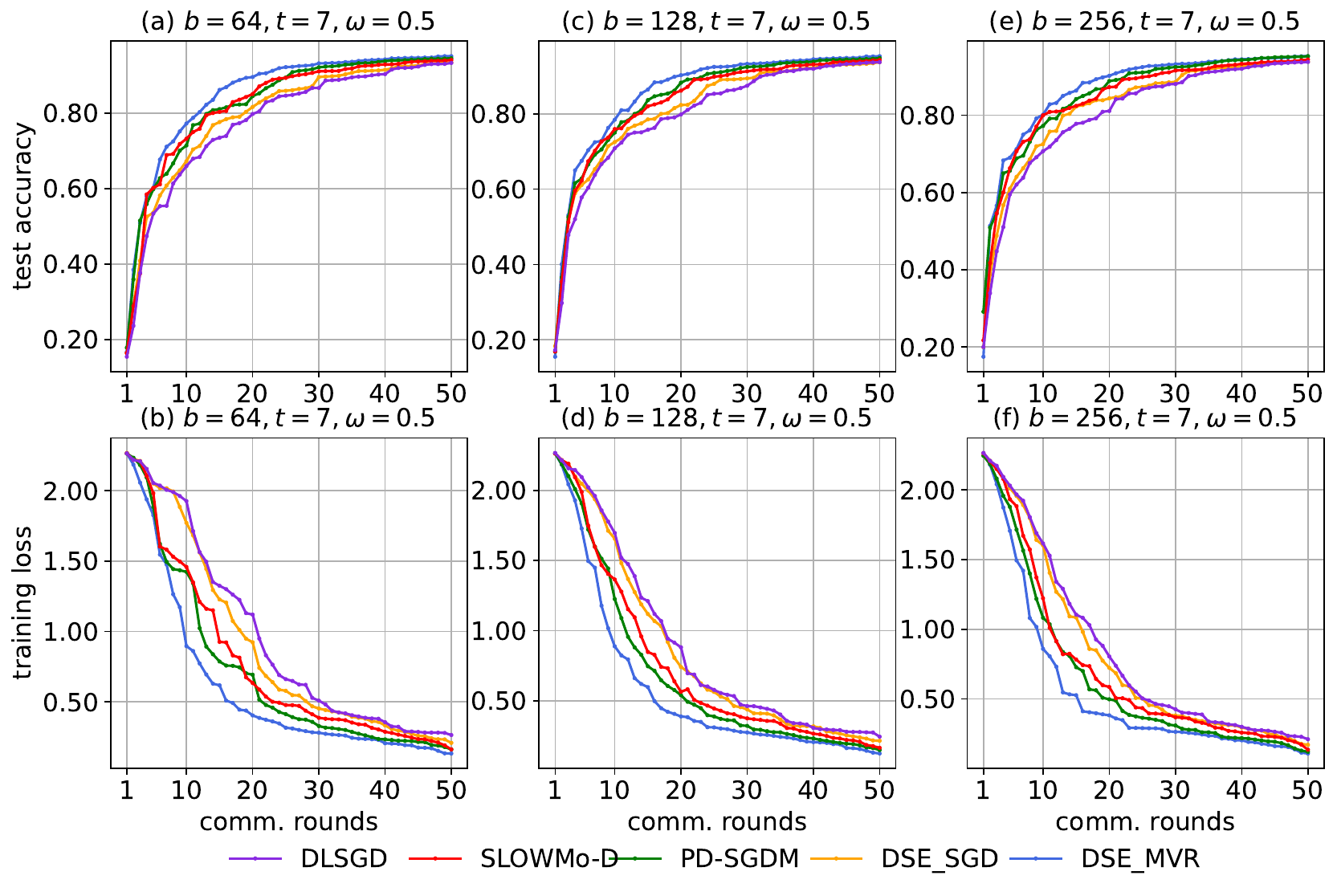}
  }
  \
  \subfigure[$\tau=20, \omega=0.5$]{
  \includegraphics[scale=0.29]{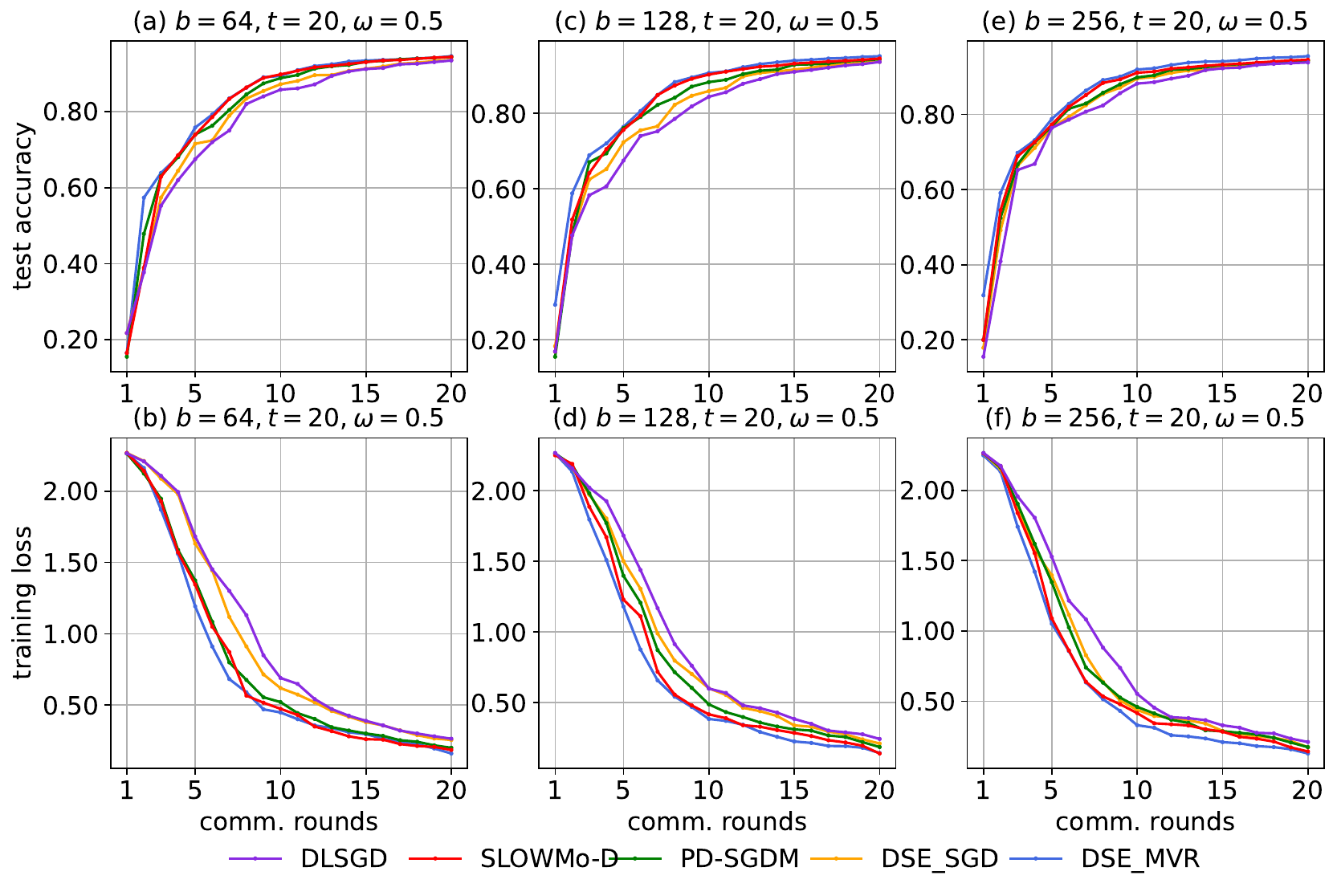}
  }
  \
  \subfigure[$\tau=3, \omega=10$]{
  \includegraphics[scale=0.29]{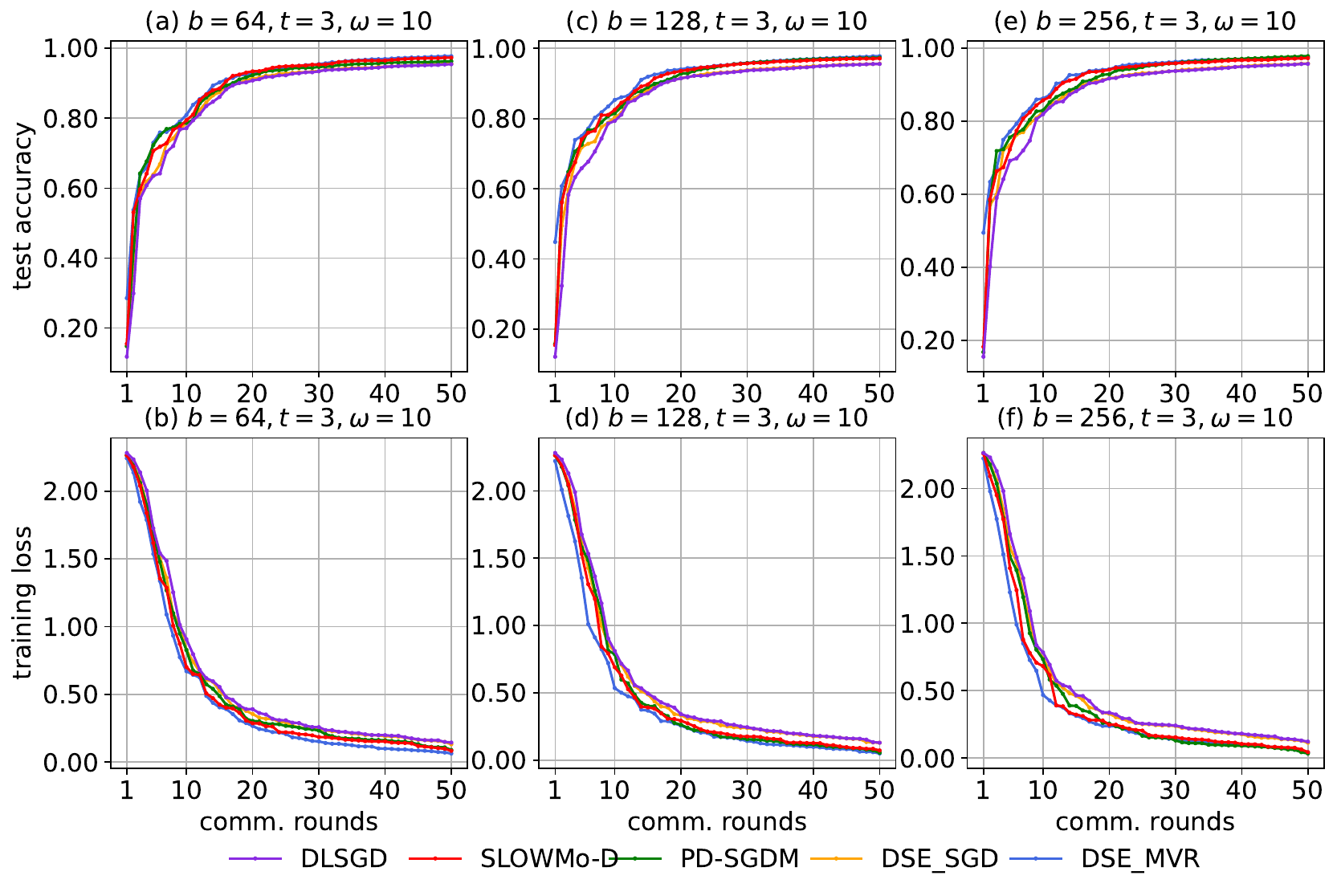}
  }
  \
  \subfigure[$\tau=7, \omega=10$]{
  \includegraphics[scale=0.29]{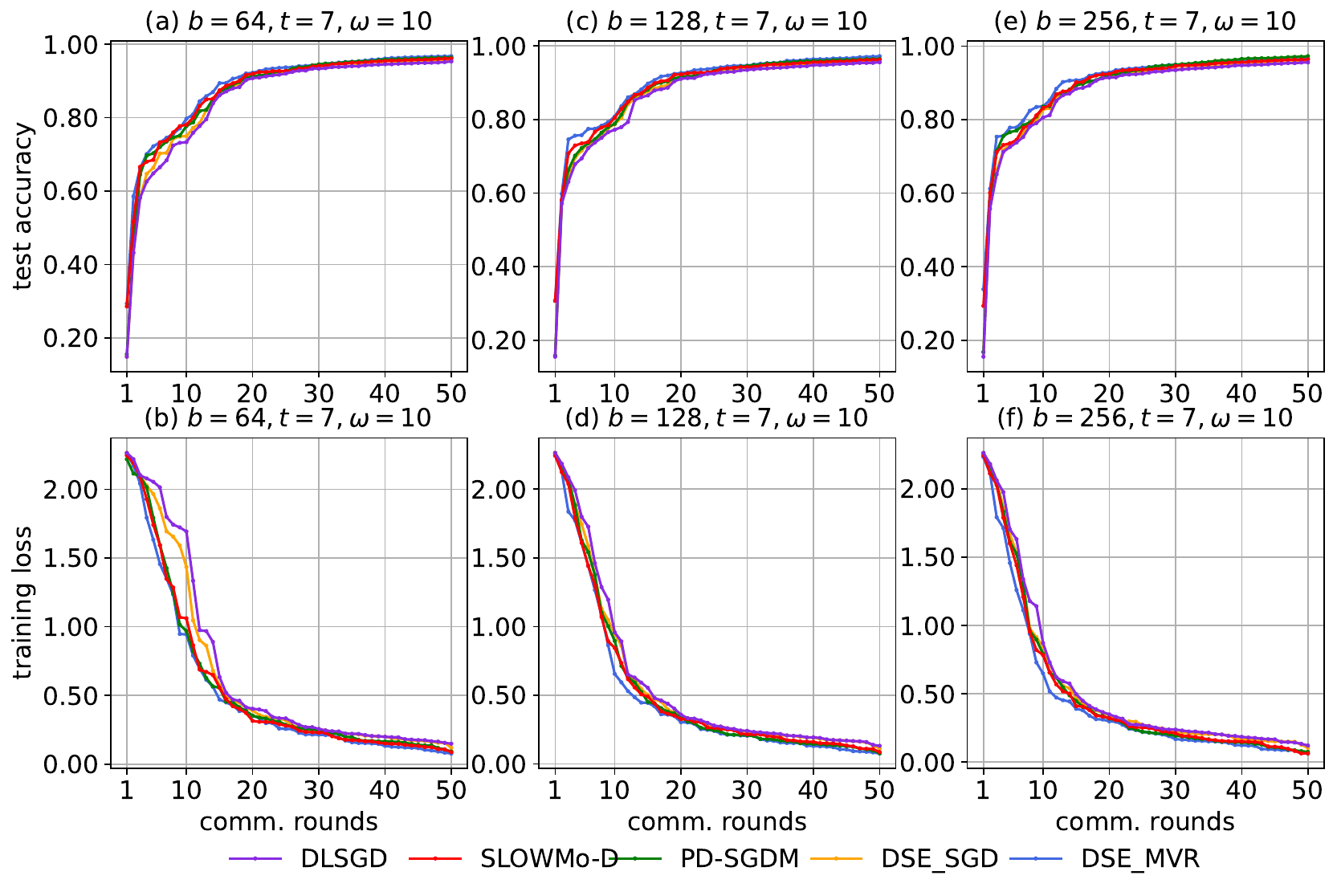}
  }
  \
  \subfigure[$\tau=20, \omega=10$]{
  \includegraphics[scale=0.29]{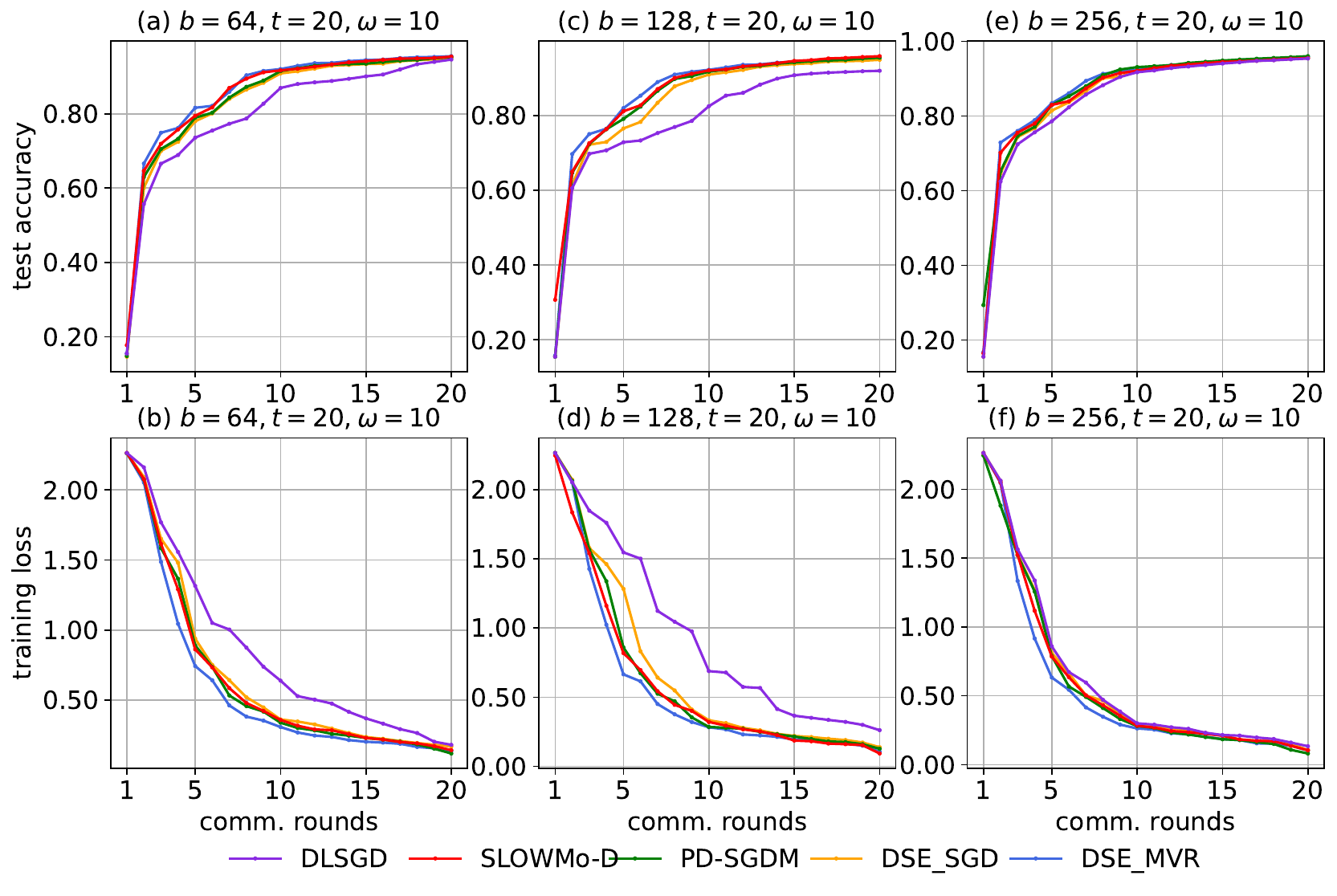}
  }
  \caption{Full learning curves w.r.t test accuracy and training loss over MNIST dataset, which are  averaged over 3 random seeds.}
  \label{fig:MNIST_full_res}
\end{figure}

\begin{figure}[htbp]
  \small
  \centering
  \subfigure[$\tau=4, \omega=0.5$]{
  \includegraphics[scale=0.29]{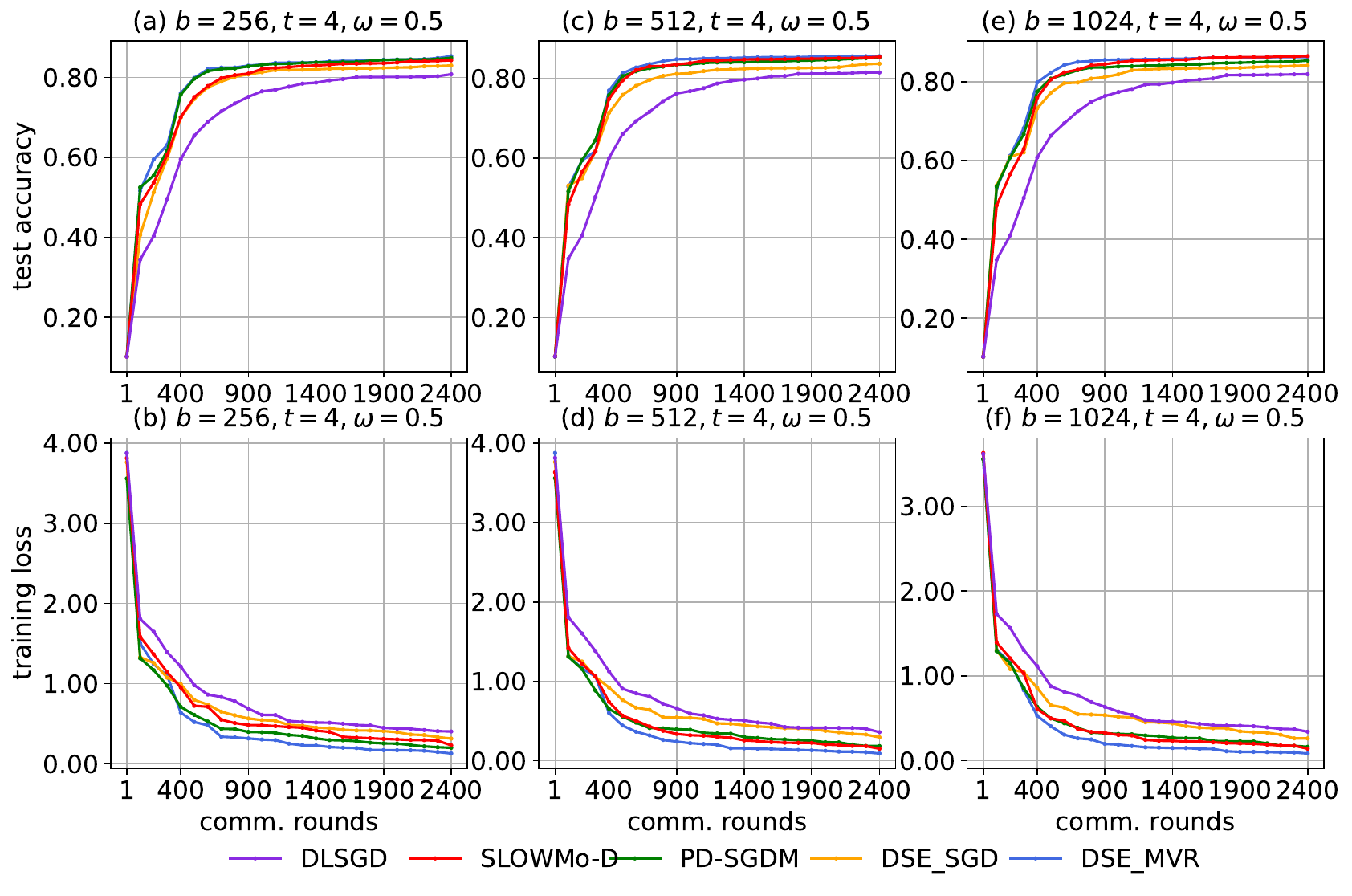}
  }
  \
  \subfigure[$\tau=8, \omega=0.5$]{
  \includegraphics[scale=0.29]{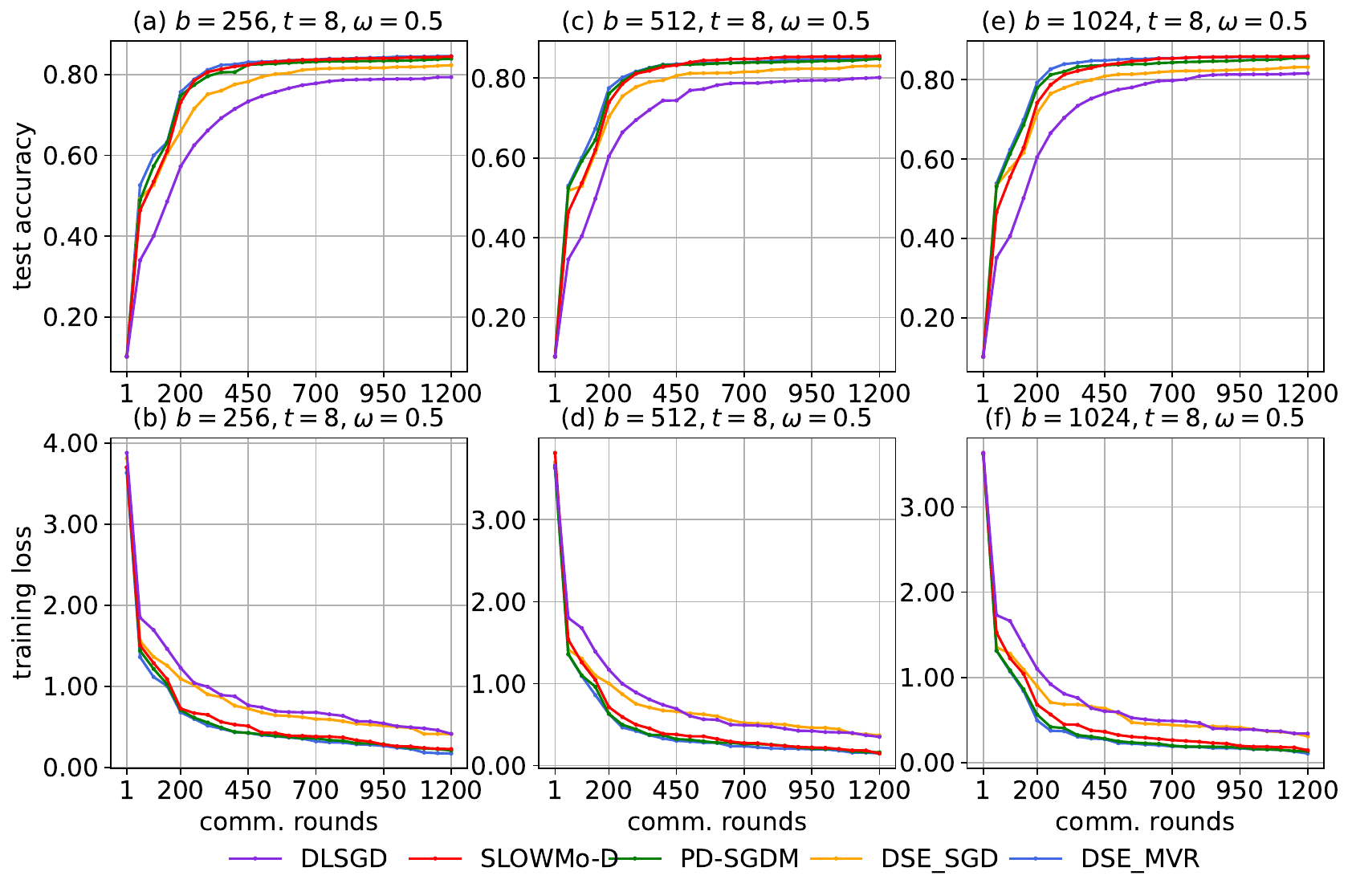}
  }
  \
  \subfigure[$\tau=20, \omega=0.5$]{
  \includegraphics[scale=0.29]{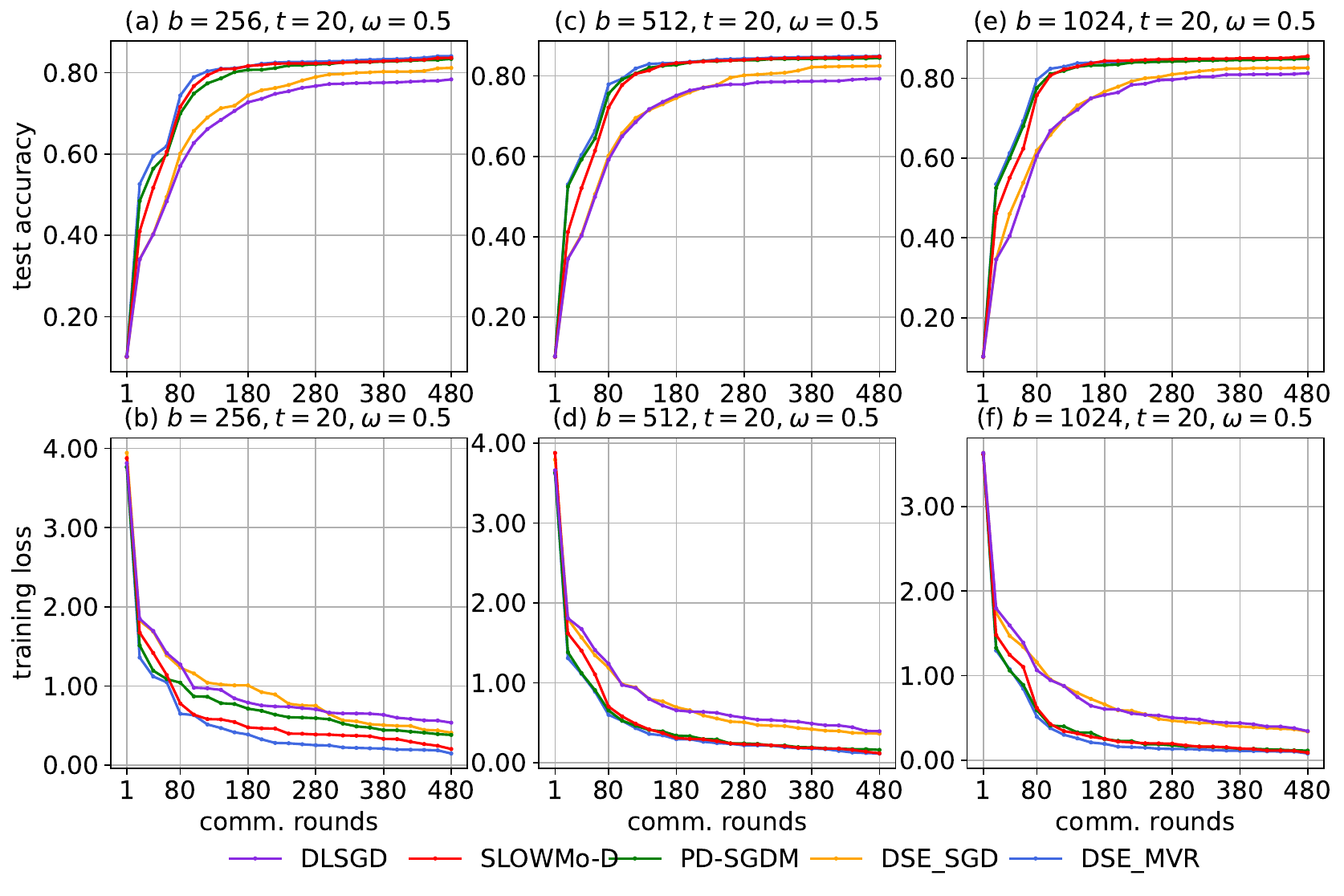}
  }
  \
  \subfigure[$\tau=4, \omega=10$]{
  \includegraphics[scale=0.29]{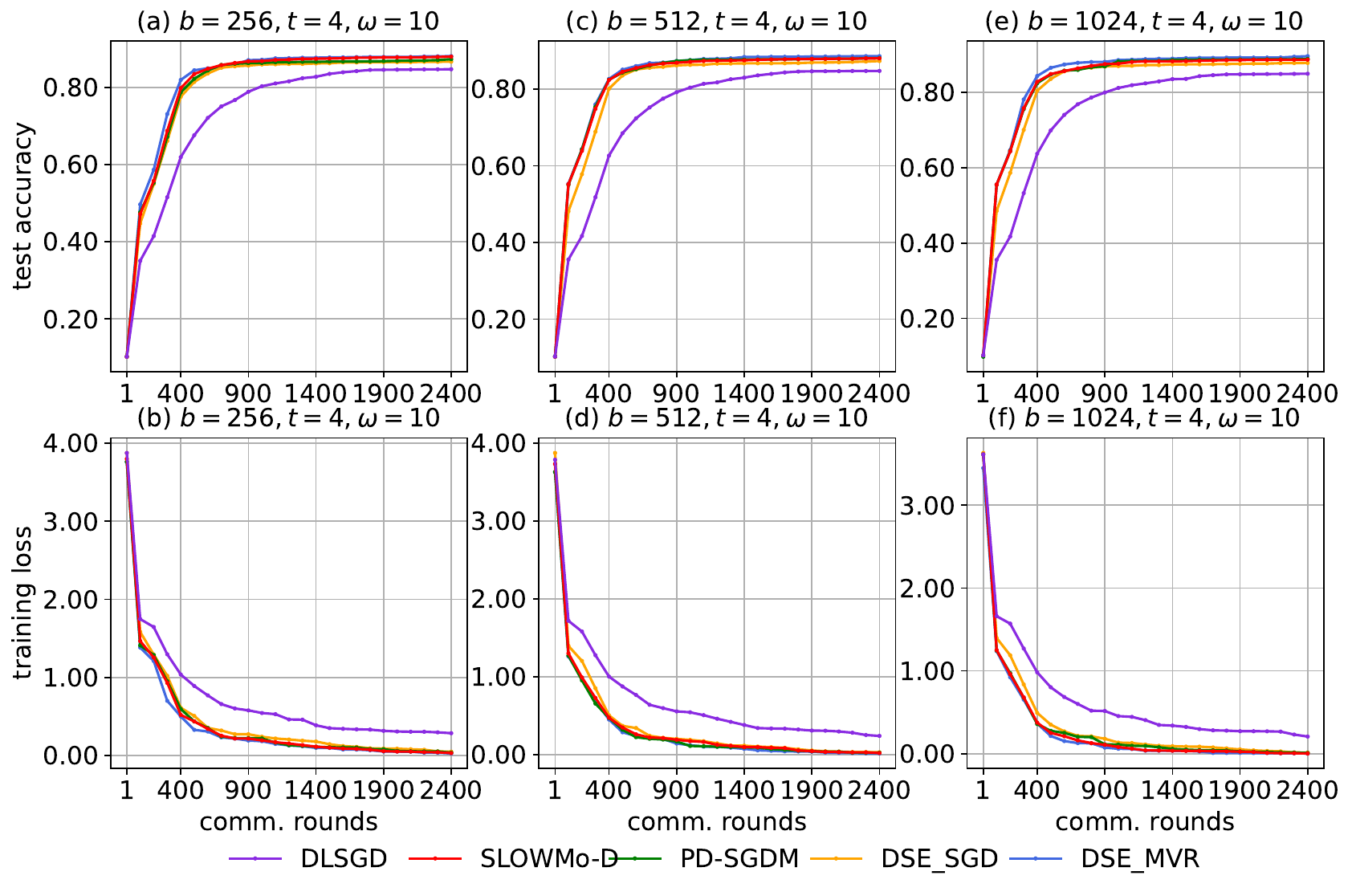}
  }
  \
  \subfigure[$\tau=8, \omega=10$]{
  \includegraphics[scale=0.29]{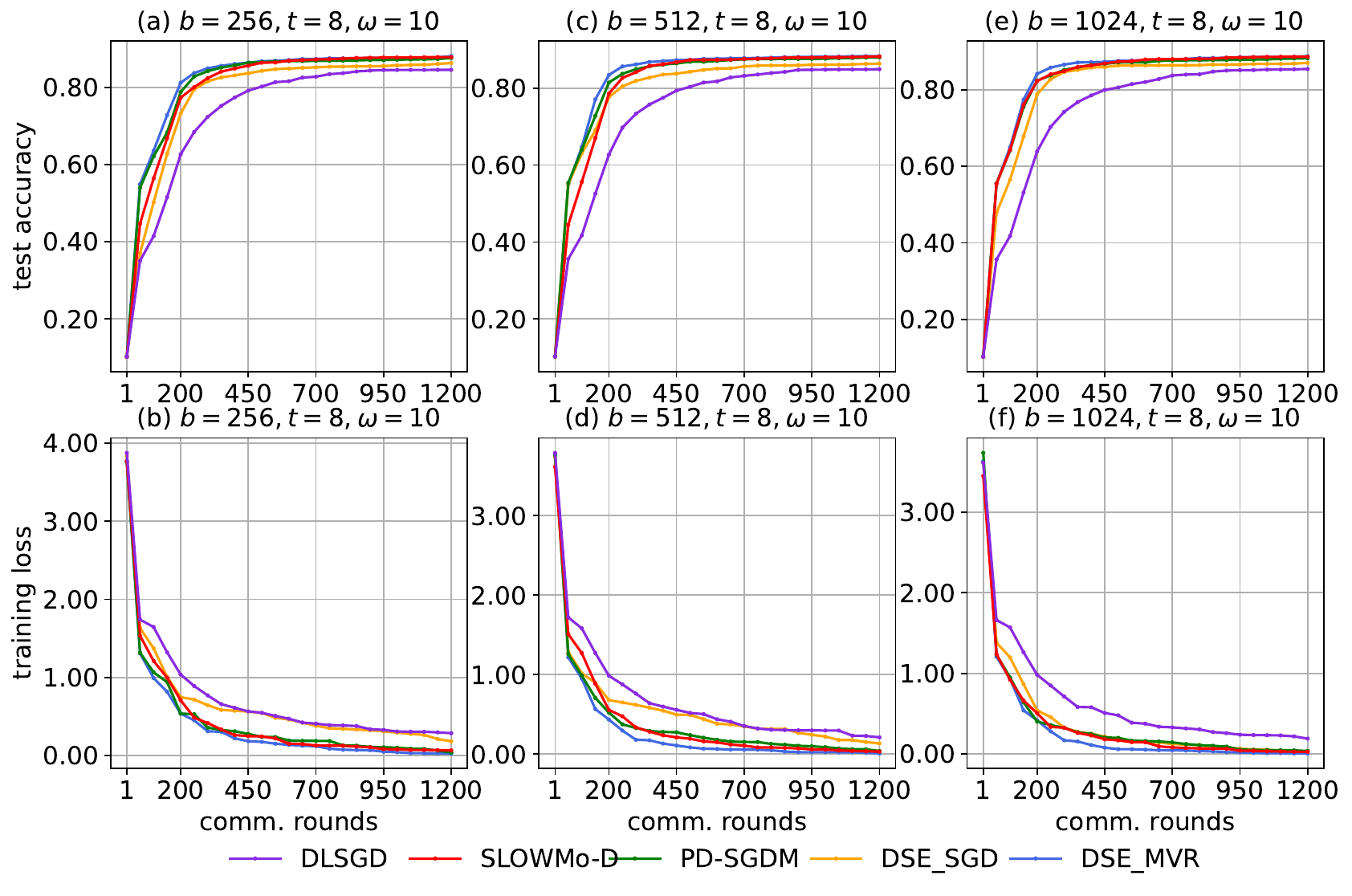}
  }
  \
  \subfigure[$\tau=20, \omega=10$]{
  \includegraphics[scale=0.29]{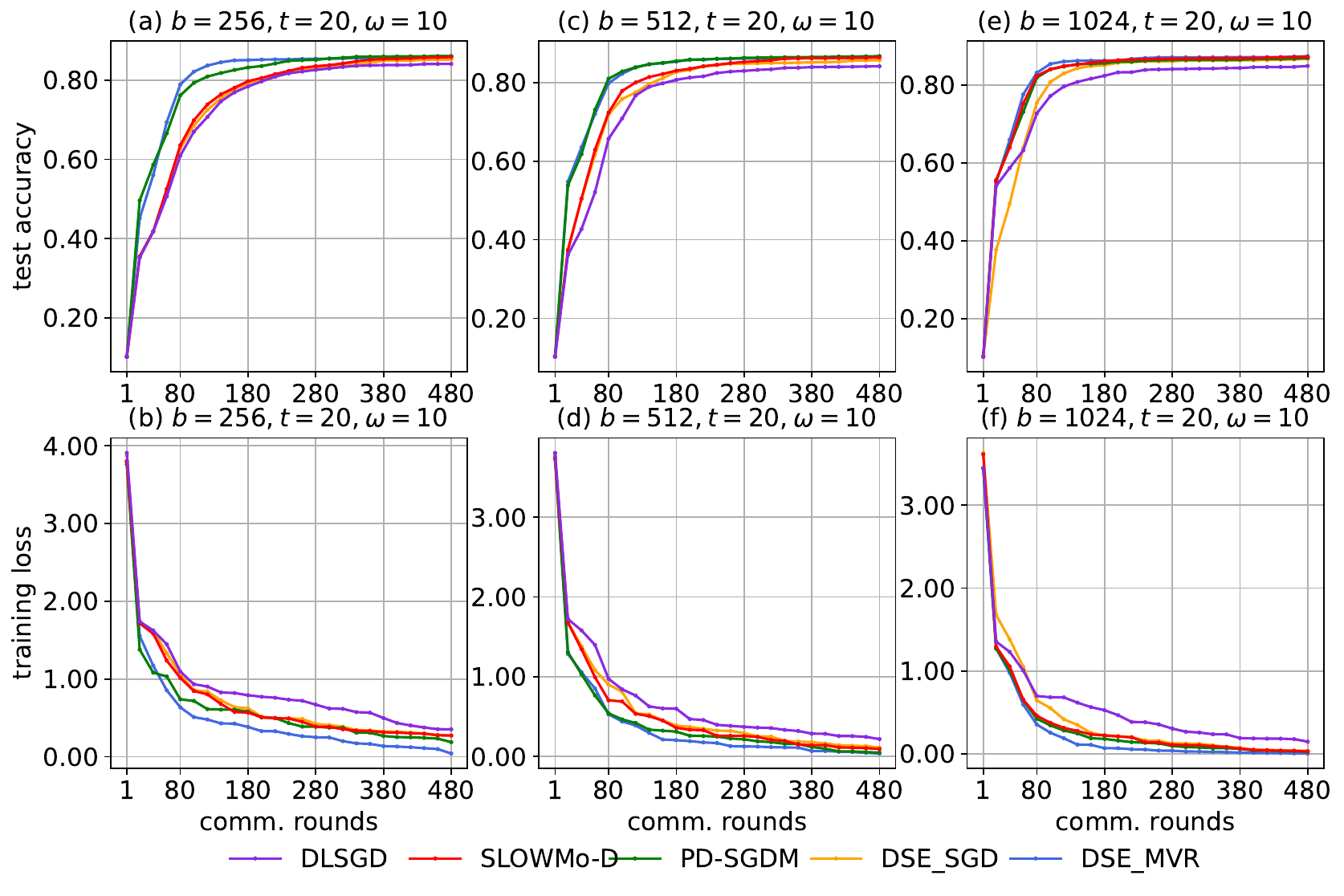}
  }
  \caption{Full learning curves w.r.t test accuracy and training loss over CIFAR-10 dataset, which are  averaged over 3 random seeds.}
  \label{fig:CIFAR10_full_res}
\end{figure}

\end{document}